\documentstyle[version2,eqsecnum,aps]{revtex}

\input psfig.tex

\begin{document}
\draft
\begin{title}
Measuring gravitational waves from binary black hole coalescences:\\
I. Signal to noise for inspiral, merger, and ringdown.
\end{title}
\author{\'Eanna \'E.\ Flanagan}
\begin{instit}
Newman Laboratory, Cornell University, Ithaca, NY 14853-5001.
\end{instit}
\author{Scott A.\ Hughes}
\begin{instit}
Theoretical Astrophysics, California Institute of Technology,
Pasadena, California 91125.
\end{instit}

\begin{date}
\newline
\end{date}

\begin{abstract}

We estimate the signal-to-noise ratios (SNRs) that one would expect to
measure from coalescing binary black hole (BBH) systems for the
following three broadband gravitational-wave observatories: initial
and advanced ground-based interferometers (LIGO/VIRGO) and space-based
interferometers (LISA).  We focus particularly on the highly
relativistic and nonlinear {\it merger} portion of the
gravitational-wave signal, which comes after the adiabatic {\it inspiral}
portion and before the {\it ringdown} portion due to the quasinormal
ringing of the final Kerr black hole.

Ground-based interferometers can do moderate SNR (a few tens),
moderate accuracy studies of the dynamics of merging black holes in
the mass range $({\rm a~few}) M_\odot$ to $\sim 2000 M_\odot$.
LISA, by contrast, can do high SNR (a few $\times\,10^4$),
high-accuracy studies of BBH systems in the mass range $10^5 M_\odot\alt
(1+z) M \alt 10^8 M_\odot$, where $z$ is the binaries' cosmological
redshift.

Our estimated SNRs suggest that coalescing black
holes might well be the first sources detected by the LIGO/VIRGO
network of ground-based interferometers.  Because of their larger
masses, they can be seen out to much greater distances
(up to $\sim 250 \, {\rm Mpc}$ for $M \alt 50 M_\odot$ for the initial
LIGO interferometers) than
coalescing neutron star binaries (heretofore regarded as the ``bread
and butter'' workhorse source for LIGO/VIRGO, visible to $\sim 30 \,
{\rm Mpc}$ by the initial LIGO interferometers).

Low-mass BBHs ($M \alt 30 M_\odot$ for the first LIGO
interferometers; $M \alt 80 M_\odot$ for the advanced; $(1+z) M \alt 3
\times 10^6 M_\odot$ for LISA) are best searched for via their
well-understood inspiral waves; more massive BBHs must be
searched for via their more poorly understood merger waves and/or
their well-understood ringdown waves.
A search for low-mass BBHs based on the inspiral waves, at a
sensitivity level roughly half way between the first LIGO
interferometers and the advanced LIGO interferometers, should
be capable of finding BBHs out to $\sim 1 \, {\rm Gpc}$.
A search for massive BBHs based on the ringdown waves can be
performed using the method of matched filters.  If one wants the
reduction in the event rate due to the discreteness of the template
family to be no more than $10\%$, then the number of independent
templates needed in the search will only be
about $6000$ or less.  Such a search with the first LIGO
interferometers should be capable of finding BBHs in the mass range
from about $100M_\odot$ to about $700 M_\odot$ out to $\sim 200 \,
{\rm Mpc}$; with advanced LIGO interferometers, from about $200 M_\odot$
to about $3000 M_\odot$ out to $z \sim 1$; and with LISA, BBHs with $10^6
M_\odot \alt (1+z) M \alt 3 \times 10^8 M_\odot$ should be visible out
to $z \agt 100$.  The effectiveness of a search based on the merger
waves will depend on how much one has learned about the merger waveforms
from numerical relativity simulations.  With only a knowledge of the
merger waves' range of frequency bands and range of temporal durations,
a search based on merger waves can be performed using a nonlinear filtering
search algorithm.  Such a search should increase the number of
discovered BBHs by a factor of roughly $10$ over those found from the
inspiral and ringdown waves.  On the other hand, a full set of merger
templates based on numerical relativity simulations could further
increase the number of discovered BBHs by a additional factor of up to
about $4$.

\end{abstract}

\bigskip
\pacs{PACS numbers: 04.80.Nn, 04.25.Dm, 04.30.Db, 95.55.Ym}

\newpage
\narrowtext \twocolumn

\section{INTRODUCTION AND SUMMARY}
\label{intro}

\subsection{Coalescences of black hole binaries}

It has long been recognized that coalescences of binary systems of two
black holes could be an important source of gravitational waves
\cite{300yrs,schutzreview}, both
for the ground based interferometric detectors LIGO \cite{ligoscience}
and VIRGO \cite{virgo} currently under construction, and also for the
possible future space-based interferometer LISA
\cite{cornerstone,lisa,LISAreport}.  The orbits of binary black holes
(BBHs) gradually decay from energy and angular momentum loss to
gravitational radiation.  Eventually,
the holes coalesce to form a final black hole.  For gravitational
radiation reaction to successfully drive the binary to merge in
less than a Hubble time, the initial orbital period must be $\alt
0.3 \, {\rm days} \, (M/M_\odot)^{5/8}$, where $M$ is the total mass of
the binary; thus the critical orbital period is of order days
for solar mass black holes, and of order years to hundreds of
years for supermassive black holes ($10^6 M_\odot \alt M \alt 10^9
M_\odot$).

The process of coalescence can be divided up into three more or less
distinct phases:
\begin{itemize}

\item
An adiabatic {\it inspiral}, during which the gravitational radiation
reaction timescale is much longer than the orbital period.  The inspiral
ends when the binary orbit becomes relativistically dynamically unstable at
an orbital separation of $r \sim 6 M$ (in units where $G=c=1$)
\cite{kww,cook}.  The gravitational waves from the inspiral carry encoded
within them the masses and spins of the two black holes, some of the orbital
elements of the binary and the distance to the binary \cite{300yrs,Kipreview}.

\item
Towards the end of the inspiral, the black holes encounter the
dynamical instability and make a gradual
transition from a radiation-reaction driven inspiral to a
freely-falling plunge \cite{kww,KipAmos,laiwiseman}, after which, even
if the radiation reaction could be turned off, the black holes would still
merge.  We will call the subsequent plunge and violent collision the
{\it merger} phase.  Gravitational waves from the merger
could be rich with information about the dynamics of relativistic
gravity in a highly nonlinear, highly dynamic regime about which
we have a poor theoretical understanding today.

\item
As the final black hole settles down to a stationary Kerr state, the
nonlinear dynamics of the merger gradually become more and more
describable as oscillations of this final black hole's quasinormal
modes \cite{pressteukolsky,chandradetweiler}.  The corresponding
emitted gravitational waves consist of a superposition of
exponentially damped sinusoids.  We will call the phase of the coalescence
for which the emitted gravitational-wave signal is dominated by the
strongest $l=m=2$ quasinormal mode signal the {\it ringdown} phase.
The waves from the ringdown carry information about the mass and spin
of the final black hole \cite{echeverria,finnmeasure}.
(Note that, for want of a better terminology,
throughout this paper we consistently use {\it coalescence} to refer
to the entire process of inspiral, merger and ringdown, and reserve
the word merger for the phase intermediate between inspiral and ringdown.)

\end{itemize}

In this paper we focus primarily on BBHs in which the masses of the
two black holes are approximately the same, although we do
also consider sources where one BH is much smaller than the other.
We consider three different classes of BBHs:

\medskip
(i) {\it Solar mass} black hole binaries, {\it i.e.}, binaries that are
formed from massive main-sequence progenitor binary stellar systems.  Such
BBHs are expected to have total masses in the range $10 M_\odot
\alt M \alt 50 M_\odot$, but not much larger than this.  The rate of
coalescence of solar-mass BBHs in the Universe is not very well known.
However, theory suggests that most of the BBH progenitor systems may
not disrupt during the stellar collapses that produce the black holes, so
that their coalescence rate could be about the same as the birth rate
for their progenitors, about $1/100,000$ years in our Galaxy, or
several per year within a distance of $200 \, {\rm Mpc}$
\cite{narayan,phinney,heuvel,tutukov,yamaoka}.  Note that this
coalescence rate is roughly the same as the expected event rate for
what has traditionally been regarded as the most promising source for
ground based interferometers, coalescences of neutron star---neutron star
(NS-NS) binaries \cite{ligoscience,Kipreview}.  The expected rate of NS-NS
coalescences is more firmly known, however, since it is based on
extrapolations from detected progenitor NS-NS systems
\cite{narayan,phinney,heuvel}.

\medskip
(ii) {\it Intermediate mass} black hole binaries, with total masses in
the range $50 M_\odot \alt M \alt ({\rm a~few}) \times 10^3 M_\odot$.
In contrast to the cases of solar mass black holes and supermassive
black holes (discussed below), there is little direct observational
evidence for the existence of black holes in this mass range.
Although there have been suggestions that the globular cluster
M15 harbors a black hole of mass $\sim 10^3 M_\odot$ \cite{postnov1},
theoretical modeling combined with the most recent HST observations
neither confirm nor rule out this possibility \cite{M15}.  Despite the
lack of observational evidence, it is plausible that black holes in this
mass range are formed in the cores of globular clusters, or in
galactic nuclei in the process of formation of a supermassive
black hole \cite{Rees}.  Simulations by Quinlan and Shapiro suggest
that black holes with $M \sim 100 M_\odot - 1000 M_\odot$ could be formed
in the evolution of dense stellar clusters of main sequence stars in
galactic nuclei \cite{QuinlanShapiro}, and that coalescences of binaries
of such black holes could be possible en route to the formation of a
supermassive black hole.

We are prompted to consider these intermediate mass BBHs by the following
consideration: even if the coalescence rate of intermediate mass
BBHs is $\sim 10^{-4}$ that of NS-NS binaries (which is thought to be
$\sim 10^{-5} \, {\rm yr}^{-1}$ in our Galaxy as discussed above),
these sources would still be seen more often than NS-NS binaries by
LIGO's initial and advanced interferometers, and thus could be the
first detected type of source.  (See Sec.~\ref{results} for further details.)

\medskip
(iii) {\it Supermassive} black holes binaries: There is a variety of
strong circumstantial evidence that supermassive black holes (SMBHs)
in the mass range $10^6 - 10^9 M_\odot$ are present in quasars and
active galactic nuclei, and also that $\sim 25\%$ -- $ 50\%$ of nearby
massive spiral and elliptical galaxies harbor quiescent SMBHs
\cite{LISAreport,BlandfordRees}.  See Ref.~\cite{Rees} for a
review of this evidence.  One of the main scientific goals of the LISA
project is to detect and monitor various processes involving SMBHs,
such as the capture of compact stars
\cite{schutzreview,LISAreport,Kipreview,HilsBender,Ericnew},
and their formation \cite{schutzreview,LISAreport}.
In particular, the coalescences of SMBH binaries that are formed in
Galaxy mergers, in which the individual SMBHs are driven together by
dynamical friction and gas accretion until gravitational radiation
reaction takes over \cite{Blandford}, have often been suggested as a
promising source for space-based interferometers
\cite{300yrs,schutzreview,LISAreport,Kipreview,Wahlquist,Haehnelt}.
Such coalescences would be detectable throughout the observable
universe with large signal to noise ratios
\cite{LISAreport,Kipreview}.  There is also observational evidence for
SMBH binaries: wiggles in the radio jet of QSO 1928+738 have been
attributed to the orbital motion of a SMBH binary
\cite{Roos}, as have time variations in quasar luminosities
\cite{sillanpana} and in emission line redshifts \cite{gaskell}.  The
overall event rate is uncertain, but could be large ($\agt 1/{\rm
yr}$), especially if the hierarchical scenario for structure formation
is correct \cite{Haehnelt}.

\subsection{Status of theoretical calculations of the
gravitational-wave signal}
\label{calcstatus}

Detailed theoretical understanding and predictions of the
gravitational waveforms $h_+(t)$ and $h_\times(t)$ produced in BBH
coalescences will facilitate both the detection of the
gravitational-wave signal, and the extraction of its useful
information.  In situations where a complete family of theoretical
template waveforms is available, it will be possible to use the
procedure of Wiener optimal filtering to search the interferometer
data streams and to detect the gravitational-wave signal
\cite{300yrs,schutz1}.  The resulting signal-to-noise ratios (SNRs)
can be larger than those obtainable without theoretical templates by a
substantial factor; see Sec.~\ref{derivesnrformulagen}.
Thus, while it is possible to detect the various phases of BBH
coalescences without theoretical
templates, such templates can greatly increase the effective range of
the interferometers and the event detection rate.  (Accurate theoretical
templates are also essential for extracting the maximum amount of
information from detected signals \cite{paperII}.)

Such theoretical template waveforms are available for the inspiral and
ringdown phases of the coalescence, but not yet for the merger phase,
as we now discuss.

For the inspiral phase of the coalescence, the gravitational waves and
orbital evolution can be described reasonably well using the
post-Newtonian approximation to general relativity.  To date, inspiral
waveforms have been calculated up to post-2.5-Newtonian order
\cite{Blanchet}, and the prospects look good for obtaining waveforms
up to post-3.5-Newtonian order \cite{Blanchet1,ordercount}.  Post-Newtonian
templates will be fairly accurate over most of the inspiral, the most
important error being a cumulative phase lag \cite{prl,ericaccuracy}.
This cumulative phase lag will not be important for searches for inspiral
waves; template phasing error will be largely compensated for by
systematic errors in best-fit values of the binary's parameters, and
the signals will still be found
\cite{prl,bala1,cutlerflan2,searchtemplates}.  By contrast, template
inaccuracies will be significant when one attempts to extract from the
data the binary's parameters.
In particular, post-Newtonian templates' errors start to become
very significant around an orbital separation of $r \sim 12 M$
\cite{PNbreakdown}, well before the end of the inspiral at the
dynamical orbital instability ($r \sim 6 M$).  Templates for the
phase of the inspiral between roughly $12 M$ and $6 M$ will most likely
have to be calculated using methods other than the post-Newtonian
approximation.  Moreover, the methods of full blown numerical
relativity cannot be applied to this ``Intermediate Binary Black
Hole'' (IBBH) phase, since the total time taken to evolve from $12 M$
to $6 M$ is about $1500 M$, too long for supercomputer simulations to
even contemplate evolving.  Alternative analytic and numerical methods
for calculating gravitational waveforms from the IBBH portion of BBH
inspirals, based on the adiabatic approximation, are under development
\cite{IBBHworkshop}; it is likely that such alternative
methods will be successfully developed and implemented before the
gravitational-wave detectors begin their measurements \cite{IBBH1}.

Waveforms from the dynamic, complicated merger phase of the
coalescence can only be obtained from numerical relativity.  Unlike
mergers of neutron star binaries, BBH mergers are particularly clean in
the sense that there is no microphysics or hydrodynamics to complicate
simulations of the evolution, and external perturbations are negligible:
the entire merger can be described as a solution to the vacuum Einstein
equation \cite{notsoclean}.  Finding that solution is not a particularly easy
task: a major computational effort to evolve the vacuum Einstein equation
for BBH mergers using massive computational resources is currently
underway, funded by the National Science Foundation's Grand Challenge
program \cite{grandchall,samreview}.

The ringdown phase of the coalescence can be accurately described using
perturbation theory on the Kerr spacetime background \cite{chandra}.
The gravitational waveforms from this phase are well understood, being
just exponentially damped sinusoids.  Thus, Wiener optimal filtering
is feasible for searches for ringdown waves.

\subsection{Purpose of this paper}
\label{whyarewehere}

The principal purpose of this paper is to estimate, in more detail
than has been done previously, the prospects for measuring
gravitational waves from the three different phases of coalescence
events (inspiral, merger and ringdown), for various different
detectors, and for a wide range of BBH masses.  We estimate in each
case the distances to which the different types of source can be seen
by calculating expected SNRs.  In particular, we determine for each
BBH mass and each detector whether a coalescence event is most
effectively detected by searching for the inspiral portion of the
signal, or the merger portion, or the ringdown portion.  We also
determine how much the availability of theoretical template waveforms
for the merger phase could increase the event detection rate.

Previous estimates of SNRs for ground-based interferometers have
focused on the inspiral \cite{300yrs,prl} and ringdown
\cite{echeverria,finnmeasure} phases, and also focused on solar-mass
BBHs.  For space-based interferometers, previous estimates of SNRs
from the merger phase \cite{LISAreport,Kipreview} were restricted to
specific mass values and did not consider the ringdown portion of the
signal.

In a companion paper, we discuss in detail the useful information carried
by the three phases of the gravitational-wave signal, and methods and
prospects for extracting this information both with and without templates
for the merger phase \cite{paperII}.

In the following subsection we describe our calculations and summarize
the assumptions underlying our estimated signal-to-noise ratios.  In
Sec.~\ref{results} we summarize our main results and conclusions,
and an outline of the paper is given in Sec.~\ref{map}.

\subsection{Estimating the signal-to-noise ratios: method and
assumptions}
\label{assumptionsec}

We calculate SNRs for three different types of interferometer:
initial and advanced ground-based interferometers (LIGO/VIRGO), and
the proposed space-based interferometer LISA.  The noise spectra of the
initial and advanced ground-based interferometers we took from
Ref.~\cite{ligoscience}, and that for LISA we obtained from
Ref.~\cite{LISAreport}.  Our approximate versions of these noise
spectra are given in
Eqs.~(\ref{noisespec}) -- (\ref{noisespecLISA}), and are
illustrated in Figs.~\ref{hcharfig} -- \ref{hcharfig2} in Sec.~\ref{egs}.

We consider the following three different signal-detection methods:

\medskip
(i) {\it Matched filtering searches:}  For those phases of the coalescence for
which a complete set of theoretical templates will be available (the inspiral,
the ringdown, and possibly the merger), the method of matched filtering
or optimal filtering can be used to search for the waves
\cite{300yrs,schutz1,helstrom,wainandzub,cutlerflan1}.
The theory of matched filtering is briefly sketched in
Sec.~\ref{derivesnrformula1}.  For any source of waves, the SNR,
$\rho$, obtained from matched filtering is related to the
gravitational waveform $h(t)$ measured by the interferometer and to
the spectral density $S_h(f)$ of the strain noise in the
interferometer via
\cite{twosided}
\begin{equation}
\label{snr0}
\rho^2 = 4 \int_0^\infty {| {\tilde h}(f)|^2 \over S_h(f)} df,
\end{equation}
where ${\tilde h}(f)$ is the Fourier transform of $h(t)$ defined by
Eq.~(\ref{fourier}).  The SNR (\ref{snr0}) depends, through the
waveform $h(t)$, on the orientation and position of the source
relative to the interferometer.  In Sec.~\ref{derivesnrformula}
we show that if we perform an rms average over source orientations and
positions (at a fixed distance), the rms SNR thus obtained depends only
on the energy spectrum $dE/df$ carried off from the source by the
gravitational waves.  The resulting relationship
between the waves' energy spectrum and
the rms angle-averaged SNR forms the
basis for most of our calculations.  It is given by
[{\it cf}.~Eq.~(\ref{snraveraged})]
\begin{equation}
\langle \rho^2 \rangle = {2 (1 + z)^2 \over 5 \pi^2 D(z)^2} \,
\int_0^\infty df \, {1 \over f^2 S_h(f)} \, {d E \over d f}[(1 +
z)f],
\label{snraveraged0}
\end{equation}
where $z$ is the source's cosmological redshift and $D(z)$ is the
luminosity distance from the source.

In order for a signal to be detected, the waves' measured SNR must be
larger than the detection threshold
\begin{equation}
\rho_{\rm threshold} \approx \sqrt{ 2 \ln( {\cal N}_{\rm start-times})
+ 2 \ln( {\cal N}_{\rm shapes})};
\label{MFthreshold}
\end{equation}
see, for example, Ref.~\cite{prl} and also Sec.~\ref{nonlinear}.
Here ${\cal N}_{\rm start-times}$ is the number of independent starting
times of the gravitational wave signal that are searched for in the
data set, determined by the total duration of the data set (of order
one year) and the sampling time.  The quantity ${\cal N}_{\rm shapes}$
is the total number of statistically independent waveform shapes
in the set of signals that one is searching for \cite{importantnote}.

\medskip
(ii) {\it Band-pass filtering searches:} For the merger phase, a complete
set of theoretical templates may not be available.  It therefore may not be
possible to use matched filtering, and other search methods will need to
be employed.  Band-pass filtering, followed by setting a detection threshold
in the time domain, is a simple method of searching an interferometer data
stream for bursts of unknown form \cite{schutz1}.  In
Sec.~\ref{derivesnrformula1} we derive an approximate relation between
the SNR obtainable from band-pass filtering, and the SNR (\ref{snr0})
obtainable
from matched filtering, for any burst of waves, namely
\begin{equation}
\left({S \over N} \right)_{\rm band-pass} \approx {1 \over \sqrt{2 T
\Delta f}} \, \left({S \over N} \right)_{\rm optimal}.
\label{bpa}
\end{equation}
Here $T$ is the duration of the burst and $\Delta f$ is the bandwidth
of the band-pass filter [{\it cf}. Eq.~(\ref{simp1})].  The
quantity $2 T \Delta f$ is the dimension of the linear space of
signals being searched for, which is roughly the same as the ``number
of cycles'' of the gravitational waveform.  We use this formula in
Sec.~\ref{detectmerger} to estimate band-pass-filter search SNRs
for the merger waves from BBH coalescences, by inserting on the right
hand side the rms angle-averaged matched-filter search SNR
(\ref{snraveraged0}), and by making estimates of $T$ and $\Delta f$.

\medskip
(iii) {\it Noise-monitoring, nonlinear filtering searches:} The traditional
view has been that the SNR (\ref{bpa}) is about the best that can be achieved
in the absence of theoretical templates; that is, that the gain in SNR
obtainable from matched filtering is approximately the square root of
the number of cycles
in the gravitational wave signal.  This view is based on the the assumption
that the search method used in the absence of templates is band-pass
filtering or something very similar.  However, we suggest in
Sec.~\ref{nonlinear}
an alternative search method, motivated by Bayesian analyses and
incorporating nonlinear filtering, which performs much better than
band-pass filtering and in some cases almost as well as matched filtering.
In essence, one monitors the noise level in the interferometer in a certain
frequency band, over short timescales, and looks for statistically
significant changes.  The noise level is estimated by calculating the quantity
\begin{equation}
{1 \over T} \int_{-T/2}^{T/2} d \tau \, s(t + \tau)^2,
\end{equation}
where $T$ is the maximum expected duration of the signal, and $s(t)$
is a suitably pre-filtered version of the interferometer data stream.

The efficiency of this noise-monitoring search method cannot usefully be
described in terms of a signal-to-noise ratio, since the detection
statistic is very non-Gaussian.  Instead, its efficiency can be
described in the following way.  Let $\rho$ denote the SNR that would
be obtained if matched filtering were possible [Eq.~(\ref{snr0})].  We use
$\rho$ as a convenient parameterization of the signal strength; as such,
it is meaningful even in situations where matched filtering cannot be
carried out.  Then, a signal will be detected with high confidence using
the noise-monitoring technique whenever $\rho$ is larger than the
threshold $\rho_*$, where to a good approximation $\rho_*$ satisfies
\FL
\begin{equation}
\rho_*^2 = 2 \ln( {\cal N}_{\rm start-times}) + {\cal N}_{\rm bins}
\ln \left( 1 + \rho_*^2 / {\cal N}_{\rm bins} \right).
\label{NLthreshold}
\end{equation}
Here ${\cal N}_{\rm bins} = 2 T \Delta f$ is the dimension of the
signal space discussed above (or the number of independent frequency bins
in the Fourier domain).  The derivation of Eq. (\ref{NLthreshold}) is
given in Sec.~\ref{nonlinear}.

The relation (\ref{snraveraged0}) forms the basis of our SNR
calculations.  We use the SNR thresholds (\ref{MFthreshold}) and
(\ref{NLthreshold}) to deduce from the SNR values the detectability
of the various parts of the gravitational-wave signal.  To calculate
the SNRs, we also need to specify the waves' energy spectra for the
three different phases of the coalescence.  As we now outline, for the
inspiral and ringdown phases the waves' energy spectrum is essentially
known, while for the merger phase we make an educated guess of $dE/df$.
Sec.~\ref{signalassumptions} gives more details.

\medskip
{\it Inspiral energy spectrum}: We use the leading order expression
for $dE/df$ obtained using Newtonian gravity supplemented with the
quadrupole formula \cite{shapteuk} [Eq.~(\ref{dEdfinspiral})].
Strictly speaking, this spectrum describes the SNR that would be
achieved by searching for Newtonian, quadrupole waves using Newtonian,
quadrupole templates.  The actual SNR obtained when searching for a
real, general-relativistic inspiral waveform using post-Newtonian
templates should deviate from this by only a few tens of percent
{\cite{spectrumcomment}}.  We terminate the spectrum at
the frequency $f_{\rm merge} = 0.02/M$ which is
(roughly) the frequency of quadrupole waves emitted at the orbital
dynamical instability at $r\sim 6 M$ \cite{kww}.  For LISA, we assume
that the measurement process lasts at most one year, and choose the
frequency at which the inspiral spectrum starts accordingly.

\medskip
{\it Ringdown energy spectrum}: The spectrum that we use
[Eq.~(\ref{dEdfqnr})] is
determined (up to its overall amplitude) by the characteristics of the
$l=m=2$ quasi-normal ringing (QNR) mode of
the final Kerr black hole.  This mode is the most slowly damped of all QNR
modes, so we expect it to dominate the last stages of the
gravitational-wave emission.  The QNR spectrum depends on three parameters:
the quasi-normal modes' frequency $f_{\rm qnr}$ and damping time
$\tau$, and the overall amplitude of the quasinormal mode signal.
Equivalently, the three parameters can be taken to be
the mass $M$ and dimensionless spin parameter
$a$ of the final black hole (which determine $f_{\rm qnr}$ and
$\tau$) and the total amount of energy radiated in the
ringdown (which determines the overall amplitude).  The spectrum is
peaked at $f=f_{\rm qnr}$ with width $\Delta f \sim 1/\tau$.

In our analyses, we (somewhat arbitrarily) assume that $a=0.98$.
It seems likely that in many coalescences the spin of the final black
hole will be close to maximal, since the total angular momentum of the
binary at the end of the inspiral is $\sim 0.9 M^2$ when the
individual black holes are non-spinning \cite{explainorb}, and the
individual black hole spins can add to this.  Exactly how close to
maximally spinning the final black hole will be is a matter that
probably will not be decided until supercomputer simulations---or
observations---settle the issue.  In any case, the ringdown SNR values
that we obtain depend only weakly on our assumed value of $a$
[{\it cf}.~Eq.~(\ref{qnrsnrII})], for fixed total energy radiated in
the ringdown.

The overall amplitude of the ringdown signal depends upon one's
delineation of where ``merger'' ends and ``ringdown'' begins, which is
somewhat arbitrary.  For equal-mass BBHs, we assume a value of the
overall amplitude that corresponds to a total radiated energy in the
ringdown of $0.03 M$---a radiation efficiency of $3\%$.  This
number is based on a back-of-the-envelope, quadrupole-formula-based
estimate of the QNR mode's amplitude when the distortion of the horizon of
the black hole is of order unity ({\it cf}.~Sec.~\ref{ringdownphase}).
Although this radiation efficiency seems rather high, there have been
numerical simulations of the evolution of distorted, spinning black holes
in which the ringdown waves carry away $\agt 3\%$ of the black hole's
total mass energy \cite{seidel}.

For non equal-mass black holes, we assume that the total energy
radiated is $F(\mu/M) 0.03 M$, where $F(\mu/M) = (4 \mu/M)^2$ and
$\mu$ is the reduced mass of the binary.  This function gives the
correct results for the equal-mass case and also gives the correct
scaling law 
in the regime $\mu \ll M$; for general mass ratios the scaling law is
probably a good approximation.

\medskip
{\it Merger energy spectrum}: Realistic merger energy spectra will vary
substantially from event to event (depending on the initial spins of the
inspiraling black holes).   Currently, we have very little concrete information
about such spectra, pending supercomputer simulations of BBH mergers.  In
Sec.~\ref{mergerphase} we describe various circumstantial pieces of evidence,
culled from the literature, relevant to merger spectrum.  Based on that
evidence, we adopt the following crude model for equal-mass BBHs: we assume
a flat spectrum $dE/df = {\rm const}$ extending from the frequency
$f_{\rm merge}=0.02/M$ of quadrupole waves at the end of inspiral to the
quasinormal ringing frequency $f_{\rm qnr} =0.13/M$, with amplitude such
that the total radiated energy in the merger is $10\%$ of the total mass
energy of the spacetime [Eq.~(\ref{dEdfmerger})].  We outline in
Sec.~\ref{mergerphase} two different ``handwaving'' arguments which suggest
that in favorable cases the merger radiation efficiency may be
as high as our assumed value of $\sim 10\%$.  One of these arguments,
due originally to Smarr \cite{smarrthesis} and explored by Detweiler
\cite{detweiler2}, is based on extrapolation of perturbation theory
results; the other argument is based on angular momentum conservation.

Our assumed radiation efficiencies of $3\%$ and $10\%$ for the
ringdown and merger phases should perhaps be interpreted as reasonable
upper bounds that could be achieved in favorable cases, rather than as
best-guess estimates.  We note that numerical simulations that have been
performed to date (which are restricted to axisymmetric situations) generally
yield lower radiation efficiencies than we have assumed \cite{baker}; moreover,
these axisymmetric simulations generally find that ringdown waves carry most
of the radiated energy.  In Sec.~\ref{mergerphase} we argue that the
radiated energy in the merger phase could be boosted by the lack of symmetry
in generic black hole mergers, and especially by the individual black holes'
spins
(if these spins are large).

For non-equal mass BBHs, we again simply reduce the energy spectrum by
the factor $F(\mu/M) = (4 \mu/M)^2$, while the upper and lower
frequencies $f_{\rm merge}$ and $f_{\rm qnr}$ are taken to be
independent of $\mu$.

\subsection{Signal-to-noise ratios: results and implications}
\label{results}

By inserting our assumed energy spectra (\ref{dEdfinspiral}),
(\ref{dEdfmerger}) and (\ref{dEdfqnr}) into Eq.~(\ref{snraveraged0}),
we obtain optimal-filtering SNRs for the three different phases
of BBH coalescences as a function of the redshifted total mass $(1+z)
M$ of the binary.  The results are shown in
Figs.~\ref{initialligosnr}, \ref{advancedligosnr} and \ref{lisasnr}
and formulae summarizing the results are given in Appendix
\ref{appsnr}.  Also in Sec.~\ref{mergernofilters} we estimate that, for
the merger waves, the number of independent frequency bins ${\cal
N}_{\rm bins}$ which characterize the signal falls roughly in the range
$10 \alt {\cal N}_{\rm bins} \alt 30$, and that a conservative upper-bound
is $\sim 60$.  We use this upper bound in Sec.\ \ref{detectmerger} to
estimate the SNR threshold (\ref{NLthreshold}) for merger waves
using noise-monitoring searches when templates are unavailable.
We discuss the implications of these SNRs and SNR thresholds in
Sec.~\ref{detection}; here we summarize our main conclusions:

\begin{itemize}

\item Ground-based interferometers can study black-hole mergers in the
mass range $({\rm a~few}) M_\odot$ to $\sim 2000 M_\odot$; LISA, by
contrast can study mergers in the mass range $10^5 M_\odot \alt (1+z) M
\alt 10^8 M_\odot$.

\item Ground-based interferometers can do moderate SNR (a
few tens), moderate accuracy studies of the dynamics of merging black
holes.  LISA, by contrast, can do high SNR (a few $\times 10^4$),
high-accuracy studies.

\item Coalescing black holes are fairly likely to be the first sources
detected by the
LIGO/VIRGO network of ground-based interferometers: because of
their larger masses, they can be seen out to much greater distances
than coalescing neutron star binaries.  (With the initial LIGO
interferometers, binary black holes with $M \alt 50 M_\odot$ can be
seen out to $\sim 250 \, {\rm Mpc}$, whereas binary neutron stars can
be seen out to $\sim 30 \, {\rm Mpc}$ \cite{rangenote}).  Estimates of
the number of BBH coalescences per NS-NS coalescence by experts in
binary evolution range from $1/10$ \cite{sternberg} to $1/300$
\cite{tutukov}; the distance gain factor for BBHs could easily compensate
for the reduced birth rate.

\item Low-mass BBHs [$M \alt 30 M_\odot$ for the first LIGO
interferometers; $M \alt 80 M_\odot$ for the advanced; $(1+z) M \alt 3
\times 10^6 M_\odot$ for LISA] are best searched for via their
well-understood inspiral waves; while more massive BBHs must be
searched for via their more poorly understood merger waves and/or
their well-understood ringdown waves.

\item When the interferometers' sensitivities has reached a level roughly
half way between the first LIGO interferometers and the advanced LIGO
interferometers \cite{rangenote}, a search based on the inspiral waves
should be capable of finding low-mass BBHs out to about $1 \, {\rm Gpc}$.

\item A search for massive BBHs based on the ringdown waves can be
performed using matched filtering.  We show in Sec.\ \ref{qnrsearches}
that the number of independent templates needed for such a search is
about $6000$ or less, assuming that one wants the event rate reduction
due to discreteness of the template family to be no more than $10\%$.
With the first LIGO interferometers, such a search should be capable of
finding equal-mass BBHs in the mass range $100 M_\odot$ to $700
M_\odot$ out to about $200 \, {\rm Mpc}$; with advanced LIGO
interferometers, BBHs with $200 M_\odot \alt M (1+z) \alt 3000
M_\odot$ should be detectable out to $z \sim 1$; and with LISA, BBHs
with $10^6 M_\odot \alt (1+z) M \alt 3 \times 10^8 M_\odot$ should
be visible out to $z \agt 100$.  These distances are reduced by a factor
$\sim (4 \mu/M)$ for non-equal-mass BBHs.

\item The effectiveness of a search based on the merger waves will
depend on how much one has learned about the merger waveforms from
numerical relativity simulations.  With only a knowledge of the merger
waves' range of frequency bands and range of temporal durations, a
search based on merger waves can be performed using the noise-monitoring
search algorithm discussed above.  Such a search should
increase the number of discovered BBHs by a factor $\sim 10$ over
those found from the inspiral and ringdown waves.  On the other hand,
a full set of merger templates based on numerical relativity
simulations could further increase the number of discovered BBHs by an
additional factor of up to $\sim 4$.

\end{itemize}

\subsection{Organization of this paper}
\label{map}

The organization of this paper is as follows.
In Sec.~\ref{derivesnrformulagen} we discuss in detail the three
methods of searching for gravitational wave signals referred to above.
In Sec.~\ref{derivesnrformula} we derive the general formula
[Eq.~(\ref{snraveraged})] discussed above for the optimally-filtered SNR
of a gravitational-wave source at a cosmological distance, in terms of
the source's emitted spectrum, redshift and luminosity distance, and
the detector's noise spectrum.  This formula is one underpinning for
the BBH SNR computations of Appendix \ref{appsnr} and
Sec.~\ref{snrsection}.  In Sec.~\ref{derivesnrformula1} we derive the
approximate formula [Eq. (\ref{simp1})] discussed above for the SNR
obtainable by band-pass filtering in the absence of theoretical
templates.  In Sec.~\ref{nonlinear} we describe our suggested
noise-monitoring search, and derive the
relevant detection threshold [Eqs.~(\ref{e3}) and (\ref{e4})].

In Sec.~\ref{threelittlepigs} we split the waveform from a coalescing
equal-mass black hole binary into three successive epochs: inspiral,
merger and coalescence.  Then we estimate the frequency bands and
spectra for each of these epochs: the merger phase in
Sec.~\ref{mergerphase} [Eqs.~(\ref{mergefreq}), (\ref{fqnrdef}),
(\ref{dEdfmerger})]; the inspiral phase in Sec.~\ref{inspiralphase}
[Eq.~(\ref{dEdfinspiral})]; and the ringdown phase in
Sec.~\ref{ringdownphase} [Eqs.(\ref{dEdfqnr}) --
(\ref{qnrenergytot})].  We have already summarized, above, the
assumptions that underlie our assumed merger and ringdown spectra, and
the uncertainties in those assumptions; we give a more detailed
description in Secs.~\ref{mergerphase} -- \ref{ringdownphase}.  A
subtlety in the derivation of the ringdown energy spectrum is
explained in Appendix \ref{ringdown3}.
In Sec.~\ref{mergernofilters} we make rough estimates of the range of
the bandwidths and temporal durations of merger waveforms.

In Sec.~\ref{noisecurves} we devise a simple piece-wise power-law
analytic fit [Eq.~(\ref{noisespec})] to the noise spectra of an
initial LIGO interferometer, an advanced LIGO interferometer, and a
space-based LISA interferometer.  This single formula, by adjustment
of its parameters, can describe all three interferometer types.

In Appendix \ref{appsnr}, we insert the interferometer noise spectra from
Sec.~\ref{noisecurves}, and the estimated BBH inspiral, merger, and
ringdown spectra from Secs.~\ref{mergerphase}, \ref{inspiralphase} and
\ref{ringdownphase}, into the general SNR formula (\ref{snraveraged}).
That formula's frequency integral is then performed to compute
the angle-averaged, optimal-filtering SNR for
each type of interferometer and each phase of BBH coalescence.  The
resulting SNRs for an initial LIGO interferometer, an advanced LIGO
interferometer, and a LISA interferometer, are expressed in terms of
the binary's total mass $M$, cosmological redshift $z$, and luminosity
distance $D$.

In Sec.~\ref{generalresults}, we graph the SNR formulae derived in Appendix
\ref{appsnr} (Fig.~\ref{initialligosnr} for an initial LIGO interferometer;
Fig.~\ref{advancedligosnr} for an advanced LIGO interferometer;
Fig.~\ref{lisasnr} for LISA).  We give intuitive insight into these SNRs in
Sec.~\ref{egs} by re-expressing the power SNR for a source as
\begin{equation}
\rho^2 = \int d(\ln f) [h_{\rm char}(f) / h_n(f)]^2,
\end{equation}
where $h_{\rm char}(f)$ is the source's ``characteristic amplitude'' as
a function of frequency, and $h_n(f)$ is the detector's rms noise in a
bandwidth equal to frequency for sources with random orientations.
We give plots of $h_{\rm char}(f)$ and $h_n(f)$ for five specific
examples of binaries --- a $10 M_\odot - 10 M_\odot$ binary at
distance $D = 200 \, {\rm Mpc}$ and a $15 M_\odot - 15 M_\odot$ binary
at cosmological redshift $z=1$ (Fig.~\ref{hcharfig}); a $50 M_\odot -
50 M_\odot$ binary at cosmological redshift $z=0.5$
(Fig.~\ref{hcharfig1}); a $5 \times 10^6 M_\odot - 5 \times 10^6
M_\odot$ binary at $z=5$ and a $2.5 \times 10^4 M_\odot - 2.5 \times
10^4 M_\odot$ binary at $z=1$ observed by LISA
(Fig.~\ref{hcharfig2}).  For each example, we explain, in terms of the
plots, why the inspiral, merger, and ringdown SNRs have the values
shown in Figs.~\ref{initialligosnr}, \ref{advancedligosnr} or
\ref{lisasnr}.

The SNR graphs in Figs.~\ref{initialligosnr},
\ref{advancedligosnr} and \ref{lisasnr} are the foundation for our
conclusions, summarized above, about what features of which binaries
should be observable with which interferometers.  A detailed discussion
of these conclusions is given in Sec.~\ref{detection}.  In
Sec.~\ref{qnrsearches}, we estimate the number of templates required
for a search for ringdown waves based on matched filtering, estimate
the SNR detection thresholds, and hence the range of the various
interferometers for ringdown waves.  In Sec.~\ref{detectmerger} we
examine the prospects for searches for BBHs via their merger waves,
both with and without templates.

\section{DERIVATION OF GENERAL FORMULAE FOR
SIGNAL-TO-NOISE RATIOS AND DETECTION THRESHOLDS}
\label{derivesnrformulagen}

In this section we discuss in detail the various signal-search methods
which were briefly described in the Introduction.  We start in
Sec.~\ref{derivesnrformula1} by deriving the approximate relation
(\ref{bpa}) between the SNR achievable
using matched filtering searches for signals and the SNR obtainable
via band-pass filtering searches.  This approximate relation is closely
related to the standard lore that the gain factor due to matched
filtering is the square root of the number of cycles in the waveform
\cite{300yrs,schutz1}, which strictly speaking is applicable only to
waveforms that are almost monochromatic.  For completeness, we discuss
in Sec.~\ref{derivesnrformula1} both the standard lore relation
[Eq.~(\ref{snrgain})] and the approximate more general relation
[Eq.~(\ref{simp1})].

In Sec.~\ref{nonlinear} we describe our proposed noise-monitoring search
method, and derive the detection threshold (\ref{NLthreshold})
discussed in Sec.~\ref{results}.  Finally, in
Sec.~\ref{derivesnrformula} we derive the general formula
[Eq.~(\ref{snraveraged})] for the angle-averaged, optimal-filtering
SNR for a gravitational-wave source, which was discussed in the
Introduction.

\subsection{Searches for gravitational-wave bursts: band-pass filtering
and matched filtering}
\label{derivesnrformula1}

Consider the situation where some arbitrary gravitational-wave burst
$h(t)$ is present in the data stream $s(t)$, so that
\begin{equation}
s(t) = h(t) + n(t),
\label{decompos1}
\end{equation}
where $n(t)$ is the noise.  If one integrates a filter
$K(t)$ against the data stream $s(t)$ to produce a number, $Y = \int
K(t) s(t) \, dt$, then
the standard definition of the signal to noise ratio is \cite{twosided}
\begin{eqnarray}
{S\over N} &=& {\mbox{expected\ value\ of\ $Y$\ when\ signal\ present}\over
\mbox{rms\ value\ of\ $Y$\ when\ no\ signal\ present}} \nonumber \\
\mbox{} & = & {\langle Y \rangle \over \sqrt{ \langle Y^2 \rangle_{s=0} }}
\nonumber \\
\mbox{} &=& {4 \int_0^\infty df \, \Re \left[ {\tilde h}(f)^* {\tilde
K}(f)\right] \over \sqrt{ 4 \int_0^\infty df \, \left| {\tilde
K}(f) \right|^2 S_h(f) }};
\label{snrgeneral}
\end{eqnarray}
see, {\it e.g.}, Refs.~\cite{helstrom,wainandzub}.  Here tildes denote
Fourier transforms according to the convention
\begin{equation}
\label{fourier}
{\tilde h}(f) = \int_{-\infty}^{\infty} e^{2 \pi i f t}\, h(t) \, dt,
\end{equation}
and $S_h(f)$ is the power spectral density of strain noise in the
detector \cite{twosided}.

Now, consider searching for a signal $h(t)$ when the only information about
it that one has is its approximate bandwidth in the frequency domain. Perhaps
the simplest search algorithm one could use to search for the signal is to
choose for $K(t)$ the following band-pass filter:
\begin{equation}
{\tilde K}(f) = e^{2 \pi i f t_{\rm start}}
	\Theta(\Delta f/2 - |f - f_{\rm char}|).
\label{bpfilter}
\end{equation}
Here $\Theta$ is the step function and $t_{\rm start}$ is the starting
time of the filter.  This filter chops out all the data in the frequency
domain except that in a bandwidth $\Delta f$ about a characteristic
central frequency $f_{\rm char}$ \cite{bpfilternote}.  Suppose that
the frequency interval has been chosen wisely, so that the signal $h(t)$
has negligible power outside the interval.  Then ${\tilde h}(f)$ can be
taken to vanish outside the chosen bandwidth, and Eqs.~(\ref{snrgeneral})
and (\ref{bpfilter}) yield
\begin{eqnarray}
\left({S \over N}\right)_{\rm band-pass\ filter} &=&  {h(t_{\rm start})
\over \sqrt{ \int_{\Delta f} df \, S_h(f)}} \nonumber \\
&\approx& \sqrt{f_{\rm char} \over \Delta
f} \ {h(t_{\rm start}) \over h_{\rm rms}(f_{\rm char})},
\label{snrbp}
\end{eqnarray}
where $h_{\rm rms}(f) \equiv \sqrt{f S_h(f)}$ is the rms
fluctuation in the noise at frequency $f$ in a bandwidth equal to $f$.
The starting time of the filter $t_{\rm start}$ is then varied to give
the maximum filter output $Y$, which is achieved at some value $t_{\rm
best}$ of $t_{\rm start}$.  At this maximum overlap time, the SNR is
given by Eq.~(\ref{snrbp}) with $t_{\rm start}$ replaced by $t_{\rm best}$.
In particular, for broadband signals for which $\Delta f \sim f_{\rm
char}$ (as opposed to $\Delta f \ll f_{\rm char}$),
Eq.~(\ref{snrbp}) simplifies to the standard result \cite{schutz1}
\begin{equation}
\left({S \over N}\right)_{\rm band-pass\ filter} \approx  {h(t_{\rm
best})  \over h_{\rm rms}(f_{\rm char})}.
\label{snrbp1}
\end{equation}
By contrast, if the shape of the signal is known one can use the well
known optimal filter ${\tilde K}(f) = {\tilde h}(f)/S_h(f)$, for which
the SNR squared is, from Eq.~(\ref{snrgeneral})
\cite{helstrom,wainandzub,twosided}
\begin{equation}
\label{snr}
\rho^2 =
\left({S \over N}\right)^2_{\rm optimal\ filter} =4 \int_0^\infty {|
{\tilde h}(f)|^2 \over S_h(f)} df.
\end{equation}

A crucial element of both optimal filtering searches and
most especially band-pass filtering searches with ground-based
interferometers is the use of coincidencing between different
interferometers to circumvent the effects of non-Gaussian noise bursts
\cite{schutz1}.  Coincidencing between the 4
interferometers in the LIGO/VIRGO network (the Hanford 2 km, Hanford 4
km, Livingston 4 km and Pisa 3 km interferometers) should be
sufficient to achieve this.  To be conservative, our assumed detection
thresholds for the SNR values (which we discuss in
Sec.~\ref{detection}) are based on combining just the two LIGO 4
km interferometers, albeit with assumed Gaussian statistics.

The standard lore approximate relationship between the
optimal-filtering SNR (\ref{snr}) and the band-pass filtering SNR
(\ref{snrbp1}) can be obtained as follows.  Consider the special case of a
waveform that is quasi-monochromatic, {\it i.e.}, of the form
\begin{equation}
h(t) = h_{\rm amp}(t) \cos\left[\Phi(t)\right],
\label{qm}
\end{equation}
where the amplitude $h_{\rm amp}(t)$ and instantaneous frequency [given
by $2 \pi f(t) = d \Phi / d t$] are slowly evolving.  Using Eq.~(\ref{snr})
and the stationary phase approximation to the Fourier transform of the
signal (\ref{qm}) yields
\begin{equation}
\rho^2 = \int d(\ln f) \, n_{\rm cyc}(f)
 \, {h_{\rm amp}\left[t(f)\right]^2 \over h_{\rm rms}(f)^2},
\label{snrgain}
\end{equation}
where $n_{\rm cyc}(f) \equiv f^2 / {\dot f}$ is the number of cycles
spent within a bandwidth $\Delta f \sim f$
centered on $f$, and $t(f)$ is the time at which the gravitational-wave
frequency is $f$.  By comparing Eqs.~(\ref{snrbp1}) and
(\ref{snrgain}) it can be seen that $n_{\rm cyc}(f)$ is the gain
factor in SNR squared for optimal filtering over band-pass filtering,
per logarithmic interval in frequency \cite{300yrs}.

For general signals $h(t)$ which are not quasi-monochromatic, Eq.~(\ref{qm})
does not apply.  However, we can derive an approximate formula for the
SNR (\ref{snrbp1}) for general signals as follows.
Approximating $S_h(f)$ to be constant in Eq.~(\ref{snr}) gives \cite{schutz1}
\begin{eqnarray}
\left({S \over N}\right)^2_{\rm optimal\ filter} &\approx& {2 \over
S_h(f_{\rm char})} \int_{-\infty}^\infty dt\,  [h(t)]^2 \nonumber \\
&\approx& 2 f_{\rm char} T \, { {\bar h}^2 \over h_{\rm rms}(f_{\rm
char})^2}
\label{snropt2}
\end{eqnarray}
where ${\bar h}$ is a rms average of $h(t)$ and $T$
is the effective duration of the signal.  Comparing
Eqs.~(\ref{snropt2}) and (\ref{snrbp}) we find that
\begin{equation}
{\left(S/N\right)_{\rm band-pass\ filter} \over \left(S/N\right)_{\rm
optimal\ filter}} \approx
{h(t_{\rm best}) \over {\bar h}}  \ {1 \over \sqrt{{\cal
N}_{\rm bins}}},
\label{simp}
\end{equation}
where the ``number of {\it a priori\/} frequency bins'' is
\begin{equation}
{\cal N}_{\rm bins} = 2 T \Delta f.
\label{nbinsdef}
\end{equation}
The reason for this terminology is as follows.  Suppose that we are
trying to detect a signal whose total duration we know to be less than
or equal to $T$, and whose spectrum we know to lie in some frequency
interval of bandwidth $\Delta f$.  When we discard the measured data
outside of these time and frequency intervals, the remaining relevant
data is described by ${\cal N}_{\rm bins} = 2 T \Delta f$ real Fourier
coefficients (or equivalently, ${\cal N}_{\rm bins}$ frequency bins).
The quantity ${\cal N}_{\rm bins}$ is the number of independent real
variables that parameterize the space of signals for which we will search.

This notion of number of {\it a priori\/} frequency bins is closely
related to the notion of number of cycles in the waveform.  The
number of waveform cycles will be approximately ${\cal N}_{\rm cyc}
\sim T f_{\rm char}$, and so for a broadband burst for which $\Delta f
\approx f_{\rm char}$ we have ${\cal N}_{\rm cyc} \approx
{\cal N}_{\rm bins}$.  There is a distinction between the two concepts,
however:  the number of cycles is intrinsic to the
signal, whereas the number of frequency bins in part characterizes our
{\it a priori\/} information or assumptions about the signal and not the
signal itself.  This is because for band-pass filtering we must choose
at the start some bandwidth $\Delta f$.  Since the true bandwidth of
the signal will vary from one signal to another and will not be known
in advance, the true bandwidth of the signal will in general be
somewhat less than $\Delta f$ if we have chosen $\Delta f$ wisely.
The quantity ${\cal N}_{\rm bins}$ depends not on the true bandwidth
of the signal but on our assumed bandwidth $\Delta f$.

The first factor on the right hand side of Eq.~(\ref{simp}) is the ratio
between the peak strain amplitude $h(t_{\rm best})$ in the time domain and
an rms value ${\bar h}$ of this  strain amplitude.  By defining the
effective duration $T$ of the signal to be given by
\begin{equation}
\int dt \, [h(t)]^2 = T \, h(t_{\rm best})^2,
\label{Tdef}
\end{equation}
this factor reduces to unity.  With this interpretation of $T$ in
Eq.~(\ref{nbinsdef}), Eq.~(\ref{simp}) reduces to the formula
\begin{equation}
{\left(S/N\right)_{\rm band-pass\ filter} \over \left(S/N\right)_{\rm
optimal\ filter}} \approx
 {1 \over \sqrt{{\cal
N}_{\rm bins}}}
\label{simp1}
\end{equation}
discussed in the Introduction.  We use this formula in
Sec.~\ref{detectmerger}.

The SNR threshold $\rho_{\rm threshold}$ appropriate for matched
filtering is approximately given by
\cite{prl}
\begin{equation}
{\rm erfc}(\rho_{\rm threshold}/\sqrt{2}) = {\epsilon \over {\cal N}_{\rm
start-times} \, {\cal N}_{\rm shapes}}
\label{MFthreshold0}
\end{equation}
which to a good approximation reduces to
\begin{equation}
\rho_{\rm threshold} \approx \sqrt{ 2 \ln( {\cal N}_{\rm
start-times}/\epsilon) + 2 \ln( {\cal N}_{\rm shapes})};
\label{MFthreshold1}
\end{equation}
[{\it cf.}~Eq.~(\ref{MFthreshold}) where the factor of $\epsilon$,
which does not have a large effect, was omitted].  Here the number
${\cal N}_{\rm shapes} = {\cal N}_{\rm shapes}(\rho_{\rm threshold})$
is the number of statistically independent waveforms with SNR $\le
\rho_{\rm threshold}$ in the set of signals to be searched for
\cite{importantnote}; Eq.~(\ref{MFthreshold1}) must be solved
self-consistently to determine $\rho_{\rm threshold}$.

\subsection{Searches for gravitational-wave bursts: noise monitoring}
\label{nonlinear}

In this section we describe our suggested noise-monitoring method
for searching for gravitational wave bursts of unknown form; more
details can be found in Ref.~\cite{ComingSoonToAJournalNearYou}.
In essence, the method consists of monitoring over short timescales
the total rms noise in the detector output in the frequency band in
which the signal is expected, and waiting for statistically
significant changes in one's estimate of the noise power.

Suppose that the maximum expected signal duration is $T$, and that the
interferometer output is $s(t)$.  We construct a quantity $Q(t)$ in
the following way.  First, focus attention on the data stream
$s(\tau)$ in the time interval $t - T/2 \le \tau \le t+T/2$.  Because
the data stream is in fact discrete and not continuous, this interval
of data can be represented by the numbers
\begin{equation}
s_j = s(t - T/2 + j \Delta t)
\end{equation}
for $0 \le j \le N_{\rm total} = T / \Delta t$, where $\Delta t$ is
the sampling time.  From Eq.~(\ref{decompos1}) we have
\begin{equation}
s_j = h_j + n_j,
\label{basicc}
\end{equation}
where $h_j$ is the gravitational wave signal and $n_j$ is the noise.
Now because the interferometer noise is colored, the noise matrix
\begin{equation}
\Sigma_{ij} \equiv \langle n_i n_j \rangle
\label{noisem}
\end{equation}
will not be diagonal.  Here, angle brackets denote ensemble averaging over
realizations of the noise.  If one performs an FFT just of this finite
stretch of data, the noise matrix on the new basis will not be diagonal
either because of aliasing effects.  However, it is possible to
diagonalize the matrix (\ref{noisem}) and change to a basis on which
the noise is diagonal.  We will denote this new basis by capital Roman
letters $I,J,K$.  The data points $s_I$ on this new basis can be chosen
to correspond approximately to frequencies $f_I = I/T$, $I =
1,-1,2,-2,\ldots$ \cite{ComingSoonToAJournalNearYou}.
Equation~(\ref{noisem}) can now be
replaced by
\begin{equation}
\langle n_I \, n_J \rangle = \delta_{IJ} \, \sigma_I^2.
\label{noisem1}
\end{equation}
The data $s_I$ extend up to some high frequency (of order several kHz)
determined by the sampling time.  We next discard all data above some
upper cutoff frequency.  (Note that we have effectively performed band-pass
filtering of the data, since the restriction to a segment of length $T$ in
the time domain removes frequency components at $f \alt 1/T$.)  Let us
denote by ${\cal N}_{\rm bins}$ the total number of data points remaining,
which will be approximately given by
\begin{equation}
{\cal N}_{\rm bins} = 2 T \Delta f
\end{equation}
where $\Delta f$ is the bandwidth of our effective band-pass filter.

In terms of this basis, matched filtering consists of calculating, for
each trial waveform shape $h_J$, the quantity
\begin{equation}
{\sum_J s_J h_J / \sigma_J^2 \over \sqrt{ \sum_J h_J^2 / \sigma_J^2}.}
\end{equation}
(We are assuming here that all the trial waveform shapes have duration
less than $T$ and in the frequency domain have most of their power
within the bandwidth $\Delta f$).  We introduce the notation $\rho_I =
h_I / \sigma_I$; then the optimal
filtering SNR (\ref{snr0}) is given by
\begin{equation}
\rho^2 = \sum_I \rho_I^2 = \sum_I {h_I^2 \over \sigma_I^2}.
\label{snr00}
\end{equation}
Thus, the quantity $\rho_I^2$ is the optimal filtering signal-to-noise
squared per data bin.  Note that throughout this subsection, we use $\rho$ as a
convenient parameterization of the signal strength, which is
meaningful even in situations where templates are not available and
where optimal filtering cannot be carried out.
Band-pass filtering (of a pre-whitened data stream) approximately
corresponds in this language to calculating the statistic
\begin{equation}
{\hat \rho}_{\rm BP} \equiv \max_J {s_J \over \sigma_J}.
\label{hh}
\end{equation}
This will have an expected value of roughly $\rho / \sqrt{{\cal N}_{\rm
bins}}$ [{\it cf.}~Eq.~(\ref{simp1})] if the signal is not sharply
peaked in frequency but instead is spread out over the bandwidth
$\Delta f$.

The statistic used to diagnose the presence of a signal is
\begin{equation}
Q(t) = - {\cal N}_{\rm bins} + \sum_{J =
- {\cal N}_{\rm bins}/2}^{J={\cal N}_{\rm bins}/2} \,
{s_J^2 \over \sigma_J^2}.
\label{Qdef}
\end{equation}
This statistic is essentially an estimate of the apparent rms
noise power in the given bandwidth over the given time interval,
up to an additive constant.  The additive constant,
$-{\cal N}_{\rm bins}$, is chosen so that when no signal is
present, $\langle Q(t) \rangle=0$ and thus $Q(t)$ random walks over
positive and negative values.  When a signal is present, $Q(t)$
will with high probability be large and positive.  One monitors $Q(t)$
as a function of time and sets a detection threshold such that when
$Q(t)$ exceeds the threshold, the probability that the data stream
contains no signal is very low (see below).  Note that this search
method constitutes a type of nonlinear filtering.

The technique of monitoring the statistic (\ref{Qdef}) is closely
related to two commonly used techniques in radio astronomy.  In
the first such technique, observers sum the power from frequency bins
which are expected to contain harmonics of the signal they are trying
to detect.  This procedure is not quite as good as coherently combining
the signal from all the frequency bins but is much easier, computationally.
The second technique \cite{periodic} is useful when one is looking for periodic
signals in a data train that is too long to Fourier transform.  One splits
the data train into some number $N$ of shorter data segments, takes
the FFT of each shorter segment, and adds the FFTs incoherently ({\it i.e.}.,
adds the individual power spectra).  Again, this is not the optimal search
method, but it is often a useful thing to do given finite computational
resources.  One difference between our suggested noise-monitoring search
method and the radio astronomy techniques is the following:  One adds the
frequency bins incoherently in the statistic (\ref{Qdef}) because the phase
relationships are unknown; in the radio astronomy contexts, one adds the
frequency bins incoherently to save on computation time.  But the
techniques themselves are conceptually very similar.

We now turn to a derivation of the efficiency and performance of our
suggested noise-monitoring search method.  It is straightforward to
show that when a
signal is present
\begin{equation}
\langle Q(t) \rangle = \rho^2
\label{e1}
\end{equation}
and that
\begin{equation}
\left< \left[Q(t) - \langle Q(t) \rangle \right]^2 \right> = 4 \rho^2
+ 2 {\cal N}_{\rm bins}.
\label{e2}
\end{equation}
When no signal is present, Eqs.~(\ref{e1}) and (\ref{e2}) continue to
hold with $\rho=0$.  These equations show that a signal should be
detectable in the regime
\begin{equation}
{\cal N}_{\rm bins}^{1/4} \ll \rho \alt {\cal N}_{\rm bins}^{1/2},
\label{domain}
\end{equation}
as well as at larger $\rho$, since in the regime $(\ref{domain})$ the
expected value (\ref{e1}) of $Q$ is large compared to its rms value
in the absence of a signal.  By contrast, a signal is detectable using
band-pass filtering only in the regime
$\rho\agt {\cal N}_{\rm bins}^{1/2}$, from Eq.~(\ref{hh}) and
associated discussion.  The noise-monitoring method thus extends the
domain of detectable signals to include the regime (\ref{domain}).

The approximate SNR threshold predicted by Eq.~(\ref{e2}) is correct
in order of magnitude, but to obtain an accurate SNR threshold it is
necessary to calculate the full probability distribution for the
statistic $Q$, since its statistical properties are very non-Gaussian.
This probability distribution is given by, from Eqs.~(\ref{basicc}),
(\ref{noisem1}) and (\ref{Qdef}),
\begin{equation}
P[Q(t) \ge Q_0] = {\Gamma[{\cal N}_{\rm bins}/2,(Q_0 + {\cal N}_{\rm
bins})/2] \over \Gamma({\cal N}_{\rm bins}/2) }
\label{probQ}
\end{equation}
where $\Gamma(\cdots,\cdots)$ is the incomplete Gamma function and
$\Gamma(\cdots)$ is the usual Gamma function.  Suppose now that we
examine a number ${\cal N}_{\rm start-times}$ of different starting
times $t$, and that
we wish to find the number $Q_0$ such that the probability (\ref{probQ})
of $Q(t)$ exceeding $Q_0$ for any $t$, in the absence of a signal, is
some small number $\epsilon\sim 10^{-4}$.  This threshold $Q_0$ is
obtained by solving
\begin{equation}
{\Gamma[{\cal N}_{\rm bins}/2,(Q_0 + {\cal N}_{\rm
bins})/2] \over \Gamma({\cal N}_{\rm bins}/2) } = { \epsilon \over {\cal
N}_{\rm start-times}}.
\label{e3}
\end{equation}
{}From Eq.~(\ref{e1}), this threshold will be exceeded by a signal
whenever the signal strength (\ref{snr00})
satisfies
\begin{equation}
\rho \ge \rho_* = \sqrt{Q_0}.
\label{e4}
\end{equation}
Equations (\ref{e3}) and (\ref{e4}) determine the threshold $\rho_*$
as a function of the parameters $\epsilon$, ${\cal N}_{\rm
start-times}$, and ${\cal N}_{\rm bins}$; we use these formulae in
Sec.~\ref{detectmerger}.

We now discuss an approximate formulae for the threshold $\rho_*$
which gives an insight into the nature of the nonlinear filtering
algorithm.  It can be shown \cite{ComingSoonToAJournalNearYou} that
for ${\cal N}_{\rm bins} \gg 1$, $\rho_*$
can be obtained to a good approximation by solving the equation
\FL
\begin{equation}
\rho_*^2 = 2 \ln( {\cal N}_{\rm start-times}/\epsilon) + {\cal N}_{\rm bins}
\ln \left( 1 + \rho_*^2 / {\cal N}_{\rm bins} \right),
\label{NLthreshold1}
\end{equation}
[{\it cf.}~Eq.~(\ref{NLthreshold}) discussed in the Introduction].
It is possible to understand the origin of Eq.~(\ref{NLthreshold1}) in
the following simple way.

The total number ${\cal N}_{\rm shapes,max}$ of waveform shapes
of duration less than or equal to $T$, with bandwidth $\Delta f$ and with
SNR $\le \rho$ which are distinguishable in the interferometer noise is
approximately given by \cite{paperII}
\begin{equation}
\ln({\cal N}_{\rm shapes,max} ) \approx {1 \over 2} {\cal N}_{\rm
bins} \ln \left( 1 + \rho^2 / {\cal N}_{\rm bins} \right).
\label{numshapes}
\end{equation}
Let us denote by ${\cal M}$ the manifold of all possible waveforms
satisfying these criteria --- duration
less than or equal to $T$, bandwidth $\Delta f$ and SNR $\le \rho$.
Not all of the waveforms in this manifold correspond to physically
interesting BBH merger waveforms; the set ${\cal M}_{\rm merge}$ of
BBH merger waveforms will form a sub-manifold of ${\cal M}$.
Consider now the hypothetical limit in which the sub-manifold
${\cal M}_{\rm merge}$ becomes so large (perhaps curving back and
intersecting itself) that, when smeared out by the interferometer noise,
it effectively fills the entire manifold ${\cal M}$.  In this limit,
the number ${\cal N}_{\rm shapes}$ of BBH merger waveform shapes for
which we are searching approaches ${\cal N}_{\rm shapes,max}$.  It is
clear that in this limit, knowledge of the waveform shapes is not useful
in searching
for the waves, and that a search method which simply seeks the maximum
overlap between the measured signal and all possible waveform shapes
(thus not requiring templates) would perform just as well as matched
filtering.  This is essentially what the nonlinear search method does.
Using this insight,
the threshold (\ref{NLthreshold1}) can be obtained simply by combining
Eqs.~(\ref{MFthreshold1}) and (\ref{numshapes}).

The above derivation is based on frequentist statistics.  In
Ref.~\cite{ComingSoonToAJournalNearYou} a Bayesian analysis is
outlined of the detection of gravitational wave signals of
unknown form which automatically identifies the statistic $Q(t)$ as
optimal, and which also approximately reproduces the detection threshold
$\rho_*$.

In practice, this search method would be combined with coincidencing
between interferometers to achieve high detection reliability and to
reduce the effects of non-Gaussian noise, as is the case with band-pass
filtering as discussed in Sec.~\ref{derivesnrformula1}.  Matched
filtering could be more efficient than the noise-monitoring
method at combating non-Gaussian noise via coincidencing,
for the following reason.  When coincidencing with templates, one can
demand that the SNR in each interferometer be above the appropriate
threshold, {\it and} that the signal-parameter values deduced in each
interferometer be consistent with each other.  For the noise-monitoring
searches, one can only demand that the SNR in each interferometer be
above the appropriate threshold.  Hence, matched filtering has more
discriminating power against situations in which all the interferometers
have moderately large non-Gaussian noise spikes somewhere in the
relevant time window.  Non-Gaussian noise may therefore
make the less-discriminating noise-monitoring search perform
somewhat worse in practice, relative to matched filtering searches, than
is indicated by the threshold (\ref{NLthreshold1}).

\subsection{Signal-to-noise ratio for matched filtering in terms of
waves' energy spectrum}
\label{derivesnrformula}

In this section we derive the relation
[Eq.~(\ref{snraveraged})] between the expected value of the
optimal-filtering SNR (\ref{snr}), and the energy spectrum of
gravitational waves emitted by the source.
In general, the SNR (\ref{snr}) for a burst of waves depends on the
details of the
gravitational waveform, on the orientation of the source with respect
to the interferometer, and on the direction to the source.  By contrast, the
quantity $\langle\rho^2 \rangle$, the average of the squared SNR over
all orientations of and directions to the source, depends only on the
total energy per unit frequency $d E/d f$ carried off from the source
by the waves.  Specifically, consider a gravitational-wave source
located at a cosmological redshift $z$ and corresponding luminosity distance
$D(z)$.  Let the locally measured frequency of the waves near the
source be $f_e$, related to the frequency $f$ measured at the interferometer
by $f = f_e/(1+z)$.  Let the locally measured energy spectrum of the
waves be $d E_e / d f_e(f_e)$.  Then the orientation-averaged SNR squared
measured at the interferometer is given by
\begin{equation}
\langle \rho^2 \rangle = {2 (1 + z)^2 \over 5 \pi^2 D(z)^2} \,
\int_0^\infty df \, {1 \over f^2 S_h(f)} \, {d E_e \over d f_e}[(1 +
z)f].
\label{snraveraged}
\end{equation}

Note that the relation (\ref{snraveraged}) refers to an angle-averaged
SNR obtained from an {\it rms average} of signal amplitudes over
different possible orientations of the source and interferometer.
This averaging convention differs from that adopted in
Refs.~\cite{300yrs,Kipreview}.  There, the angle-averaged SNR is
defined to be a cube root of an average of cubed signal amplitudes,
rather than an rms average.  That ``cube root of a mean cube''
averaging method is the appropriate method for calculating the
expected event detection rate \cite{300yrs}.  As a result, the SNR
formulae used in Refs.~\cite{300yrs,Kipreview} are a factor of
$\sqrt{3/2}$ larger than the formulae used in this paper,
the factor of $\sqrt{3/2}$ being an approximation to the effect of
the different angle-averaging methods.

We now turn to the derivation of Eq.~(\ref{snraveraged}). Consider
first the case where the source of gravitational waves is sufficiently
close that cosmological effects can be neglected.  Let the source be
at a distance $r$ from the detector and at a location
$(\theta,\varphi)$ on the sky.  Let $(\iota,\beta)$ denote the
direction towards the detector (spherical polar angles) with respect
to a set of Cartesian axes centered at and determined by the source.
Let the two independent polarizations of the strain amplitude at the
interferometer be $h_+(t,r,\iota,\beta)$ and
$h_\times(t,r,\iota,\beta)$, and let the polarization angle be $\psi$.
Then the response of the interferometer will be $h(t) + n(t)$, where
$n(t)$ is the interferometer noise, and
\begin{eqnarray}
\label{decompos}
h(t) &=& F_+(\theta,\varphi,\psi) h_+(t,r,\iota,\beta) \nonumber \\
&+& F_\times(\theta,\varphi,\psi) h_\times(t,r,\iota,\beta).
\end{eqnarray}
Here $F_+$ and $F_\times$ are the interferometer beam pattern functions,
given in, {\it e.g.}, Ref.~\cite{300yrs}.  The dependence of the Fourier
transformed waveform ${\tilde h}_+$ on $r$ is clearly of the form
\begin{equation}
\label{rout}
{\tilde h}_+(f,r,\iota,\beta) = H_+(f,\iota,\beta) / r
\end{equation}
for some function $H_+$; we define $H_\times(f,\iota,\beta)$
similarly.  Combining Eqs.~(\ref{snr}), (\ref{decompos}) and (\ref{rout})
gives
\FL
\begin{equation}
\label{snr1}
\rho^2(r,\theta,\varphi,\psi,\iota,\beta) = {4 \over r^2}
\int_0^\infty {| F_+ H_+ + F_\times H_\times
|^2 \over S_h(f)} df.
\end{equation}

We now average over all of the angles $\theta,\varphi,\psi,\iota$ and
$\beta$. The average over polarizations and over the sky location
gives $\langle F_{+}^2\rangle=\langle F_{\times}^2\rangle=1/5$,
$\langle F_{+}F_{\times}\rangle=0$ \cite{300yrs}, where the meaning of
the angular brackets is given by, for example,
\begin{equation}
\langle F_+^2 \rangle \equiv {1 \over 4 \pi} \int d
\Omega_{\theta,\varphi} \int_0^\pi {d \psi \over \pi} \,
F_+(\theta,\varphi,\psi)^2.
\label{angleaveragedefine}
\end{equation}
{}From Eq.~(\ref{snr1}) this gives
\begin{equation}
\langle \rho^2 \rangle = {4
\over 5 r^2} \int_0^\infty \, {H(f)^2 \over S_h(f)} df,
\label{averagesnr}
\end{equation}
where
\begin{equation}
H(f)^2 \equiv {1 \over 4 \pi} \int
d\Omega_{\iota,\beta} \left(|H_+(\iota,\beta)|^2
+|H_\times(\iota,\beta)|^2\right) .
\label{H2def}
\end{equation}

We now express the energy spectrum $d E / d f$ of the waves in terms
of the quantity $H(f)^2$.  The local energy flux is
\begin{equation}
{dE\over dA\,dt} = {1\over16\pi}{\overline{ \left[{
\left({\partial h_{+}\over\partial t}\right)^2 +
\left({\partial h_{\times}\over\partial t}\right)^2}\right]}},
\end{equation}
where the overbar means an average over several cycles of the
wave.  Switching to the frequency domain using
Parseval's theorem,
inserting a factor of two to account for the folding of negative
frequencies into positive, and using $ |{\tilde h}_{+,\times}(f)|^2 d
A = |{\tilde H}_{+,\times}(f)|^2 d \Omega$ gives
\begin{equation}
\FL {dE\over d \Omega df} = {{\pi}f^2\over2}
\left({ |{\tilde H}_+(\iota,\beta)|^2 +
|{\tilde H}_\times(\iota,\beta)|^2}\right).
\label{dEdOmegadf}
\end{equation}
Combining Eqs.~(\ref{averagesnr}), (\ref{H2def}) and
(\ref{dEdOmegadf}) now yields
\begin{equation}
\left< \rho^2 \right> = {2 \over 5 \pi^2 r^2} \, \int_0^\infty df \,
\int d\Omega \, {1 \over f^2 S_h(f)} \, {d E \over d\Omega d f}[f].
\label{snraveraged1}
\end{equation}
This is essentially Eq.~(\ref{snraveraged}) with $z=0$ and $D(z) =r$,
the limiting form that applies when cosmological effects are
neglected.

Consider now sources at cosmological distances.  We can generalize
Eq.~(\ref{snraveraged1}) to incorporate redshift effects as follows:
(i) Observe that Eq.~(\ref{snraveraged1}) is valid for arbitrary
bursts of gravitational waves provided that we interpret the quantity
$$
{1 \over r^2} \, {d E \over d \Omega d f}
$$
as the locally measured energy flux $d E / d A d f$.  (ii) Use the
fact that the number of gravitons per unit solid angle per unit frequency
is conserved for propagation in a Friedmann-Robertson-Walker
background in the geometric optics limit:
\begin{equation} {d E \over d\Omega d f}(f) = {d E_e \over d\Omega d
f_e}[(1 + z)f].
\end{equation}
Here $f_e$ is the frequency at the source and
$f = f_e/(1+z)$ is the frequency at the detector.  (iii) The
conversion factor at the detector from energy per unit solid angle to
energy per unit area is just $(1+z)^2/D(z)^2$, where $D(z)$ is the
luminosity distance~\cite{draza}.  Hence
\begin{equation} {d E \over d A d f} =
{(1 + z)^2 \over D(z)^2} {d E_e \over d\Omega d f_e}[(1 + z)f].
\end{equation}
Combining this with Eq.~(\ref{snraveraged1}) now yields
Eq.~(\ref{snraveraged}).


An alternative derivation of the cosmological modifications to
Eq.~(\ref{snraveraged1}), which yields a useful rule of thumb for
understanding redshift effects, is the following.  Suppose that an
arbitrary source of waves is specified by parameters $\theta_i$ for $1
\le i \le n$.  In the local wave zone the gravitational waveform is of
the form
\begin{equation}
h(t) = {H(t;\theta_i) \over r},
\end{equation}
for some function
$H$, where $r$ is the distance to the source.  After propagating
through a FRW background the waveform becomes
\begin{equation}
h(t) =
{H(t;\theta_{i,{\rm redshifted}}) \over D(z)},
\label{generalredshift}
\end{equation}
where if $\theta_i$ has dimension $({\rm length})^p$ or $({\rm
mass})^p$, then $\theta_{i,{\rm redshifted}}$ is just $(1+z)^p
\theta_i$.  Equation (\ref{generalredshift}) can be derived from the
general solution for wave propagation in FRW backgrounds in the
geometric optics limit given in Ref.~\cite{KipLesHouches}.  The fact
that the waveform must depend on and reveal the ``redshifted'' source
parameters is clear: all source parameters with dimension that we can
measure from the waves must essentially come from timescales in the
waveform, since the wave amplitude is dimensionless.  All these
timescales will be redshifted.

There is a corresponding simple rule for how the SNR is affected by
cosmological effects.  Suppose that when cosmological effects are
neglected, the SNR is of the form
\begin{equation}
\rho = {F(\theta_i) \over r},
\end{equation}
for some function $F$.  Then the correct SNR is just
\begin{equation}
\rho = {F(\theta_{i,{\rm redshifted}}) \over D(z)}.
\end{equation} Applying this rule to Eq.~(\ref{snraveraged1}) yields
Eq.~(\ref{snraveraged}).

\section{THE GRAVITATIONAL-WAVE SIGNAL\\
FROM COALESCING BLACK HOLES}
\label{signalassumptions}

In this section we describe our assumptions concerning the
gravitational-wave signal from BBH mergers, and the evidence that
underlies those assumptions.  We start in subsection
\ref{threelittlepigs} by discussing the approximate splitting of the
total waveform into the inspiral, merger, and ringdown epochs.  In
Sec.~\ref{mergerphase} we discuss the various circumstantial pieces of
evidence, culled from the literature, about the energy spectrum and
the total energy radiated during the nonlinear merger phase.  Our
assumed merger energy spectrum is given by
Eq.~(\ref{dEdfmerger}), with parameters given by
Eqs.~(\ref{mergefreq}), (\ref{fqnrdef}) and (\ref{epsilonmergerdef}).
The inspiral energy spectrum is discussed in Sec.~\ref{inspiralphase}
and specified in Eq.~(\ref{dEdfinspiral}).
Section~\ref{ringdownphase} discusses the spectrum and likely overall
amplitude of the ringdown signal; the spectral shape is given in
Eq.~(\ref{dEdfqnr}) and our assumption about the overall energy radiated
in Eq.~(\ref{qnrenergy}).  In Sec.~\ref{mergernofilters} we
estimate the SNR obtainable in
band-pass filtering searches for waves from the merger phase.

\subsection{The three phases of the gravitational-wave signal}
\label{threelittlepigs}

As discussed in the Introduction, the coalescence and its
associated gravitational-wave signal can be  divided into three
successive epochs in the time domain: inspiral, merger, and
ringdown.  Physically, the inspiral consists of the portion of the
coalescence in which the black holes are separated bodies that
gradually lose energy and angular momentum, slowly spiraling
towards one another.  By contrast, the merger consists of that portion
of the coalescence in which the dynamics are highly nonlinear and must be
treated by numerical relativity.  With this in mind, it is useful
to define the end of inspiral as the time and frequency \cite{freqtime}
at which numerically generated templates start to be needed.  Up to
this time, post-Newtonian templates, possibly supplemented with IBBH
templates, will be used ({\it cf}.~Sec.~\ref{calcstatus}).

After inspiral, we expect there to be a lot of complicated merger
dynamics that at present are not well understood.  The system will
gradually settle down to a Kerr black hole; the last gravitational
waves we expect to see are those produced by the quasi-normal
ringing modes of this merged black hole.  It is clear that there will
be a smooth
transition in the gravitational waveform from the merger portion to
the ringdown portion, as the effects of nonlinearities become less
and less important with time.  As this happens, the signal should become
increasingly well approximated by a linear combination of exponentially
decaying sine waves.  This is the behavior that has been seen in numerical
simulations of, for example, head-on collisions~\cite{headon,suenheadon}.
At late times, the $l=m=2$ mode will dominate over other quasi-normal
modes, for two reasons which are of comparable importance: (i) The
$l=m=2$ mode is the most slowly damped of all the QNR
modes \cite{chandradetweiler}, and (ii) during coalescence, the binary
will have a rotating shape roughly corresponding to spheroidal harmonic
indices $l=m=2$, and thus this mode will be preferentially
excited \cite{detweiler1}.  We define the ringdown as beginning when the
waveform becomes dominated by the $l=m=2$ QNR mode; the merger
thus contains those portions of the waveform where other modes
and/or non-linear mode-mode couplings are important.  Clearly there
is some arbitrariness in the exact time at which the ringdown starts,
related to the accuracy we require of the fit of the waveform to the
ringdown signal.

Note that, by definition, the three phases of the signal are disjoint
in the time domain.  It does not follow, though, that they should be
disjoint in the frequency domain: their energy spectra might overlap.
However, it is at least approximately true that the inspiral and merger
and disjoint in both time and frequency.  Let $t_{\rm merge}$ be the
time at which inspiral is said to end and merger to begin.  Then, since
the inspiral waves chirp monotonically and since the adiabatic
approximation is just beginning to break down at the end
of the inspiral, the inspiral spectrum is confined, to good approximation,
to the frequency region $f < f_{\rm merge}$.  Here, $f_{\rm merge}$
is the frequency of the inspiral waves at $t = t_{\rm merge}$.
We shall further assume that the merger waves' spectrum is confined to
the frequency regime $f>f_{\rm merge}$; this assumption should be
valid to a  moderately good approximation.  (We discuss below
estimates of the value of $f_{\rm merge}$.)

One particular component of the gravitational-wave signal,
the Christodoulou memory~\cite{memory}, {\em will} violate this
assumption.  The memory component of the gravitational-wave signal
has most of its power below $f_{\rm merge}$ in the frequency domain,
but accumulates gradually during the inspiral, merger and ringdown in
the time domain.  The memory waves from BBH coalescences will probably
not be detectable with ground based interferometers, but very probably
will be detectable with LISA \cite{daniel}.  We will neglect the memory
component of the waves in our analysis, since it will not be as easy to
detect as the components we do discuss.

\subsection{Energy spectrum of the emitted gravitational radiation
from the merger phase}
\label{mergerphase}

The total amount of energy radiated in BBH mergers, and its
distribution in frequency, is highly uncertain because detailed
numerical calculations of these mergers have not yet been made.
In this subsection, we discuss what little evidence there is about the energy
radiated, and describe our crude model of the spectrum.  We assume
a uniform distribution in frequency from the lower frequency
${f_{\rm merge}}$ mentioned above and some upper frequency
${f_{\rm high}}$.

The total amount of energy radiated during a BBH coalescence will be some
fraction $\epsilon$ of the total mass $M = m_1 + m_2$ of the system:
\begin{equation}
E_{\rm radiated} = \epsilon M.
\end{equation}
The fraction $\epsilon$ will depend only on the mass ratio $m_1/m_2$,
on the initial spins ${\bf S}_1$ and ${\bf S}_2$ of the two
black holes, and on the
initial direction ${\hat {\bf L}}$ of the orbital angular momentum
\cite{lessvars}:
\begin{equation}
\epsilon = \epsilon\left({m_1 \over m_2},{{\bf S}_1 \over M^2},{{\bf
S}_2 \over M^2},{\hat {\bf L}}\right).
\label{dependson}
\end{equation}
We can very roughly divide up this fraction as
\begin{equation}
\label{division}
\epsilon = \epsilon_{\rm inspiral} + \epsilon_{\rm merger} +
\epsilon_{\rm ringdown},
\end{equation}
according to the amounts of energy radiated in the three different
epochs of the waveform.  There is clearly some arbitrariness in this
division of the radiated energy into three pieces, which is related to
the choice of the frequency ${f_{\rm merge}}$ at which the inspiral is
said to end and the merger to begin \cite{freqtime}, and also to the
choice of the time $t_{\rm qnr}$ at which the merger is said to end and
the ringdown to begin.

We now discuss methods of estimating $f_{\rm merge}$, $\epsilon_{\rm
merger}$, and $\epsilon_{\rm ringdown}$.  Roughly speaking, we want to
choose the interface between ``inspiral'' and ``merger'' to be at
that point where post-Newtonian templates cease to be useful and where
numerically generated templates will need to be used.  Numerically
generated initial data sets of two orbiting black holes provide some
guidance about this interface.  Cook \cite{cook} compares the properties
of such initial data sets to the predictions of post-Newtonian theory at
second post-Newtonian order.  He finds, for example, that the discrepancy
in the binding energy will be $\sim 5\%$ when the gravitational-wave
frequency (twice the orbital frequency) is $f\sim 0.02/M$, where $M$ is
the total mass of the system, and will be $\sim 15\%$ at $f
\sim 0.05/M$ \cite{cook}.

Towards the end of the inspiral, the black
holes make a gradual transition from a radiation-reaction driven
inspiral to a freely-falling plunge \cite{kww,KipAmos,laiwiseman},
after which, even if the
radiation reaction could be turned off, the black holes would still merge.
This will occur roughly at the last stable circular orbit (LSCO).
It would seem to make sense to choose ${f_{\rm merge}} = f_{\rm LSCO}$, the
frequency at which this occurs.  Strictly speaking, the
concept of an LSCO makes sense only in the test
particle limit ($m_1 / m_2 \ll 1$).  Various somewhat {\it ad hoc} methods
have been used to estimate $f_{\rm LSCO}$ outside this regime, but
the results differ by factors $\agt 2$.  Cook's initial data set analysis
together with the calculation of an ``effective
potential'' yields the estimate $f_{\rm LSCO} \sim 0.055/M$ for equal
mass black holes \cite{cook}.
In post-Newtonian theory, the LSCO can be defined by
artificially turning off the radiation reaction terms in the equations
of motion.  Using this method, Kidder, Will and Wiseman estimate that
$f_{\rm LSCO} \sim 0.02 / M$ \cite{kww}. They
use hybrid equations of motion which are accurate to post-2-Newtonian
order for comparable mass ratios, and which are exact in the test
particle limit.  The value they obtain varies by less than $\sim 20\%$
as the mass ratio is varied.  Note that the frequency $0.02/M$ is close to
the value $6^{-3/2} / (\pi M)$ obtained from the usual LSCO at $r = 6 M$
in the test particle limit (here $r$ is the usual Schwarzschild radial
coordinate).  Finally, earlier analyses by Blackburn and Detweiler used a
variational principle together with the assumption of periodic
solutions to Einstein's equations to obtain the approximate lower bound
$f_{\rm LSCO} \agt 0.06/M$ \cite{blackburndet}.

Note that all of these estimates are for equal mass, non spinning
black holes; the value of the frequency $f_{\rm LSCO}$ can also vary
by factors $\agt 2$ if the black holes are spinning or have
different masses.

In light of this uncertainty, we somewhat arbitrarily adopt the
conservative value of
\begin{eqnarray}
{f_{\rm merge}} &=&  {0.02 \over M} \nonumber \\
\mbox{} &=& 205 \, {\rm Hz} \left( { 20 M_\odot \over M}\right)
\label{mergefreq}
\end{eqnarray}
as the lower cutoff frequency for the energy spectrum.  This (low)
value of ${f_{\rm merge}}$ is conservative in the sense that we can be
reasonably sure that numerically generated templates will not be needed
before $f={f_{\rm merge}}$.  On the other hand, it may optimistically
overestimate the merger SNR by increasing the number of cycles in what
we define as our merger waveform at the expense of the number of cycles
in the inspiral.

We next discuss our choice of upper frequency shutoff for the merger energy
spectrum.   As discussed above, we define the end of the merger to occur at
that time $t_{\rm qnr}$ after which the waveform can be fit fairly
accurately by the $l=m=2$ QNR mode signal.  The merger and the ringdown
will therefore be disjoint in the time domain, but not necessarily in the
frequency domain.  It seems likely, however, that an approximate upper bound
for the frequencies carrying appreciable power during the merger is the
quasinormal ringing frequency itself.  This is supported to some extent by
calculations of the energy spectrum of waves produced in the test particle
limit---see Fig. 2 of Ref.\ \cite{detweiler2}.  The energy spectrum $dE/df$
in that case is peaked near the quasinormal mode frequency $.06/M$ of
a non-rotating black hole, and has fallen by 2 orders of magnitude at $.14/M$.
Although this analysis does not incorporate any nonlinear effects, there
is some evidence that the energy spectrum in the equal mass case will be
qualitatively similar---namely, the head-on collision of black holes
in full numerical relativity {\em has} been computed, and the resulting
waveforms are qualitatively similar to those predicted from perturbation
theory \cite{headon,suenheadon}.  It is not clear how relevant the
head-on collision case is to the gradual inspiral; but on the other
hand, there is no other guidance available at the present time.  We
will therefore use as our upper shutoff frequency for the merger
spectrum the
frequency $f_{\rm qnr}$ of the $l=m=2$ quasinormal mode.  This
frequency will depend on the unknown dimensionless spin parameter
$a$ of the final Kerr black hole (its spin angular momentum divided
by $M^2$). We choose to use the value of $a$ for which the frequency is
highest: in the limit $a \to 1$, the QNR frequency becomes $0.9/(2 \pi
M)$ (more than twice the value for a Schwarzschild black hole)
\cite{chandra}.  Thus, our upper cutoff frequency is
\begin{eqnarray}
f_{\rm high}  = {f_{\rm qnr}} &=& {0.13 \over M} \nonumber \\
\mbox{} &=& 1430 \, {\rm Hz} \, \left( {20 M_\odot \over M} \right).
\label{fqnrdef}
\end{eqnarray}
Our reasons for assuming a high value of $a$ are discussed in
Sec.~\ref{ringdownphase} below.

Finally, consider the total amount of energy
\begin{equation}
E_{\rm rad} = (\epsilon_{\rm merger} + \epsilon_{\rm ringdown}) M
\end{equation}
radiated during the final merger and ringdown.  To estimate this we
consider first an extrapolation due to Smarr \cite{smarrthesis,headon}.
The idea is to calculate the energy radiated in the test particle
approximation using black-hole perturbation theory.
The result will be of the form $E_{\rm rad} = k \mu^2 / M$, where $k$
is a dimensionless constant, $\mu$ is the mass of the test particle,
and $M$ is the total mass.  Now
simply replace $\mu$ by the reduced mass $m_1 m_2/ (m_1 + m_2)$ of the
two black holes.  This extrapolation  works exactly in Newtonian
gravity supplemented by the quadrupole formula, in the sense that the
dynamics of two bodies of
comparable masses can be deduced from the corresponding dynamics
in the extreme mass ratio limit.  Surprisingly, it also works to
within $\sim 20\%$ to predict the total energy radiated in the head-on
collision of two equal-mass black holes
\cite{detweiler2,headon,suenheadon,eardley}. Thus, one might consider
applying this extrapolation to inspiral-preceded mergers.  Such a
calculation would require a determination of the radiated energy from
the final plunge of a test particle, {\it i.e.}, the solution of a
coupled radiation reaction/radiation generation problem outside of the
adiabatic regime (although the coupling is important only in a brief
transitional phase and essentially sets the initial conditions for an
approximately geodesic plunge \cite{KipAmos}).  This calculation has
not yet been done.  However, there is
an alternative approximate method to estimate the amount of energy
radiated in the test particle limit.  The innermost, marginally bound
(unstable) circular orbits should be a fair approximation to the
motion during the final plunge.  Detweiler \cite{detweiler2} showed that
the amount of energy radiated per orbit by a test particle in such an orbit
is of the form $E_{\rm rad} = k \mu^2 / M$,
where $k$ varies from $0.65$ at $a=0$ to $2.8$ at $a=0.95$,
where $a$ is the spin parameter of the black hole.
Assuming that there will be $\agt 1$ effective orbit during the final
plunge, Detweiler estimated that
\cite{detweiler2}
\begin{equation}
0.03 M \, F(\mu/M) \alt E_{\rm rad} \alt 0.2 M \, F(\mu/M),
\label{Eest0}
\end{equation}
where
\begin{equation}
F(\mu/M) = \left({4 \mu \over M}\right)^2
\label{reductionfactor}
\end{equation}
which is unity in the equal-mass case $\mu = M/4$.

Arguments based on conservation of angular momentum also suggest an
approximate lower bound on the radiated energy of about $0.1 M$ for
equal-mass BBHs in the most favorable cases, as we now outline.
Roughly speaking, the system's angular momentum divides up as
\begin{equation}
{\bf S}_1 + {\bf S}_2 + {\bf L}_{\rm orb} = {\bf L}_{\rm rad} +
{\bf S}_{\rm final},
\label{conserve}
\end{equation}
where ${\bf S}_1$ and ${\bf S}_2$ are the spins of the two black holes
just before the final plunge, ${\bf L}_{\rm orb}$ is the orbital
angular momentum just before the plunge, ${\bf L}_{\rm rad}$ is the
angular momentum carried off by the gravitational waves, and
${\bf S}_{\rm final}$ is the spin of the final Kerr black hole.
This splitting of the total angular momentum of the spacetime
into various pieces is really only well defined only in a post-Newtonian
type of limit; however, the effects of this ambiguity are presumably
not important for the purposes of our crude, approximate
argument.  We now specialize to the most favorable case in which
${\bf S}_1$, ${\bf S}_2$ and ${\bf L}_{\rm orb}$ are all aligned.
We take the orbital angular momentum at the beginning of merger
to be $|{\bf L}_{\rm orb}| \approx 0.9 M^2$, the value predicted
by Cook's initial data sets at $f = 0.02/M$ \cite{cook}.  We also
assume that both inspiraling black holes are rapidly spinning, so
that $|{\bf S}_1| \approx |{\bf S}_2| \approx (M/2)^2$.  Equation
(\ref{conserve}) then yields
\begin{equation}
|{\bf L}_{\rm rad}| \agt 0.4 M^2,
\label{bound1}
\end{equation}
since $|{\bf S}_{\rm final}| \le M^2$.

Next, we use that fact that the energy $E_{\rm rad}$ and the angular
momentum $L_{\rm rad}$ carried off by gravitons of frequency $f$ and
azimuthal multipole order $m$ are related by
\cite{proof}
\begin{equation}
E_{\rm rad} = 2 \pi f L_{\rm rad} / m.
\label{EradLrad}
\end{equation}
If we estimate $f$ as $(f_{\rm merge} + f_{\rm qnr})/2$, and make the
admittedly optimistic assumption that most of the radiation is
quadrupolar, we obtain from Eqs.~(\ref{bound1}) and (\ref{EradLrad})
the crude estimate for the total radiated energy \cite{infact}
\begin{equation}
E_{\rm rad} \agt 0.1 M.
\label{Eest1}
\end{equation}
This estimate includes both merger and ringdown radiated energies; we
need to subtract out the ringdown portion to obtain the energy
contained in the merger portion of the waveform.  Below we estimate
$\sim 0.03 M$ to be an approximate upper bound for the ringdown
energy, and hence most of the energy (\ref{Eest1}) should be radiated in
the merger waves.

There is an additional, separate argument one can make which indicates
that most of the energy (\ref{Eest1}) should be radiated  as
merger waves and not as ringdown waves.  As noted by Eardley and
Hirschmann \cite{eardleyhirschmann}, any system with $J > M^2$ cannot
evolve to $J< M^2$ by
radiating quadrupolar waves at the ringing frequency $f_{\rm qnr} \sim 1
/ (2 \pi M)$ of a near-extremal Kerr black hole.  This is because at
this high frequency, too much mass-energy is radiated per unit angular
momentum radiated; Eq.~(\ref{EradLrad}) with $m=2$ and with $f =
f_{\rm qnr}$ yields $\Delta J = \Delta (M^2)$.  Hence, since the final
black hole
must have $J < M^2$, a substantial amount of the radiation must be
emitted at lower frequencies.

Based on the estimates (\ref{Eest0}) and
(\ref{Eest1}), and on the estimated upper bound
$\sim 0.03 M$ mentioned above for the ringdown radiated energy,
we perhaps optimistically take $0.1 M$
to be the energy radiated during merger in the equal-mass case (which
corresponds to about $0.13 M$ total radiated energy in the merger and
ringdown).  For non equal-mass BBHs we assume that the radiated energy is
reduced by the factor (\ref{reductionfactor}), so that
\begin{equation}
E_{\rm merger} = \epsilon_{\rm merger} \, F(\mu/M) \, M = 0.1 F(\mu/M) M.
\label{epsilonmergerdef}
\end{equation}

The rather high radiation efficiency of $0.1$ that we are assuming is
probably most plausible in the context of rapidly spinning coalescing
black holes.  In particular, if the spins and the orbital angular
momentum are somewhat misaligned, the coalescence will be something
like ``two tornados with orientations skewed to each other, embedded
in a third, larger tornado with a third orientation''
\cite{Kipreview}.  Intuitively one would expect that such systems have
more ``settling down'' to do to get to the final Kerr black hole, and that
correspondingly the nonlinear, highly dynamical phase should last
longer and/or produce more radiation.  Also, the potential barrier that
surrounds the final black hole (which normally tends to reflect back
into the black hole the dominant waves of frequency $f \sim 1/{\rm
a~few~times}\,M$)
presumably will effectively not be present during the violent phase of
a merger in which the spins and orbital angular momentum are of comparable
magnitude and are misaligned.

We note that coalescences which radiate as much energy as the estimate
(\ref{epsilonmergerdef}) will also likely radiate a substantial amount
of linear momentum, and the consequent recoil of the final
black hole could correspond to a kick velocity that is a moderate
fraction of the speed of light.

Finally, consider the shape of the energy spectrum $dE / df$
between ${f_{\rm merge}}$ and ${f_{\rm qnr}}$.  For simplicity, and
for lack of evidence in favor of anything more specific, we choose
a flat spectrum:
\begin{eqnarray}
\label{dEdfmerger}
{dE\over df} &=& {{\epsilon_m \,M \, F(\mu/M)}
\over{{f_{\rm qnr}} - {f_{\rm merge}}}} \ \Theta(f -
{f_{\rm merge}}) \Theta({f_{\rm qnr}} -f)
\nonumber \\
\mbox{} &=& 0.91 M^2 F(\mu/M)\ \Theta(f -
{f_{\rm merge}}) \Theta({f_{\rm qnr}} -f),
\end{eqnarray}
where $\epsilon_m = \epsilon_{\rm merger} = 0.1$ and $\Theta$
is the step function.

Our assumed merger spectrum has the desirable feature that it is
reasonably consistent with the inspiral energy spectrum at
$f= f_{\rm merge}$.  This can be seen as follows.  We do not expect a
very large change in
$dE/df$ at the interface between inspiral and merger, for the
following reason: throughout the inspiral, the energy spectrum can
be written \cite{300yrs}
\begin{equation}
{dE \over df}(f) \propto n_{\rm cyc}(f) \, h[t(f)]^2,
\end{equation}
where $n_{\rm cyc}(f)$ is the number of cycles spent near frequency
$f$, and $h[t(f)]$ is the wave amplitude at the time $t(f)$ when the
waves' frequency is $f$ [see also Eqs.~(\ref{snrgain}) and
(\ref{snrOPT})].  In the transition from inspiral to plunge,
the amplitude $h[t(f)]$ should change only moderately, but
$n_{\rm cyc}(f)$ could conceivably change by a large amount: from
its pre-plunge value to $\sim 1$.  It turns out that the number
of cycles does not change very drastically: at the interface between
inspiral and merger,
\begin{eqnarray}
n_{\rm cyc}(f) &=& {f^2 \over {\dot f}} \approx {5 \over 96 \pi} \left(
\pi {\cal M} f \right)^{-5/3} \nonumber \\
\mbox{} &\sim& 6.7\left[ {f \over  f_{\rm merge} }\right]^{-5/3}.
\label{ncycinspiral}
\end{eqnarray}
(Here, ${\cal M} = \mu^{3/5} M^{2/5}$ is the so-called chirp mass.)
Hence, we do not expect $dE/df$ to change too much.

We now show that our assumed values of $dE/df$ from the end of the
inspiral and from the beginning of the merger are reasonably close
to one another, in agreement with the above expectation.
Using the approximation that the motion is an adiabatic evolution from one
circular orbit to another, the energy radiated during inspiral can
be written
\begin{equation}
\left( {d E \over d f}\right)_{\rm radiated} = {d \over df} E_{\rm
circ}(f)\left[1 + O(1/n_{\rm cyc}(f))\right],
\label{adiabaticapprox}
\end{equation}
where $E_{\rm circ}(f)$ is the binding energy of a particle in a circular
orbit at orbital frequency $f/2$ (recall that the gravitational-wave
frequency is twice the orbital frequency).  At our estimated plunge
frequency $f_{\rm merge}=.02/M$, the hybrid equations of motion of
Kidder, Will and Wiseman~\cite{kww} predict in the equal mass case that
\begin{equation}
{d E_{\rm circ} \over d f}(f = 0.02 /M) = 0.88 M^2,
\label{est1}
\end{equation}
while Cook's initial data sets predict that \cite{cook}
\begin{equation}
{d E_{\rm circ} \over d f}(f = 0.02 /M) = 0.41 M^2.
\label{est2}
\end{equation}
The true value of $dE_{\rm circ}/df$ is presumably somewhere between
these two values.  We would expect the value of $(d E / df)_{\rm
radiated}$ to be smaller than this, because the approximation
(\ref{adiabaticapprox}) is breaking down at this point---some of the
gravitational binding energy is being converted to inward radial
kinetic energy instead of being radiated.  This picture is roughly
consistent with our assumption of $dE / df = 0.91 M^2$ for $f \ge
0.02/M$, given that Eqs.~(\ref{est1}) and (\ref{est2}) apply to the
non-spinning case, and we are interested in the rapidly spinning case
for which $dE / df$ could be a few times larger.

\subsection{Energy spectrum of the radiation from the inspiral phase}
\label{inspiralphase}

The standard quadrupole formula prediction for the inspiral energy
spectrum is (see, {\it e.g.}, Ref.~\cite{MTW})
\begin{equation}
{d E \over d f} = {1 \over 3} \pi^{2/3} \mu M^{2/3} f^{-1/3}.
\label{dEdfinspiral}
\end{equation}
This approximate formula is adequate to estimate the SNR obtained
from optimal filtering of the inspiral waveform.  The formula will be
accurate to within a few tens of percent up to
$f=f_{\rm merge}$ \cite{spectrumcomment}.  Using
Eq.~(\ref{dEdfinspiral}) to estimate the SNR effectively assumes that
both the gravitational-wave signal and the theoretical templates used
to cross-correlate with the data stream are given by the quadrupole
approximation (\ref{dEdfinspiral}).  The SNR we calculate using this
lowest order formula will be approximately the same as that found by
cross-correlating real signals [which incorporate higher order
corrections to Eq.~(\ref{dEdfinspiral})] against sufficiently accurate
theoretical templates.  As outlined in Sec.~\ref{calcstatus},
the required template accuracy should be achievable by
post-Newtonian expansions \cite{searchtemplates,owen}, perhaps
supplemented with alternative techniques for the latter, high
frequency part of the
signal at $0.01/M \alt f \alt 0.02/M$ (the IBBH regime).  We assume that the
inspiral energy spectrum shuts off at $f = f_{\rm merge} = 0.02 / M$, as
discussed in Sec.~\ref{mergerphase}.

\subsection{Energy spectrum of the radiation from the ringdown phase}
\label{ringdownphase}

As we have discussed, the ringdown portion of the gravitational-wave
signal is that portion in the time domain that can be fit fairly
accurately by an exponentially decaying sinusoid corresponding to
the most slowly damped $l=m=2$ quasinormal mode of the final black
hole.  The shape of the corresponding energy spectrum is well
understood: it is just a resonance curve (although see Appendix B for
discussion of a subtlety in the applicability of the concept of the
waves' energy spectrum for calculating ringdown SNRs).  The overall
amplitude of the energy spectrum, however, is
not well understood.  In this section we discuss our estimate of
the overall amplitude for the ringdown signal.

The QNR gravitational waveforms $h_+(t,\iota,\beta)$ and
$h_\times(t,\iota,\beta)$ are given by \cite{echeverria}
\begin{eqnarray}
h_{+} - i h_{\times} &=& {{\cal A} M\over r}\,{_2S^2_2}(\iota,\beta,a)
\nonumber \\
&& \times \exp[-2i\pi{f_{\rm qnr}} t - t/{\tau} + i \varphi_0],
\label{qnrwaveform}
\end{eqnarray}
for $t>0$.  Here we have chosen $t=0$ to be the time of the start of
the ringdown, $M$ is the mass of
the black hole, $a M^2$ is its
spin, and $\varphi_0$ is a constant
phase.  The quantities $\iota$ and $\beta$ are spherical polar coordinates
centered on the black hole [{\it cf}.~Sec.~\ref{derivesnrformula}],
${_2S^2_2}(\iota,\beta,a)$ is
a spin weighted spheroidal harmonic whose angle averaged rms value is
\begin{equation}
\left({1\over4\pi}\int d\Omega \,|{_2S^2_2(\iota,\beta,a)}|^2\right)^{1/2}=
	{1\over\sqrt{4\pi}},
\label{spheroidal}
\end{equation}
and ${\cal A}$ is a dimensionless coefficient that describes the
magnitude of the perturbation when the ringdown begins.  The
quantities ${f_{\rm qnr}}$ and ${\tau}$ are the frequency and damping
time, respectively, of the QNR mode of the black hole.  The quality
factor $Q$ of the mode is given by $Q = \pi \tau f_{\rm qnr}$.

As mentioned in the Introduction, there is a mapping between the values
of $M$ and $a$, and of $f_{\rm qnr}$ and $\tau$.  This mapping has
been explored by Leaver \cite{unpubleaver} and Echeverria
\cite{echeverria}.  Leaver studied in depth the damping times and
ringing frequencies of various QNR modes for various Kerr black hole
masses and spins.  By numerically solving the Teukolsky equation, he
produced extensive catalogs of the variation of $\tau$ and
$f_{\rm qnr}$ with black hole mass $M$ and dimensionless spin $a$
\cite{unpubleaver}.  From that data for the $l=m=2$ mode,
Echeverria {\cite{echeverria}} produced the
following analytic fits, which are good to within roughly 5\%:
\begin{eqnarray}
{f_{\rm qnr}} &\approx& \left[1-0.63(1-a)^{3/10}\right] \ {1 \over 2 \pi M}
\nonumber \\
\mbox{} \, &=& \left[1-0.63(1-a)^{3/10}\right] \ \left({37\,M_\odot\over
M}\right)865\,{\rm Hz}\nonumber\\
Q &\approx& 2 \, (1-a)^{-9/20}.
\label{fqnrandQ}
\end{eqnarray}
Note that $Q$ diverges as $a \to 1$ but that $f_{\rm qnr}$ tends towards
a constant.

The energy spectrum for the QNR waveform (\ref{qnrwaveform}) is derived in
Appendix \ref{ringdown3} and is given by
\begin{eqnarray}
{dE\over df} &=& {{\cal A}^2M^2 f^2\over32\pi^3 \, \tau^2}
	\bigg\{
{1 \over \left[(f-{f_{\rm qnr}})^2 +(2\pi{\tau})^{-2}\right]^2} \nonumber \\
\mbox{} &&+
{1 \over \left[(f+{f_{\rm qnr}})^2 +(2\pi{\tau})^{-2}\right]^2} \bigg\}
\label{dEdfqnr} \\
&\approx& {1\over 8} {\cal A}^2 Q M^2 f_{\rm qnr} \, \, \delta(f -
f_{\rm qnr}) \left[1 + O(1/Q)\right].
\label{dEdfqnrapprox}
\end{eqnarray}
Approximating the energy spectrum by a delta function as in
Eq.~(\ref{dEdfqnrapprox}) will often
(but not always) provide a fairly good approximation to the SNR; see
Appendix \ref{ringdown3} for more details.

The values of the spin $a$ of the final black hole (and
correspondingly, $Q$) and also of the amplitude ${\cal A}$ will depend
on the initial parameters of the system as in Eq.~(\ref{dependson}).
This dependence is very poorly understood at present.
We expect the final black hole to be rapidly spinning since, as
explained in Sec.~\ref{mergerphase}, the total angular momentum
of the binary at the end of the inspiral is $\sim 0.9 M^2$ when even
the individual black holes are non-spinning \cite{explainorb}, and the
individual black hole spins can add to this.  Moreover, the individual
black holes in the binary may typically have been spun up to near
maximal rotation by an accretion disk \cite{kipbhspin}.
For definiteness, we somewhat arbitrarily take $a=0.98$, which
corresponds from Eq.~(\ref{fqnrandQ}) to $Q = 12$ and $f_{\rm qnr} =
0.13/M$.  The final ringdown SNR values we obtain vary only weakly with
our assumed value of $a$ [{\it cf}.~Eq.~(\ref{qnrsnrII})], for
fixed total energy radiated in the ringdown.

Although the value of the overall amplitude ${\cal A}$ is uncertain, we
can estimate a upper bound on ${\cal A}$ for equal mass BBHs in the
following way.  When the $l=m=2$ quasinormal mode becomes
the dominant mode present, the merged body will be a distorted
Kerr black hole.  The horizon's cross section, seen looking down on this
black hole from above, will be a rotating oval rather than a circle.  We
can quantify the distortion by the ratio of the polar circumference
about the long axis of this oval to the polar circumference about the
short axis.  Let ${\cal A}_2$ denote the perturbation amplitude such
that the ratio of circumferences is, say, $2:1$.  It is fairly clear that
nonlinear couplings between the $l=m=2$ mode and other modes will be
important for ${\cal A} \agt {\cal A}_2$; thus, at this $2:1$ distortion
ratio , the signal will not be well approximated by just the $l=m=2$
mode.  Hence, ${\cal A}_2$ is a reasonable upper bound for the true
amplitude ${\cal A}$.

We could in principle calculate the upper bound ${\cal A}_2$ by writing
the spacetime metric as
$$
g_{ab} = g_{ab}^{\rm KERR} + {\cal A}_2 \,h_{ab}^{\rm QNR}
$$
where $g_{ab}^{\rm KERR}$ is the Kerr metric and $h_{ab}^{\rm QNR}$ is
the $l=m=2$ quasinormal mode whose asymptotic form at large $r$ is
given by Eq.~(\ref{qnrwaveform}), and by calculating from this
metric the ratio of circumferences \cite{noteguage}.  For this paper,
we used a much less sophisticated method to estimate ${\cal A}_2$.
Using the quadrupole formula, we examined the radiation produced by
a solid body that is distorted with the same $2:1$ circumference ratio
as the merged black hole, and obtained the estimate ${\cal A}_2
\sim 0.4$~\cite{Aestimate}. Setting our waveform amplitude ${\cal A}$
equal to this upper bound yields an rms angle-averaged waveform $h =
(0.4/\sqrt{4\pi})(M/r) = 0.1 (M/r)$ at the beginning of ringdown.  From
Eq.~(\ref{dEdfqnr}), the corresponding radiated energy is
\begin{equation}
E_{\rm ringdown} \approx {1\over8} {\cal
A}^2 M^2 f_{\rm qnr} Q \approx 0.03 M.
\label{qnrenergytot}
\end{equation}
As mentioned in the Introduction, comparable ringdown radiation
efficiencies $\sim 3\%$ have been seen in numerical simulations of the
evolution of distorted, spinning black holes \cite{seidel}.

To summarize, our assumed parameter values for the black hole
dimensionless spin parameter $a$ and for the dimensionless amplitude
parameter ${\cal A}$ for equal-mass BBHs are
\begin{eqnarray}
a &=& 0.98 \nonumber \\
{\cal A} &=& 0.4.
\label{qnrvals}
\end{eqnarray}
These imply the following values for the quasi-normal ringing
frequency $f_{\rm qnr}$, the quality factor $Q$ and
the fraction of total mass-energy radiated during ringdown
$\epsilon_{\rm ringdown} = E_{\rm ringdown}/M$:
\begin{eqnarray}
f_{\rm qnr} &=& {0.13 \over M} \nonumber \\
Q &\simeq& 12 \nonumber \\
\epsilon_{\rm ringdown} &=& 0.03.
\label{qnrenergy}
\end{eqnarray}
For non equal-mass BBHs, we assume that ${\cal A}$ and $\epsilon_{\rm
ringdown}$ are reduced by the factor (\ref{reductionfactor}).

\subsection{Number of independent frequency bins for the merger phase}
\label{mergernofilters}

In Sec.~\ref{derivesnrformula1} we showed that for any burst of
gravitational waves, the band-pass filtering SNR is smaller than the optimal
filtering SNR by a factor of approximately
\begin{equation}
\sqrt{{\cal N}_{\rm bins}} = \sqrt{2 T \Delta f}.
\label{factor}
\end{equation}
Here $T$ is the duration of the merger and $\Delta f$ is the band-pass
filter's bandwidth, which should be approximately the largest expected
signal bandwidth.  The product ${\cal N}_{\rm bins}$, which is roughly the
number of independent frequency bins (or Fourier coefficients) needed
to describe the signal, also determines how well matched filtering searches
perform compared to noise-monitoring searches, as explained in
Sec.\ \ref{nonlinear}.  In this section, we obtain a rough estimate of
the parameter ${\cal N}_{\rm bins}$  for the merger gravitational waves.

First consider the bandwidth $\Delta f$.  Our assumed bandwidth
for the merger signal is $\Delta f = f_{\rm qnr} - f_{\rm merge}
\approx f_{\rm qnr}$ [since $f_{\rm merge} \ll f_{\rm qnr}$;
{\it cf}.~Eqs.~(\ref{mergefreq}) and (\ref{fqnrdef})].  We cannot,
however, we completely confident that all signal power in the
merger will lie at frequencies below $f_{\rm qnr}$, so a more
appropriate choice might be $\Delta f \sim 2 f_{\rm qnr}$.
Also the quasinormal ringing frequency $f_{\rm
qnr}$ depends on the dimensionless spin parameter $a$ of the final
black hole as given by Eq.~(\ref{fqnrandQ}).
Choosing the highest possible value, $f_{\rm qnr} = 1 / (2 \pi M)$],
yields
\begin{equation}
\Delta f \sim {1 \over \pi M}.
\label{Deltafestimate}
\end{equation}

Turn, now, to the effective duration $T$ of the signal, which is
defined by Eq.~(\ref{Tdef}).  Clearly this parameter will vary
considerably from event to event and its value is highly uncertain.
To get a feeling for the range of possible values, let us consider the
type of coalescence discussed in Sec.~\ref{mergerphase}, where
both inspiraling black holes are nearly maximally spinning, with
their spins and the orbital angular momentum of the binary all
aligned.  In this favorable case, as argued in Sec.~\ref{mergerphase},
at the end of inspiral near $r \sim 6 M$ the binary has an excess
$\sim 0.4 M^2$ of total angular momentum that it needs to shed before
it can settle down to its final Kerr state.  Thus it seems reasonable to
suppose that the two black holes will be somewhat centrifugally hung up
and continue to orbit for several cycles before their event horizons
merge.  In other words, the orbital dynamical instability may be
suppressed by the aligned spins, so that what we call merger (for
data-analysis and computational purposes) may dynamically
resemble the inspiral.  The duration of the merger in this case may
therefore be long.  By contrast, for inspiraling non-spinning black
holes, there is probably no excess angular momentum that needs to
be radiated after the orbital dynamical instability is reached, and so
it is plausible that the two black holes merge rather quickly.  In such
a case, most of the emitted energy could come out in the ringdown
waves rather than in merger waves, as has typically been seen in
numerical simulations.

To estimate the duration $T$ for the first type of scenario, we
assume that the gravitational-wave luminosity $dE/dt$ during
the merger phase is roughly the same as the luminosity at the
start of the ringdown, $2 \epsilon_r M / \tau$.  (Here, $\epsilon_r =
0.03$ is the ringdown radiation efficiency, and $\tau$ is the damping
time of the QNR mode).  Combining this with a total energy radiated
during the merger of $\epsilon_m M$ (with $\epsilon_m = 0.1$)
yields the estimate
\begin{equation}
T = {1 \over 2} {\epsilon_m \over \epsilon_r} \tau.
\label{Testimate}
\end{equation}
Clearly this estimate will become invalid for high values of $\tau$;
in that limit, the high quality factor of the QNR mode causes a low
quasi-normal ringing luminosity, whereas there is no reason for the
merger luminosity to be comparably low.  Nevertheless, we insert our
assumed value of $\tau$, which corresponds to a quality factor of
$Q=12$ [Eqs.~(\ref{fqnrandQ}) and (\ref{qnrvals})] into
Eq.~(\ref{Testimate}) to obtain $T \sim  50 M$.  Combining this
with Eqs.~(\ref{simp1}) and (\ref{Deltafestimate}) yields the estimate
\begin{equation}
\sqrt{{\cal N}_{\rm bins}} \sim \sqrt{30} \sim 5.
\end{equation}
For inspiraling Schwarzschild black holes, on the other hand, $T$ may
not be much larger than $\tau = 10 M$ (assuming $a = 0.5$ say), giving
$\sqrt{{\cal N}_{\rm bins}} \sim \sqrt{6}$.

The factor $\sqrt{{\cal N}_{\rm bins}}$ is thus likely to lie in the
range $2 \alt \sqrt{{\cal N}_{\rm bins}} \alt 5$.  We adopt the
estimate $\sqrt{{\cal N}_{\rm bins}} = 4$ in Sec.~\ref{detectmerger}
below to estimate the reduction in SNR resulting from using band-pass
filters instead of optimal templates.  We use the conservative, large
value ${\cal N}_{\rm bins} = 60$ in Sec.~\ref{detectmerger} to
estimate detection thresholds for noise-monitoring searches for signals.

\section{INTERFEROMETER NOISE CURVES}
\label{noisecurves}

In this section we describe our approximate, piecewise power law,
analytic model for the noise curves for the initial LIGO
interferometers, for the advanced LIGO interferometers, and for the
LISA interferometer.  We express our model in terms of the
dimensionless quantity $h_{\rm rms}(f) \equiv \sqrt{f S_h(f)}$, where
$S_h(f)$ is the one sided power spectral density of the interferometer
noise \cite{twosided}.  The quantity $h_{\rm rms}(f)$ is the rms
fluctuation in the noise at frequency $f$ in a bandwidth $\Delta f =
f$ \cite{300yrs}. Our approximate model for the noise spectrum is
\begin{equation}
\label{noisespec}
h_{\rm rms}(f) = \left\{ \begin{array}{ll} \infty & \mbox{ $f <
        f_s$,}\nonumber\\ h_m\,(\alpha f/f_m)^{-3/2} & \mbox{
        $f_s\le f < f_m/\alpha$}\nonumber\\ h_m &
        \mbox{ $ f_m/\alpha\le f < \alpha f_m $}\nonumber\\
        h_m \left[f / (\alpha f_m)\right]^{3/2} & \mbox{ $\alpha f_m
	<f$.}\\ \end{array} \right.
\end{equation}
This noise curve scales like $f^{3/2}$ at high frequencies and like
$f^{-3/2}$ at low frequencies, and has a flat portion at intermediate
frequencies.  The noise curve depends on four parameters: (i) A
lower shutoff frequency $f_s$ below which the interferometer noise
rapidly becomes very large and can be approximated to be infinite.
For the ground based interferometers, this low-frequency shutoff is
due to seismic noise, while for LISA it is due to accelerometer noise
(Ref.~\cite{LISAreport}, p. 23). (ii) A frequency $f_m$, which is the
location of the center of the flat portion of the spectrum.  (iii) A
dimensionless parameter $h_m$, which is the minimum value of $h_{\rm
rms}(f)$.  (iv) A dimensionless parameter $\alpha$ which determines the
width in frequency of the flat portion of the noise curve.  We approximate
the noise curves by piecewise power laws in this way for calculational
convenience.

For the initial and advanced LIGO interferometers, we determined
best-fit values of the parameters $f_s$, $f_m$, $h_m$ and $\alpha$ by
fitting to the noise curves given in Ref.~\cite{ligoscience}.
(Note that Fig.~7 of Ref.~\cite{ligoscience} is a factor of 3 too
small from $\sim 10 \,{\rm Hz}$ to $\sim 70 \, {\rm Hz}$.  This error
does not appear in Fig. 10 of that reference \cite{factor3error}.)  The
resulting parameter values are:
\begin{equation}
\label{noisespecLIGOinitial}
\left. \begin{array}{lll} f_s &=& 40 \, {\rm Hz} \nonumber \\
		  f_m &=& 160 \, {\rm Hz} \nonumber \\
		  \alpha &=& 1.4 \nonumber \\
		  h_m &=& 3.1 \times 10^{-22}
        \end{array} \right\} \,\,\,
\begin{array}{l} {\rm INITIAL~LIGO} \nonumber \\
		 {\rm INTERFEROMETER},
\end{array}
\end{equation}
and
\begin{equation}
\label{noisespecLIGOadvanced}
\left. \begin{array}{lll} f_s &=& 10 \, {\rm Hz} \nonumber \\
		  f_m &=& 68 \, {\rm Hz} \nonumber \\
		  \alpha &=& 1.6 \nonumber \\
		  h_m &=& 1.4 \times 10^{-23}
        \end{array} \right\} \,\,\,
\begin{array}{l} {\rm ADVANCED~LIGO} \nonumber \\
		 {\rm INTERFEROMETER}.
\end{array}
\end{equation}
For ground-based interferometers, the $f^{-3/2}$ portion of our
approximate formula (\ref{noisespec}) is the thermal suspension noise
and the $f^{3/2}$ portion models the laser shot noise.  The
intermediate portion smoothes the transition between these two pieces
of the spectrum \cite{noisecomment}.

For the space-based LISA interferometer, we determined
best-fit values of the parameters $f_m$, $h_m$ and $\alpha$ by
fitting to the noise curve given in Ref.~\cite{Kipreview}, and
obtained the lower cutoff frequency $f_s$ from Ref.~\cite{LISAreport}.
The resulting parameter values are:
\begin{equation}
\label{noisespecLISA}
\left. \begin{array}{lll} f_s &=& 10^{-4} \, {\rm Hz} \nonumber \\
		  f_m &=& 3.7 \times 10^{-3} \, {\rm Hz} \nonumber \\
		  \alpha &=& 5.5 \nonumber \\
		  h_m &=& 5.8 \times 10^{-22}
        \end{array} \right\} \,\,\,
\begin{array}{l} {\rm LISA} \nonumber \\
		 {\rm INTERFEROMETER}.
\end{array}
\end{equation}
Our piecewise power-law model fits the
true noise curve less accurately for LISA than for the LIGO
interferometers, but it is still a fairly good approximation.

The sensitivity of LISA at the lower end of its frequency window may
be degraded somewhat by a background of gravitational waves from white
dwarf binaries \cite{LISAreport}.  We neglect this issue here as this
white dwarf noise level is fairly uncertain.

\newpage

\section{SIGNAL-TO-NOISE RATIOS}
\label{snrsection}

In this section we calculate the angle-averaged SNRs for
the three different stages of the coalescence (inspiral, merger,
and ringdown) for initial LIGO interferometers, for advanced LIGO
interferometers, and for LISA.  We do so for black hole binaries at
various distances and of various masses.  These SNR values are obtained
by substituting the energy spectra (\ref{dEdfinspiral}),
(\ref{dEdfmerger}) and (\ref{dEdfqnrapprox}) estimated in
Sec.~\ref{signalassumptions}, together with the interferometer
noise curves (\ref{noisespecLIGOinitial}),
(\ref{noisespecLIGOadvanced}) and (\ref{noisespecLISA}),
into the SNR formula (\ref{snraveraged}) derived
in Sec.~\ref{derivesnrformula}.  The analytic
calculations are
carried out in Appendix \ref{appsnr}; in this section we summarize
the results in three graphs
(Figs.~\ref{initialligosnr}, \ref{advancedligosnr} and \ref{lisasnr}), and
discuss their implications.

We start in subsection \ref{egs} by graphing the waves' energy spectra
and the interferometers' noise spectra for several different examples
of black hole binaries, in order to illustrate the factors determining
the SNR values.  The general results are presented in subsection
\ref{generalresults}.

\subsection{Specific examples}
\label{egs}

We start by rewriting the general formula (\ref{snraveraged}) for the
SNR in a more useful form.  If we define the characteristic
gravitational-wave amplitude
\begin{equation}
h_{\rm char}(f)^2 \equiv {2 (1+z)^2 \over  \pi^2 D(z)^2} \,
{dE\over df}[(1+z)f],
\label{hampdef}
\end{equation}
then from Sec.~\ref{derivesnrformula} it can be seen that the SNR
squared for an optimally oriented source is
\begin{equation}
\rho^2_{\rm optimal~orientation} = \int d(\ln f) \,
{h_{\rm char}(f)^2\over h_{\rm rms}(f)^2},
\label{snrOPT}
\end{equation}
where $h_{\rm rms}(f) = \sqrt{f S_h(f)}$.
However, Eq.~(\ref{snraveraged}) shows that the angle-averaged SNR
squared is a factor of $5$ smaller than the optimal value
(\ref{snrOPT}), so we can rewrite Eq.~(\ref{snraveraged}) as
\begin{equation}
\langle\rho^2\rangle = \int d(\ln f) \,
{h_{\rm char}(f)^2\over h_n(f)^2},
\label{snrSB}
\end{equation}
where $h_n(f) \equiv \sqrt{5} h_{\rm rms}(f)$.  We interpret the
quantity $h_n$ as the rms noise appropriate for waves from random
directions with random orientations \cite{hnexplain}.
The usefulness of the quantities $h_{\rm char}(f)$ and $h_n(f)$ is
that plotting $h_{\rm char}(f)$ and $h_n(f)$ for various sources
illustrates [from Eq.~(\ref{snrSB})] the possible SNR values and
the distribution of SNR squared with frequency (see,
{\it e.g.}, Ref.~\cite{300yrs}).

\def\DISPFIGONE{
{\psfig{file=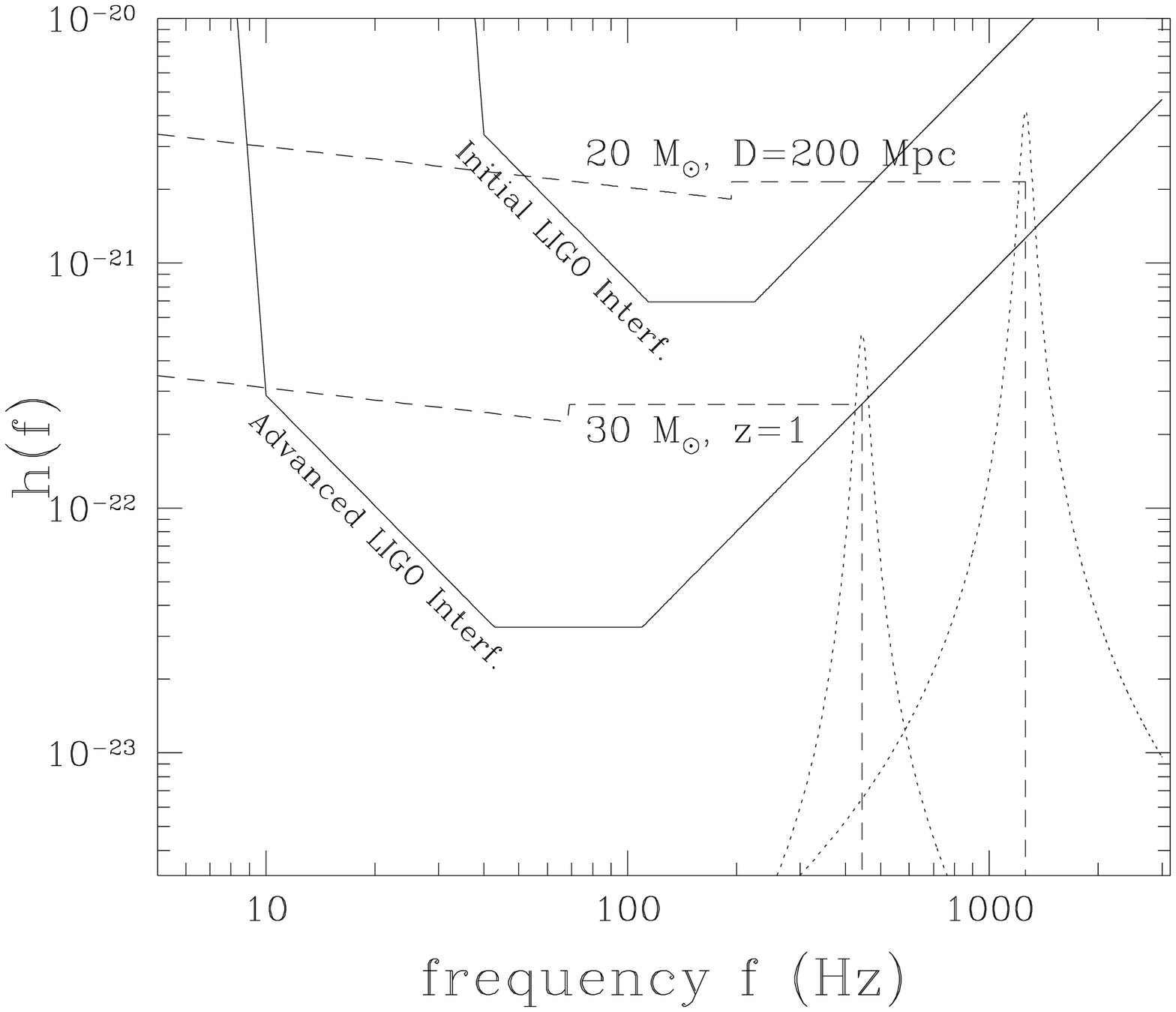,height=9.5cm,width=9cm}}
{\vskip -1.0cm}
\figure{
{An illustration of the relative magnitudes of our estimates of the
{\it inspiral}, {\it merger} and {\it ringdown} energy spectra in
two different cases.  The solid lines are the rms noise amplitudes
$h_n(f) \equiv \sqrt{5 f S_h(f)}$ for our assumed model
(\ref{noisespec}) of the
LIGO initial and advanced interferometer noise spectra.  The dashed and dotted
lines show the characteristic amplitude $h_{\rm char}(f)
\propto \sqrt{d E / df}$ of the waves, defined by Eq.~(\ref{hampdef}).
The definition of $h_{\rm char}$
is such that the signal-to-noise ratio squared for a randomly
oriented source is given by
$ (S/N)^2 = \int d (\ln f) \, \left[h_{\rm char}(f)/h_n(f)\right]^2.$
The upper dashed and dotted lines correspond to a
binary of two $10 M_\odot$ black holes, coalescing at a distance of $D
= 200$ Mpc.  The sloped portion of the dashed line
is the inspiral part of
the energy spectrum, which gives an SNR for the initial (advanced)
interferometer noise curve of $2.6$ ($84$).  The flat portion
is our crude model of the merger part of the energy spectrum, which
gives an SNR of $2.1$ ($16$).  The dotted line is our estimate of
the  energy
spectrum from the ringdown, which gives an SNR of $0.1$ ($0.86$).  An event
rate of very roughly $ 3 \, {\rm yr}^{-1}$ is expected for mergers inside
this distance \cite{narayan,phinney}.  The lower dashed and dotted lines
correspond to a binary of two $15 M_\odot$ black holes at a
redshift of $z=1$ (or at a luminosity distance of $D = 4.6$ Gpc;
the cosmological parameters $\Omega_0 =1$ and $H_0 = 75 \, {\rm km\,
s^{-1} Mpc^{-1}}$ were assumed).  In
this case the inspiral, merger and ringdown SNR values for the initial
(advanced) interferometers are
$0.08$, $0.42$, and $0.07$ ($6.6$, $7.2$ and $0.5$) respectively.
Black hole binaries with constituents this massive will be
visible to great distances, making them a possible important source,
depending on the very uncertain event rate.  The
signal-to-noise ratio from the merger phase is enhanced for these
massive distant sources, in part because the combination of
cosmological redshift plus lower intrinsic frequency (due to higher
mass) brings the merger part of the energy spectrum down to lower
frequencies at which the interferometer noise is smaller.}
\label{hcharfig}}
}

\DISPFIGONE

In Fig.~\ref{hcharfig}, we show the rms noise amplitude $h_n(f)$ for
our model (\ref{noisespec}) of the initial and advanced LIGO
interferometer noise curves, together
with the characteristic amplitude $h_{\rm char}(f)$ for two different
examples of BBH coalescences, namely a coalescence of total mass $20
\,M_\odot$ at a distance of $D = 200\,{\rm Mpc}$, and a $30 \,M_\odot$
coalescence at a redshift of $z=1$.  (We assume that the cosmological
parameters are $\Omega_0 = 1$ and $H_0= 75\,{\rm km\,s^{-1}\,Mpc^{-1}}$.)
In each case, the sloped portion of the dashed $h_{\rm char}$ line is
the inspiral signal, the flat
portion is our crude model of the merger, and the separate dotted
portion is the ringdown.  Note that the ringdown and merger overlap in
the frequency domain since (as we have defined them) they are disjoint
in the time domain.  By contrast, the inspiral and merger are
approximately disjoint in both the frequency and time domains, since
the adiabatic approximation is just breaking down at the end of the
inspiral \cite{freqtime}.


Notice that in both cases, $20 \, M_\odot$ and $30 \, M_\odot$, the
waves' characteristic amplitude $h_{\rm char}(f)$ is rather larger
than $h_n(f)$ for most of the merger spectrum for the advanced
interferometers, indicating the detectability of the
merger waveform when optimal filtering can be used.  In particular,
note that the waves should be quite visible to the advanced
interferometers for the $30\,M_\odot$ binary
even though it is at a cosmological distance.  Even if such binaries
are rare, they are visible to such great distances that measurements
can be made of a huge volume of the Universe, making them a
potentially very important and interesting source.  Cosmological
binaries have an enhanced SNR in part because the cosmological
redshift moves their frequency spectrum down closer to LIGO's optimal
band; this redshift enhancement effect can sometimes be so strong that
for two identical mergers at different distances, the more distant
merger can have the larger SNR.

Figure \ref{hcharfig} also shows that, of these two example BBH
coalescences, only the nearby one at a distance of $D = 200 \, {\rm
Mpc}$ would be detectable by the initial interferometers.  Such
coalescences may yield an interesting event rate for the initial
interferometers---as discussed in the Introduction, a coalescence
rate of very roughly $3 \, {\rm yr}^{-1}$ is expected for BBH
coalescences in this mass range ($M \alt 30 M_\odot$) inside $200 \,
{\rm Mpc}$ \cite{narayan,phinney}, although there is considerable
uncertainty in this estimate.

A qualitatively different, possibly important type of source for the
initial LIGO interferometers (and also for the advanced
interferometers) is the coalescence of black hole binaries with masses
of order $M \sim 100 \, {\rm M_\odot}$, as we have discussed in the
Introduction.  (We call such binaries intermediate mass BBHs, as
distinct from solar mass or supermassive binaries.)  In
Fig.~\ref{hcharfig1} we show the characteristic amplitude $h_{\rm
char}(f)$ for a hypothetical BBH coalescence of total mass $100
\,M_\odot$ at a redshift of $z = 0.5$, corresponding to a luminosity
distance of $D = 2.2\,{\rm Gpc}$.  Note in particular that the initial
LIGO interferometer noise curve has best sensitivity at $\sim 200 \,
{\rm Hz}$ just where the (redshifted) ringdown frequency is located.
We discuss further in Sec.~\ref{detection} the range and
possible detection rates for the initial LIGO interferometers for this
type of source.

\def\DISPFIGTWO{
{\psfig{file=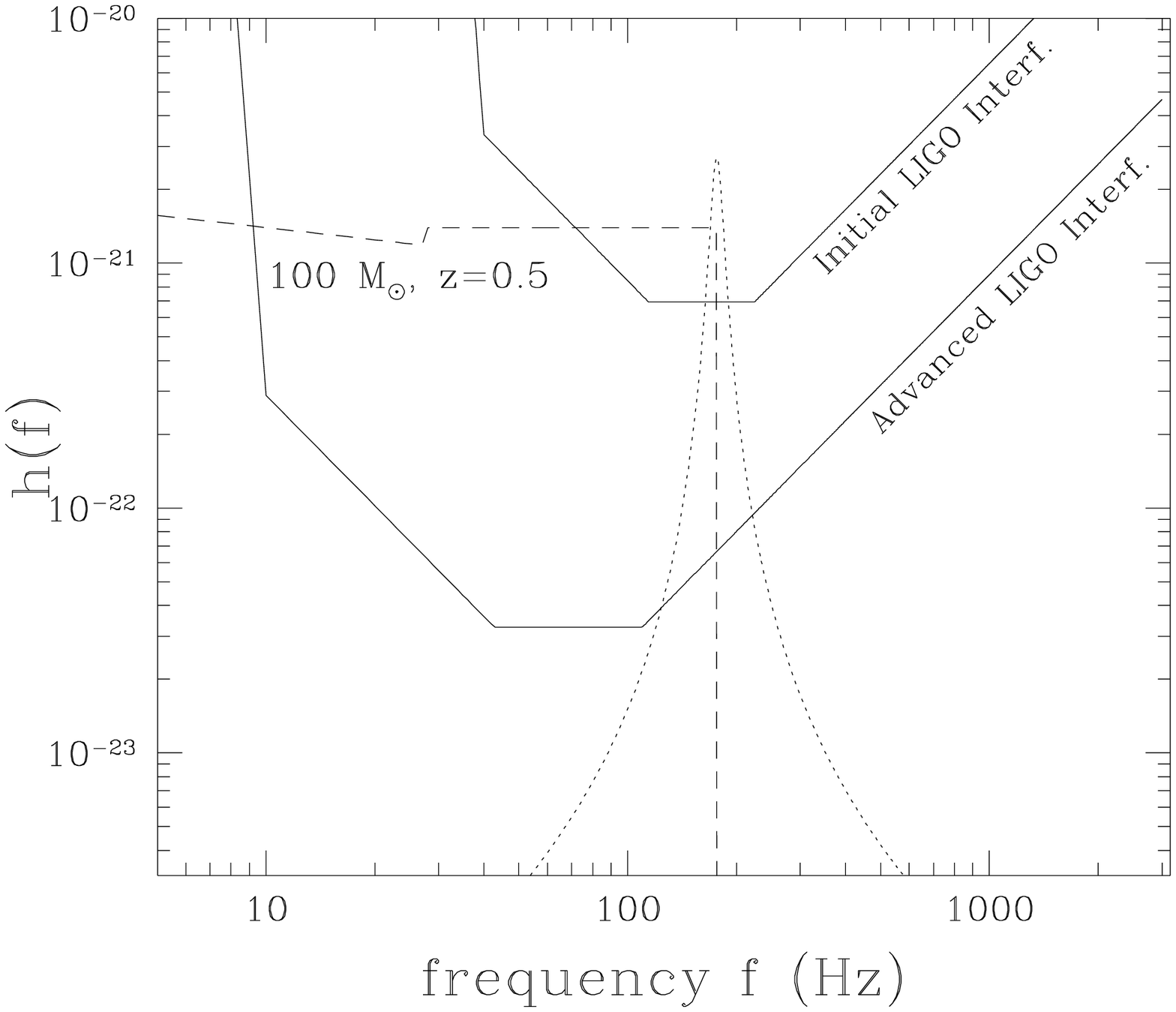,height=9.5cm,width=9cm}}
{\vskip -1.0cm}
\figure{
{A merger of a binary consisting of two $50 \, M_\odot$ black holes at
a redshift
of $z=0.5$, together with the rms noise amplitudes $h_n(f)$ for both
the initial and advanced interferometer noise curves for LIGO (see caption of
Fig.~\ref{hcharfig}).  The signal to noise ratios for the inspiral,
merger, and ringdown stages are about $0$, $1.7$ and $1.0$ respectively
for the initial interferometer noise level, and about $11$, $52$ and $11$
respectively for advanced interferometers.
}
\label{hcharfig1}}
}

\DISPFIGTWO

Turn, now, to the detection of supermassive BBH
signals by the
possible future space-based detector LISA \cite{cornerstone,lisa}.
LISA can study BBH mergers with far
higher accuracy and resolution than the ground based interferometers,
because the SNR values are typically much higher ($\agt 10^3$).
Our calculation of the inspiral SNR values for LISA
differs from the other SNR calculations in the following way.
In the frequency integral of Eq.~(\ref{snraveraged}),
if one were to integrate over the whole
frequency domain allowed by the noise curve (as is done for the LIGO initial
and advanced interferometers), one would in some cases obtain the SNR for a
measurement which lasts several hundred years, which is unrealistic.
Thus, it is necessary when calculating inspiral SNRs for LISA to restrict
the integral over frequency in Eq.~(\ref{snraveraged}) to the domain that
corresponds to, say, one year of observation; see Appendix \ref{appsnr}.

Figure \ref{hcharfig2} shows our approximate model
[Eqs.~(\ref{noisespec}) and (\ref{noisespecLISA})] of LISA's projected
noise spectrum, together with the gravitational-wave amplitude $h_{\rm
char}(f)$ for the inspiral, merger and ringdown stages of two
different BBH coalescences: a BBH of total mass $10^6 M_\odot$ at a
redshift of $z=5$, and a BBH of total mass $5 \times 10^4 M_\odot$ at
a redshift of $z=1$.  The $10^6 M_\odot$ BBH enters the LISA waveband
at $f=f_s \simeq 10^{-4} \, {\rm Hz}$ roughly one week before the final
merger. The SNRs obtained in this case from the inspiral, merger and
ringdown signals are approximately $1800$, $4600$ and $1700$
respectively.  The $5 \times 10^4 M_\odot$ BBH enters the LISA
waveband about twenty years before the final merger.  The SNR obtained
from the last year of the inspiral signal, from $f\simeq 1.6 \times
10^{-4} \, {\rm Hz}$ to $f \simeq 4 \times 10^{-2} \, {\rm Hz}$ is
approximately $900$, while the merger and ringdown SNRs are about
$70$ and $4$ respectively.

\def\DISPFIGTHREE{
{\psfig{file=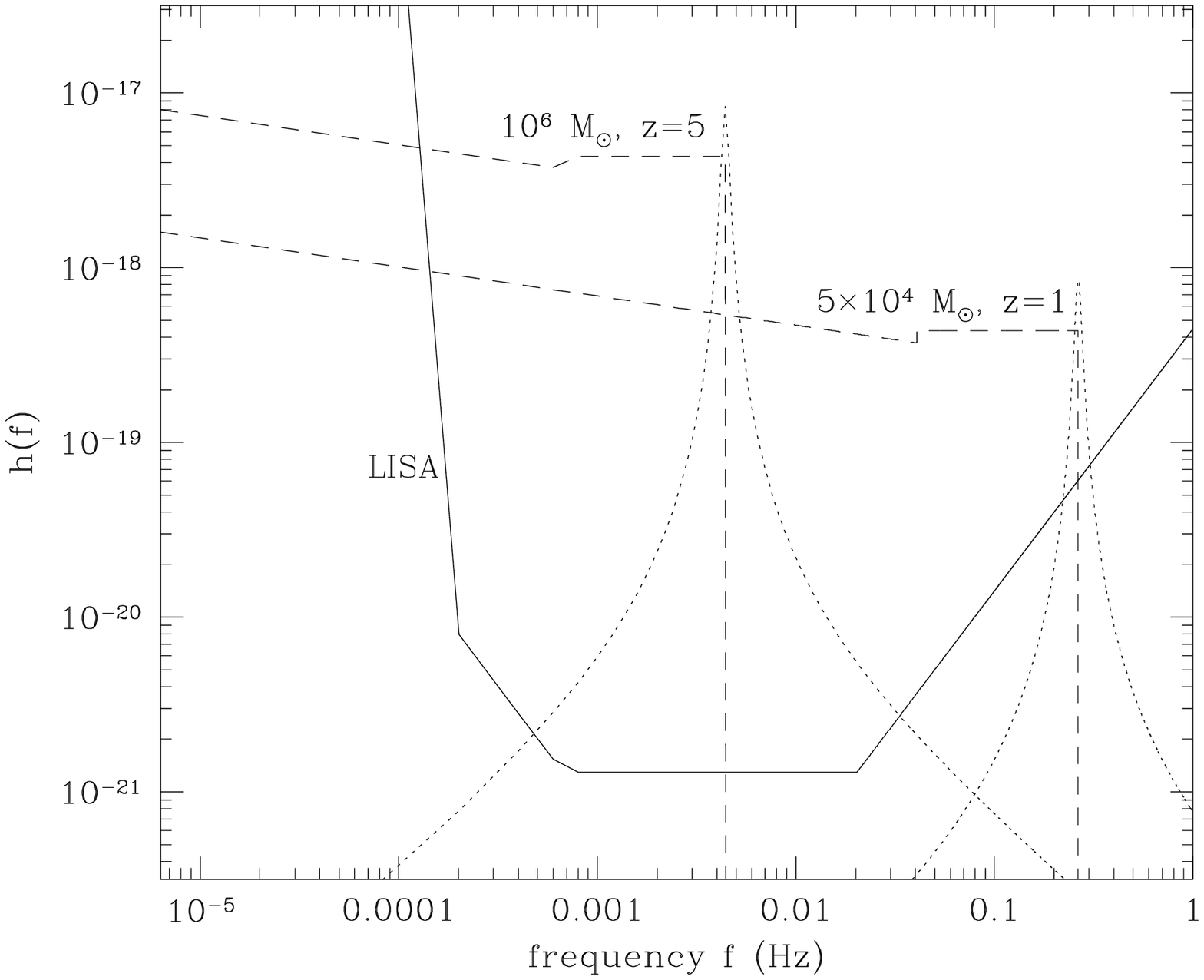,height=9.5cm,width=9cm}}
{\vskip -1.0cm}
\figure{
The noise spectrum $h_n(f)$ of the possible future space-based
detector LISA (Laser Interferometer Space Antenna), together with the
characteristic amplitude $h_{\rm char}$ for the various stages of two
different equal-mass binary black hole coalescences (see caption of
Fig.~\ref{hcharfig}).
The first coalescence is that of a binary of total mass $10^6 \,
M_\odot$ at a redshift of $z=5$.  The inspiral signal of this binary
enters the LISA waveband at $f \simeq 10^{-4} \, {\rm Hz}$ about one week
before the final merger, and the signal-to-noise ratios from the
inspiral, merger and ringdown are about $1800$, $4600$ and $1700$
respectively.  The second coalescence is that of a binary of total mass
$5 \times 10^4 M_\odot$ at a redshift of $z=1$, which enters the LISA
waveband about twenty years before the final merger.  In this case a
signal-to-noise ratio of approximately $900$ would be obtained for the
last year of inspiral (from $f\simeq 1.6\times10^{-4}\,{\rm Hz}$ to $f
\simeq 4\times10^{-2}\,{\rm Hz}$).  The signal-to-noise ratios from the
merger and ringdown in this case would be about $70$ and $4$.
\label{hcharfig2}}
}

\DISPFIGTHREE

\subsection{The general signal-to-noise ratio results}
\label{generalresults}

We now turn from these specific examples to the dependence of the
SNR values on the mass of and distance to the binary in general.  In
Appendix \ref{appsnr} we obtain analytic formulae for the SNR values
for the three phases of BBH coalescences, and for the various
interferometers.  In this section we plot the results for equal-mass
BBHs, which are shown in
Figs. \ref{initialligosnr},
\ref{advancedligosnr} and \ref{lisasnr}.  The inspiral and merger
curves in these figures (except for the LISA inspiral curves; see
Appendix \ref{appsnr}) are obtained from Eqs.~(\ref{inspsnrformula2}) and
(\ref{mergesnrformula2}) of Appendix \ref{appsnr}, while the ringdown
curves are obtained by numerically integrating Eq.~(\ref{dEdfqnr}) in
Eq.~(\ref{snraveraged}).

\def\DISPFIGFOUR{
{\psfig{file=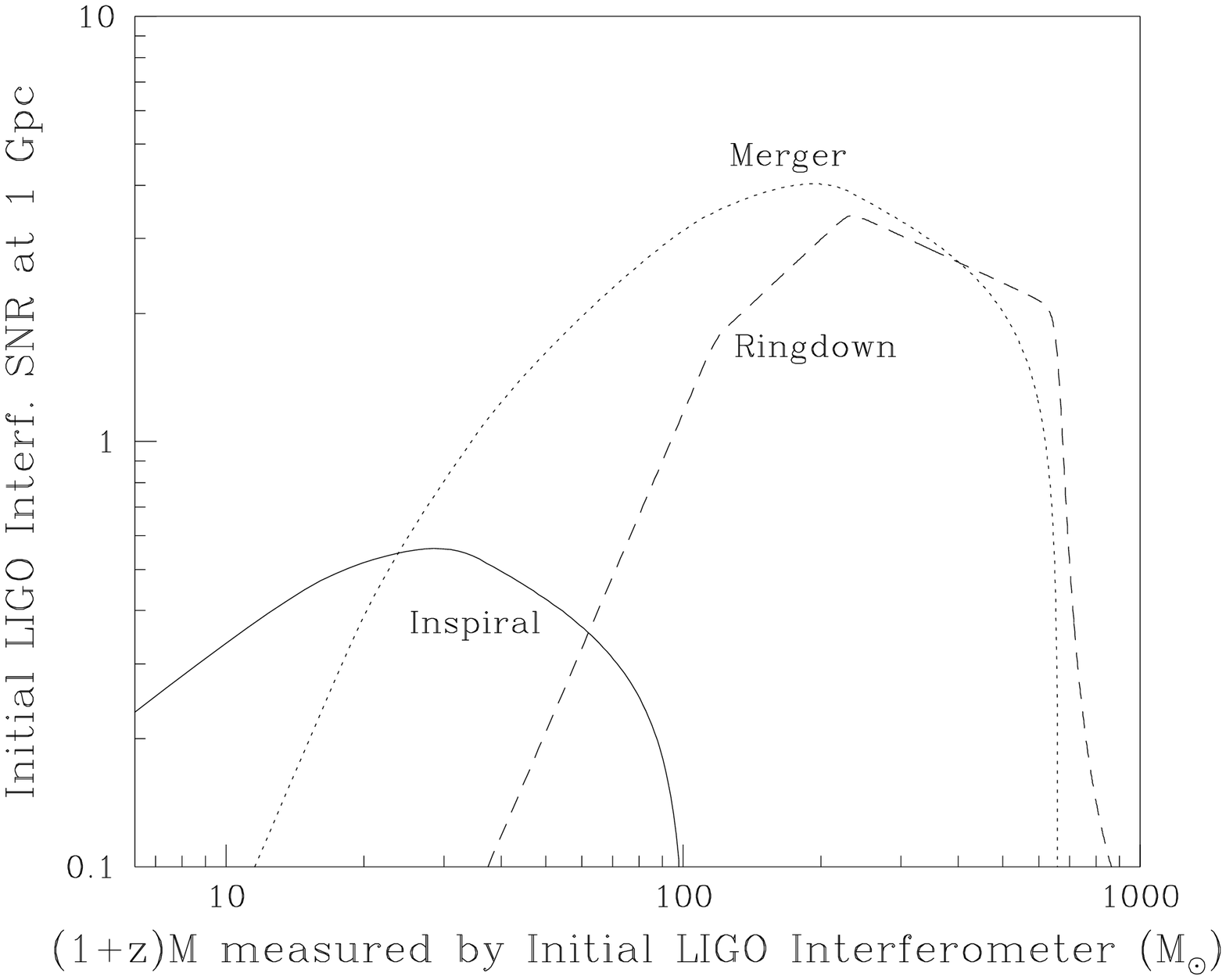,height=9.5cm,width=9cm}}
{\vskip -1.0cm}
\figure{
{The signal to noise ratio (SNR) for equal-mass
black hole-black hole coalescences detected by LIGO initial
interferometers, assuming Wiener optimal filtering, as a
function of the redshifted mass $(1+z) M$ of the final
black hole, at a luminosity distance of $1 \, {\rm Gpc}$.  For fixed
redshifted mass, the SNR values are inversely proportional to
luminosity distance.  The solid, dotted, and dashed curves are the SNR
values from the inspiral, merger and ringdown portions of the signal
respectively.  For non equal-mass binaries, the inspiral SNRs will be
reduced by the factor $\sim \sqrt{4\mu/M}$, while the merger and ringdown SNRs
will be reduced by $\sim 4 \mu/M$; thus the inspiral signal will
be enhanced relative to the merger and ringdown signals.
This plot indicates that it is possible that
an important source for the initial LIGO interferometers may be the
coalescences of binary black holes where
the black holes have masses in the range of several hundred solar
masses, which would be visible out to almost $1 \, {\rm Gpc}$.
For such sources, the inspiral portion of the signal
would not be detectable, and the waves would have to be detected
using either the ringdown or merger portions.}
\label{initialligosnr}}
}
\DISPFIGFOUR

The SNR values for the initial LIGO interferometers are shown in
Fig.~\ref{initialligosnr}.  This figure shows that it is possible that
an important source for the initial LIGO interferometers may be the
coalescences of binary black holes where the black
holes have masses in the range of several hundred solar masses, which
would be visible out to almost $1 \, {\rm Gpc}$.  For such sources,
the inspiral portion of the signal would not be detectable, and one
would need to search for the ringdown or merger portions of the signal
in order to detect the waves.  See Sec.~\ref{detection} for
further discussion.  The event rate for such high mass BBHs is very
uncertain; see Ref.~\cite{QuinlanShapiro} for a possible formation
scenario.

Intermediate mass BBHs with $\mu \ll M$ ({\it e.g.}, $m_1 \sim 10\,
M_\odot$, $m_2 \sim 500 M_\odot$) are presumably much more common in the
Universe than the intermediate mass BBHs with $\mu \sim M$ discussed
above.  The SNRs for such mixed binaries will be much lower, however.
As seen in Appendix A, the merger and ringdown SNR's scale as
$(\mu/M)^2$, while the inspiral ringdown scales as $\mu/M$.  The
difference in scaling occurs because the duration of the merger and
ringdown signals are approximately independent of $\mu$, but the
duration of the inspiral signal scales as $1/\mu$.

For BBH coalescences of total mass less than or of order
20 $M_\odot$, the results of Fig.~\ref{initialligosnr} can be
summarized by the following approximate formulae for the initial
LIGO interferometers [c.f.~Eqs.~(\ref{inspsnrformula2}),
(\ref{mergesnrformula2}), and (\ref{qnrsnr2}) of Appendix \ref{appsnr}]:
\begin{eqnarray}
& &\left( {S \over N} \right)_{\rm inspiral}\sim 2.8 \left({200\,{\rm
	Mpc}\over D(z)}\right) \left({{(1+z)M}\over18\,M_\odot}\right)^{5/6}
	\nonumber \\
& &\left( {S \over N} \right)_{\rm merger} \sim 1.5 \left({\epsilon_m\over
	0.1}\right)^{1/2} \left({200\,{\rm Mpc}\over D(z)}\right)
	\left({{(1+z)M}\over18\,M_\odot}\right)^{5/2} \nonumber \\
& &\left( {S \over N} \right)_{\rm ringdown} \sim 0.1
	\,\left( {\epsilon_r \over 0.03} \right)^{1/2}
	\left({200\,{\rm Mpc}\over D(z)}\right) \left({{(1+z)M}
	\over18\,M_\odot}\right)^{5/2},\nonumber\\
\label{lowmassinitial}
\end{eqnarray}
where $\epsilon_m$ is the fraction of the total mass-energy radiated in
the merger signal and $\epsilon_r$ is the corresponding fraction for
the ringdown signal.

Figure \ref{initialligosnr} also shows that the inspiral signals from
low mass BBH mergers with $M \alt 30 M_\odot$ should be visible to $\sim 200 \,
{\rm Mpc}$ (the SNR detection threshold is $\sim 5$ \cite{prl}).
The ground-based interferometers will, over a period of years,
gradually be improved from the initial sensitivity levels to the
advanced sensitivity levels \cite{ligoscience}.  Roughly half way
between the initial and advanced interferometers, the range of the
detector system for $M \alt 30 M_\odot$ BBHs will be $\sim 1 \, {\rm
Gpc}$.  As explained in the Introduction, it is thought to be
plausible that most BBH progenitor systems do not disrupt during the
stellar collapses that produce the black holes, so that their
coalescence rate could be about the same as the birth rate for their
progenitors, about $1/100,000$ years in our Galaxy, or several per
year within a distance of $200 \, {\rm Mpc}$
\cite{narayan,phinney,heuvel,tutukov,yamaoka}.  If this is the case,
BBHs should be detected early in the gradual process of
interferometer improvement.

It is also conceivable that BBHs could be seen closer than $200 \,
{\rm Mpc}$.  High mass stars may have been much more numerous among
population III stars due to the lower metallicity at high redshift, so
there could be a large population of BBHs that are remnants of
population III stars living in galactic halos.  Galactic nuclei could
also produce a larger population of BBHs than normal stellar
evolution in non-nuclear regions \cite{QuinlanShapiro}.  Either of
these scenarios could increase event detection rate and hasten the
first detections of BBHs.

\def\DISPFIGFIVE{
{\vskip 0.5cm}
{\psfig{file=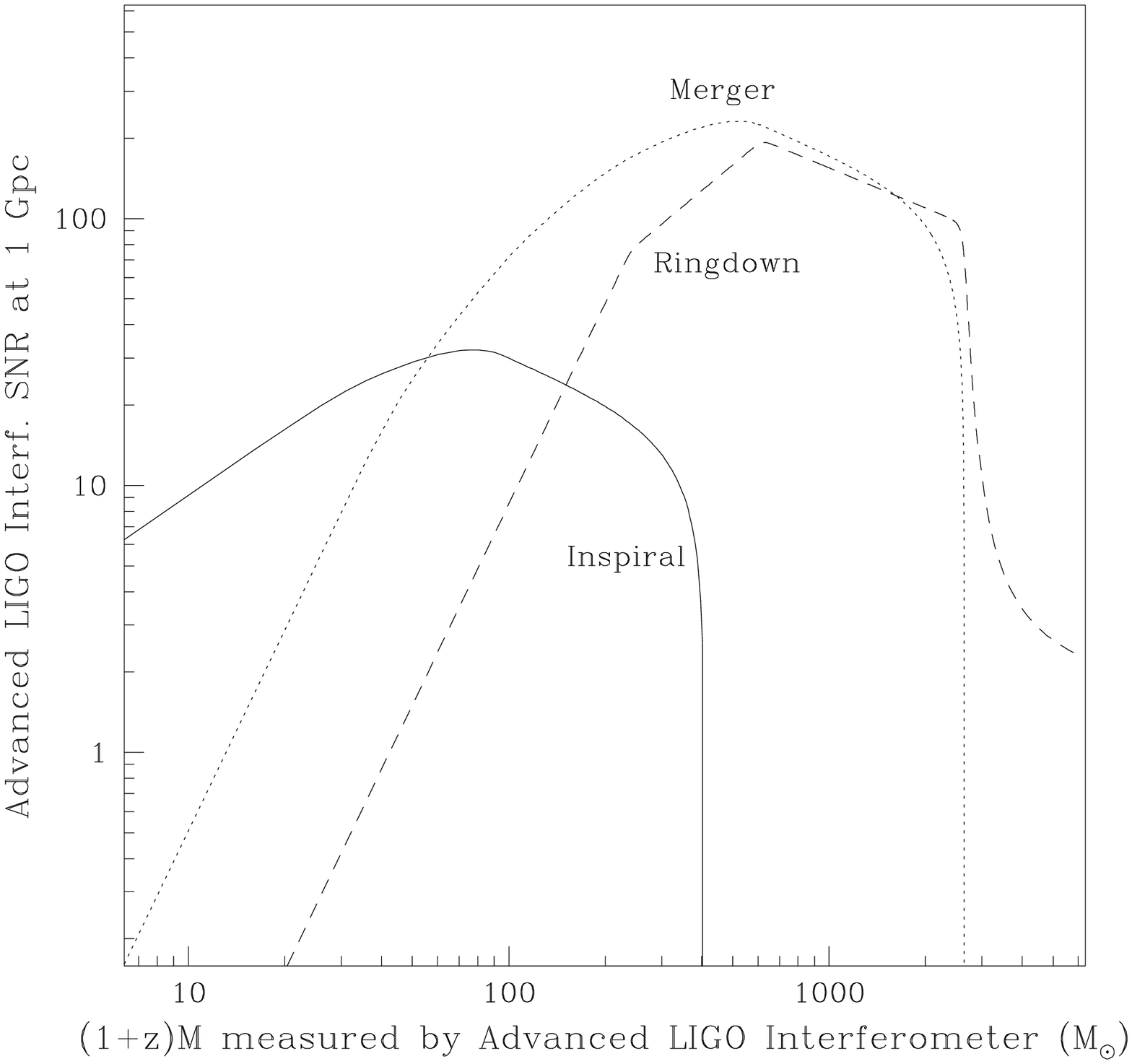,height=9.5cm,width=9cm}}
{\vskip 0.0cm}
\figure{
{The SNR values for advanced LIGO interferometers for the inspiral
(solid line), merger (dotted line) and ringdown (dashed line) phases
of equal-mass BBH coalescences; see caption of
Fig.~\ref{initialligosnr}.  For fixed
redshifted mass, the SNR values are inversely proportional to
luminosity distance.
For values of the redshifted final mass lower than
$\sim 60 \, M_\odot$ the inspiral SNR is largest, while for
larger BBH systems the merger and/or ringdown portions of the signal
dominate.}
\label{advancedligosnr}}
}

\DISPFIGFIVE

Figure \ref{advancedligosnr} shows the SNR values for the advanced
LIGO interferometers.  For BBH coalescences of total mass less than or
on the order of a few tens of solar masses, the results in
Fig.~\ref{advancedligosnr} can be summarized by the following
approximate formulae for the advanced LIGO interferometers
[c.f.~Eqs.~(\ref{inspsnrformula2}), (\ref{mergesnrformula2})
and (\ref{qnrsnr2}) of Appendix \ref{appsnr}]:
\begin{eqnarray}
& &\left( {S \over N} \right)_{\rm inspiral} \sim 27 \left({1\,{\rm Gpc}\over
D(z)}\right) \left({{(1+z)M}\over37\,M_\odot}\right)^{5/6} \nonumber \\
& &\left( {S \over N} \right)_{\rm merger} \sim 13 \left({\epsilon_m\over
	0.1}\right)^{1/2} \left({1\,{\rm Gpc}\over D(z)}\right)
	\left({{(1+z)M}\over37\,M_\odot}\right)^{5/2} \nonumber \\
& &\left( {S \over N} \right)_{\rm ringdown} \sim 0.7
	\,\left( {\epsilon_r \over 0.03} \right)^{1/2}
	\left({1\,{\rm Gpc}\over D(z)}\right) \left({{(1+z)M}\over37\,M_\odot}
	\right)^{5/2}.\nonumber\\
\label{lowmassadv}
\end{eqnarray}
Our estimate (\ref{lowmassadv}) for the ringdown SNR is roughly consistent
with previous estimates by Finn \cite{finnmeasure}.

{}From Fig.~\ref{advancedligosnr} it can be seen that for advanced LIGO
interferometers, equal-mass BBH inspirals will be visible out to $z
\sim 1/2$ for
the entire range of masses $10 M_\odot \alt (1+z) M \alt 300 M_\odot$.
Thus, there is likely to be an interesting event rate.  Indeed, the
SNRs will be high enough even for rather large distances that it should
be possible to extract each binary's parameters with reasonable
accuracy~\cite{Kipreview}. By contrast, the ringdown SNR is fairly
small except for the largest mass systems.  For very massive
binaries or binaries that are closer than 1 Gpc, advanced
interferometers may measure some fairly good ringdown SNRs, which
would allow fairly good estimates of the mass and spin of the final
black hole~\cite{echeverria,finnmeasure}.

Figure \ref{lisasnr} shows the SNR values obtainable from the three phases
of BBH coalescences by LISA: the last year of inspiral, the merger and the
ringdown.  We also show the SNR value obtainable from one year of integration
of the inspiral signal one hundred years before the merger, and a
similar curve for one thousand years before the merger.
This figure shows that LISA will be able to perform very high accuracy
measurements of BBH mergers (SNR values $\agt 10^3$) essentially throughout
the observable Universe ($z \alt 10$) in the mass range $10^6 \, M_\odot
\alt M \alt 10^9 \, M_\odot$.  As discussed in the Introduction, there is
likely to be an interesting event detection rate.  The SNR curves in
Fig.~\ref{lisasnr} for measurements one hundred and one thousand years
before merger show that many inspiraling BBHs that are far from merger
should be detectable by LISA as well.  If the merger rate of SMBH binaries
turns out to be about one per year throughout the observable Universe, then
at any given time one would expect roughly one thousand SMBH binaries to be
a thousand years or less away from merger.  LISA will be able to monitor
the inspiral of such binaries (if they are of sufficiently low mass) with
moderate to large SNR \cite{Kipreview}.

Finally, it should be noted that the relative magnitude of the
merger and ringdown SNR values is somewhat uncertain.  As we have
discussed in Sec.~\ref{signalassumptions}, we have assumed a total
radiated energy of $0.1 M$ in the merger portion of the signal,
and $0.03 M$ in the ringdown portion, a ratio of $3:1$.  In reality,
it may turn out that in individual cases the ratio is as high as $10$
or as low as $ \alt 1$.  It may even turn out to be the case that for
most coalescences, the ringdown portion of the waveform carries most
of the radiated energy of the combined merger/ringdown regime
(depending possibly on the distribution of initial spins).  Thus, the SNR
values shown in Figs.~\ref{initialligosnr}, \ref{advancedligosnr} and
\ref{lisasnr} should merely be taken as illustrative.

We now turn to a discussion of the implications of the above
signal-to-noise ratios for the required search strategies for the various
types of signals, and for the distances to which these types of signals can be
detected.

\def\DISPFIGSIX{
{\vskip 0.5cm}
{\psfig{file=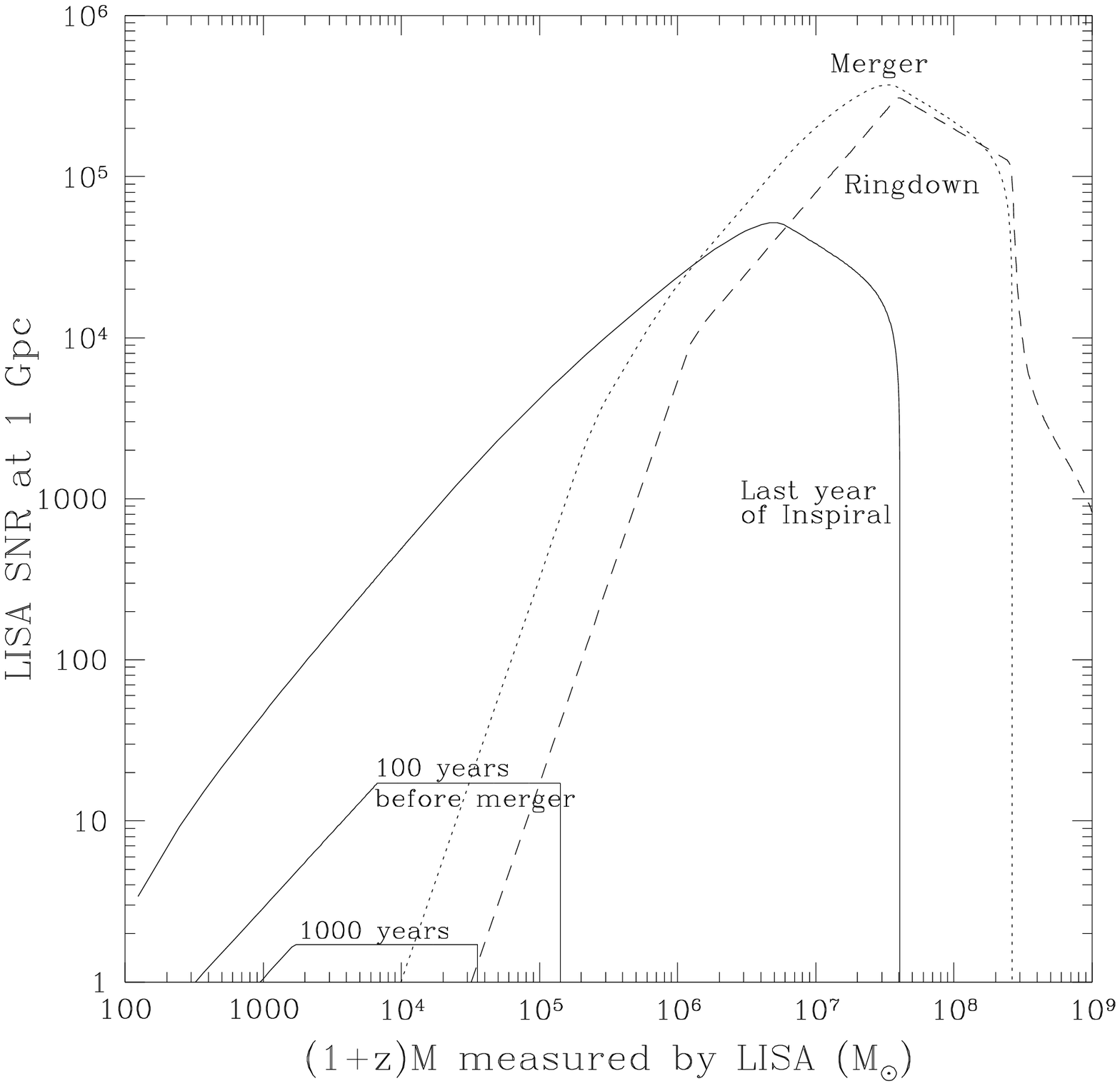,height=9.5cm,width=9cm}}
{\vskip 0.0cm}
\figure{
{The signal to noise ratio (SNR) for equal-mass black hole-black hole
coalescences detected by LISA, assuming Wiener optimal filtering, as a
function of the redshifted mass $(1+z) M$ of the final
black hole, at a luminosity distance of $1 \, {\rm Gpc}$.  For fixed
redshifted mass, the SNR values are inversely proportional to
luminosity distance.  The dotted and dashed curves are the SNR
values from the merger and ringdown portions of the signal, respectively.
The upper solid curve is the SNR value that would be obtained from
measuring the last year of the inspiral signal.  For values of the
redshifted final mass lower than about $10^6\, M_\odot$,
the last-year-inspiral SNR is largest, while for larger BBH systems
the merger and/or ringdown portions of the signal dominate.  Also shown
(lower solid curves) are the SNRs that would be obtained from one year of
integration of the inspiral signal at one hundred and one thousand years
before the final merger.  If the rate of SMBH coalescences within $z \alt
(\rm a\ few)$ is roughly one per year, then one would expect roughly one
thousand SMBH binaries to be a thousand years or less away
from merger.  This plot shows that LISA will be able to measure the
inspiral of such binaries (provided they are of sufficiently low mass)
with moderate to large SNR \cite{Kipreview}.
}
\label{lisasnr}}
}

\DISPFIGSIX

\section{IMPLICATIONS FOR DETECTABILITY OF\\
THE GRAVITATIONAL-WAVE SIGNAL}
\label{detection}

One of the reasons that coalescences of compact objects are such good
sources for gravitational-wave detectors is that the inspiral portion
of the signal is very predictable, so that optimal filtering may be
used in searching for the waves \cite{300yrs}.  As we have discussed,
optimal filtering enhances the achievable SNR values by a factor of roughly
$\sqrt{{\cal N}_{\rm cyc}}$, where ${\cal N}_{\rm cyc}$ is the number
of cycles in
the waveform in the frequency band of the detectors.  For neutron
star-neutron star (NS-NS) coalescences, ${\cal N}_{\rm cyc}$ will be on the
order of several thousand, while for BBH coalescences it will be on
the order of several hundred \cite{prl}.  Thus, for NS-NS
coalescences, and also for BBH coalescences with total mass a few tens
of solar masses or less, the inspiral portion of the gravitational-wave
signal will be used to {\it detect} the entire waveform.  In
these cases, it is not necessary to {\it search} for the merger and
ringdown portions of the waveform, since it will be known roughly
where in the interferometer data stream they are expected to lie.
However, there is no guarantee in these cases that the merger and ringdown
portions will be visible in the noisy data stream.  A key issue is how
visible these portions are and how much information can be extracted
from them.  We discuss this issue in detail in paper II \cite{paperII}.

For larger mass black hole systems, however, our signal to noise ratio
calculations show that the merger and ringdown SNR values can be
larger than the inspiral SNR
values.  For equal-mass BBHs, this will occur whenever $(1+z) M \agt
30 M_\odot$ for the initial LIGO interferometers, and whenever $(1+z)
M \agt 60 M_\odot$ for the advanced LIGO interferometers.  Moreover,
the inspiral SNR  completely shuts off for large enough $(1+z)M$,
as can be seen from Figs.~\ref{initialligosnr} and \ref{advancedligosnr}.
The reason for this shutoff is that at large enough values of $(1+z) M$,
the entire inspiral energy spectrum lies at frequencies below the ``seismic
wall'' shutoff in the interferometer noise curve.  The mass cutoff
occurs at $(1+z) M \sim 100 M_\odot$ for the initial interferometers,
and at $(1+z) M \sim 400 M_\odot$ for the advanced interferometers.
Admittedly, BBH binaries of total mass $\gg 20 M_\odot$ may well be
very much more rare than BBH binaries of $\sim 20 M_\odot$; however,
they will be visible to such great distances that there may be an
interesting event rate.  Moreover, for the initial LIGO interferometers,
the mass scale $\sim 30 M_\odot$ at which the inspiral SNR becomes
much smaller than the merger/ringdown SNRs is not terribly high.

In such high mass cases for which the merger and inspiral SNRs exceed
the inspiral SNR, it will be necessary to perform a search for the
merger and/or ringdown portions of the signal, independently of any
searches for inspiral signals, in order that all possible events be
detected.  If one seeks to detect the waves merely by optimal
filtering for the inspiral waveform, some fraction of the events will
be missed which otherwise might have been detectable.  In fact, it may
very well turn out that merger signals from BBH
coalescences could be the dominant source for the initial LIGO
interferometers.

Therefore a search in real time should be performed in the data stream
for ringdown and merger signals from high mass, BBH mergers.  One
might imagine that the gravitational waves would generally be easier
to detect by searching for the merger signal, since we have estimated
that the SNR values for the merger phase are typically a factor of a few
larger than those for the ringdown ({\it cf.}~Figs.~\ref{initialligosnr} and
\ref{advancedligosnr}).  There are several factors that complicate this
conclusion, however.  On the one hand, the ringdown's waveform shape is better
understood, which makes it easier to produce search templates and hence
easier to detect the signal.  On the other hand, the ratio between the
merger and ringdown SNRs is really quite uncertain, as discussed in
Sec.~\ref{generalresults}, and so it is plausible that the merger SNR will
be larger than we have indicated relative to the ringdown SNR.  In any
case, the ratio between merger and ringdown SNRs will presumably vary
a lot from event to event.  Thus, it would seem that searches will be necessary
for {\it both} types of signal in the data steam, at least for the mass
range in which the ringdown SNR is expected to exceed the inspiral
SNR.  [From Sec.~\ref{snrsection} we estimate this mass range to be
$(1+z) M \agt 200 \, M_\odot$ for the advanced interferometers, and
$(1+z) M \agt 60 M_\odot$ for the initial interferometers.]

We summarize the discussion of this section by displaying the optimum
search strategies for various mass ranges for the three different
interferometers.  In each case below, the mass range marked
merger refers to matched filtering searches for merger signals.  If
merger templates are available, then in the indicated mass ranges
merger searches will probably be more successful than inspiral or
ringdown searches; the question mark is a reminder that templates for the
merger may not be available, and that consequently such searches might
not be possible.
\begin{equation}
\label{searchspecLIGOinitial}
\left. \begin{array}{lll} {\rm INSPIRAL} &:& 1 M_\odot \alt M \alt 60
		M_\odot \nonumber \\
		{\rm RINGDOWN} &:& 60 M_\odot \alt M \alt 1000 M_\odot
\nonumber \\
		{\rm MERGER \,(?)}  &:& 30 M_\odot \alt M \alt 1000 M_\odot
\nonumber \\
        \end{array} \right\} \,\,\,
\begin{array}{l} {\rm LIGO} \nonumber \\
		 {\rm INITIAL} \nonumber \\
		 {\rm INTERF.}
\end{array}
\end{equation}

\begin{equation}
\label{searchspecLIGOadvanced}
\left. \begin{array}{lll} {\rm INSPIRAL} &:& 1 M_\odot \alt M \alt 200
		M_\odot \nonumber \\
		{\rm RINGDOWN} &:& 200 M_\odot \alt M \alt 3000 M_\odot
\nonumber \\
		{\rm MERGER \, (?)}  &:& 80 M_\odot \alt M \alt 3000 M_\odot
\nonumber \\
        \end{array} \right\} \,\,\,
\begin{array}{l} {\rm LIGO} \nonumber \\
		 {\rm ADVANCED} \nonumber \\
		 {\rm INTERF.}
\end{array}
\end{equation}

\begin{equation}
\label{searchspecLISA}
\left. \begin{array}{lll} {\rm INSPIRAL} &:& 10^3 M_\odot \alt M \alt 10^7
		M_\odot \nonumber \\
		{\rm RINGDOWN} &:& 10^7 M_\odot \alt M \alt 10^9 M_\odot
\nonumber \\
		{\rm MERGER \, (?)}  &:& 2 \times 10^6 M_\odot \alt M
\alt 10^9 M_\odot
\nonumber \\
        \end{array} \right\} \,\,\,
\begin{array}{l} {\rm LISA} \nonumber \\
		 {\rm INTERF.}
\end{array}
\end{equation}

\subsection{The detectability of high mass black-hole\\
coalescences via the ringdown signal}
\label{qnrsearches}

Consider first the search for ringdown signals.  In this case, since
the shape of the signal is known up to several unknown parameters, it
will be feasible to implement a search based on the method of matched
filters, analogous to what is being planned for detecting mergers of
NS-NS and
low mass NS-BH and BBH binaries using the inspiral waveform
\cite{prl,searchtemplates,owen,krolaknew}.  In other words,
a bank of templates of the form (\ref{qnrwaveform}), for different
values of the ringdown frequency and damping time, can be integrated
against the (suitably pre-filtered) interferometer output.
This will be feasible for both the initial and advanced LIGO
interferometers, and also for LISA.

The number ${\cal N}_{\rm templates}$ of required templates
\cite{importantnote} can be estimated by combining
the formalism developed by Owen \cite{owen} and the results of
Echeverria and Finn on the expected accuracy of measurement of of the
ringdown frequency and damping time \cite{echeverria,finnmeasure}.
Using Eqs. (4.15) of Ref.~\cite{finnmeasure} and Eqs.~(2.23) and
(2.28) of Ref.~\cite{owen} we find that the metric defined by Owen on
the space of parameters is given by \cite{details}
\begin{equation}
ds^2 = {1\over8 Q^2} dQ^2 + {Q^2 \over 2 f_{\rm qnr}^2} df_{\rm qnr}^2,
\label{metric}
\end{equation}
where $Q$ is the quality factor.  The formula (\ref{metric}) for the
Owen metric is valid only in the high $Q$ limit; it has corrections of
order $1/Q^2$.  Moreover, the formula is also only valid when the
noise spectrum $S_h(f)$ does not vary significantly within the
resonance bandwidth $\Delta f \sim f_{\rm qnr}/Q$.   Therefore estimates
obtained from Eq.~(\ref{metric}) for the number of template
shapes required for ringdown searches will only be accurate to within
factors of a few, but this is adequate for our purposes.

Using Eq.~(2.16) of Ref.~\cite{owen} we find that that the number of
required templates is approximately
\begin{equation}
{\cal N}_{\rm templates} \approx {1 \over 8} Q_{\rm max} (1 - MM)^{-1} \ln
\left[ M_{\rm max} \over M_{\rm min}\right],
\label{ringNtemplates}
\end{equation}
where $Q_{\rm max}$, $M_{\rm min}$ and $M_{\rm max}$ are the extremal
values of the quality factor and of the black hole mass that define
the range of signal searches.  The quantity $MM$
in the formula (\ref{ringNtemplates}) is the {\it minimal match}
parameter introduced by Owen; its meaning is that the event detection
rate obtainable by a lattice of templates that have a certain minimal
match $MM$ is smaller than the ideal event detection rate (obtainable
from an infinitely dense lattice of templates) by the factor $(MM)^3$
\cite{owen}.  We assume $MM = 0.97$ as in Ref.~\cite{owen}, which
corresponds to a $10 \%$ loss in event rate due to the coarse
gridding of the template space, and take $Q_{\rm max} = 100$ [which
by Eq.~(\ref{fqnrandQ}) corresponds to $1-a \simeq 10^{-4}$].  For the
initial and advanced LIGO
interferometers, the mass range to be searched corresponds to roughly
$M_{\rm min} \simeq 1 M_\odot$ and $M_{\rm max} = 5000 M_\odot$;
this yields from Eq.~(\ref{ringNtemplates})
\begin{equation}
{\cal N}_{\rm templates} \alt 4000.
\label{TheGoodNews}
\end{equation}
This is a rather small number of template shapes compared to the
number expected to be necessary for inspiral searches\cite{owen}, and
therefore the ringdown search should be fairly easy to implement.  A
similarly small number of required template shapes (${\cal N}_{\rm templates}
\alt 6000$) is obtained for LISA assuming  $M_{\rm min} \sim 10^3 M_\odot$
and $M_{\rm max} \sim 10^9 M_\odot$.

We next discuss the distance to which BBH mergers should be detectable
via their ringdown signals.  As explained in Sec.~\ref{nonlinear}, a rough
estimate of the appropriate signal-to-noise ratio threshold for
detection using one interferometer is \cite{exp}
\begin{equation}
\rho_{\rm threshold} \sim \sqrt{ 2 \ln [ {\cal N}_{\rm templates} T / (\epsilon
\Delta t)]}
\label{thr}
\end{equation}
where $T$ is the observation time, $\Delta t$ is the sampling time and
$\epsilon=10^{-3}$ is as defined in Section~\ref{nonlinear}.
In fact coincidencing between the 4 different interferometers in the
LIGO/VIRGO network will be carried out, in order to increase detection
reliability and combat non-Gaussian noise (see
Sec.~\ref{derivesnrformula1}).  If the noise were exactly
Gaussian, the appropriate detection criterion would be to demand that
\begin{equation}
\sum_j \rho_j^2 \ge \rho_{\rm threshold}^2,
\end{equation}
where the sum is over the different SNRs obtained in each interferometer.
In order to combat non-Gaussian noise, the detection criterion will be
modified to require approximately equal SNRs in each interferometer:
\begin{equation}
\rho_j \ge \rho_{\rm threshold}/\sqrt{2}\,\,\,\,{\rm for~all~}j.
\end{equation}
We have chosen a factor of $\sqrt{2}$ here to be conservative; it
corresponds to combining the outputs of just two interferometers (say,
the two LIGO 4km interferometers) instead of four interferometers.

Taking $T = 10^7 s$ and $\Delta t = 1 \, {\rm ms}$ yields the estimate
$\rho_{\rm threshold}/\sqrt{2} \approx 6.0$ for the initial and
advanced LIGO interferometers.  Therefore, from
Fig.~\ref{initialligosnr}, we see that the initial LIGO
interferometers should be able to see ringdowns from equal-mass BBHs
in the mass range $100
M_\odot \alt M \alt 700 M_\odot$ out to $\sim 200 \, {\rm Mpc}$, if
the radiation efficiency $\epsilon_r$ is as large as we have
estimated.  The advanced LIGO interferometers, by contrast, should see
ringdowns in the mass range $200 M_\odot \alt (1+z) M \alt 3000
M_\odot$ out to $z \sim 1$ (from Fig.~\ref{advancedligosnr}).
For non equal-mass BBHs, these distances are reduced by the factor
$\sim (4 \mu / M)$.

For LISA, the detection threshold is given by Eq.~(\ref{thr}).
Although LISA does incorporate several partially independent
interferometers, we have used the noise spectrum (\ref{noisespecLISA})
which is the effective noise spectrum that applies to the LISA
detector as a whole \cite{LISAreport}.  Thus it is consistent to treat
LISA as one interferometer.  Taking $T = 10^7 \, {\rm s}$ and $\Delta
t = 1 \, {\rm s}$, and using the value ${\cal N}_{\rm shapes} = 6000$
estimated above yields $\rho_{\rm threshold} \approx 7.5$.  Hence, from
Fig.~\ref{lisasnr}, LISA should see ringdowns in the mass range $10^6
M_\odot \alt (1+z) M \alt 3 \times 10^8 M_\odot$ out to $z \agt 100$.

\subsection{The detectability of high mass black hole \\
coalescences via the merger signal}
\label{detectmerger}

We next discuss the feasibility of searches for the merger signal.  As
we have explained, this will be most necessary when the merger SNR is
larger than both the inspiral and ringdown SNRs by factors of a few
(since the fractional loss in event detection rate, if searches for the
merger signal are not carried out, is the cube of the ratio of the SNR
values).

Consider first the ideal situation in which theoretical template
waveforms are available, so that matched filtering can be used in
searches.  From Figs.~\ref{initialligosnr} and \ref{advancedligosnr}
it can be seen that the merger SNR values are
larger than the inspiral/ringdown values by a factor of up to $\sim
4$, in the mass ranges $30 M_\odot \alt M \alt 200 M_\odot$ for initial LIGO
interferometers and $100 M_\odot \alt M \alt 400 M_\odot$ for advanced
LIGO interferometers.  More precisely, in this mass range,
\begin{equation}
{ \left( {S \over N} \right)_{\rm merger} \over {\rm max} \left[
 \left( {S \over N} \right)_{\rm inspiral} \,\, , \,\,
 \left( {S \over N} \right)_{\rm ringdown}\, \right]} \, \,
\alt 4 \left( \epsilon_m \over 0.1 \right)^{1/2} \,
\left( \epsilon_r \over 0.03 \right)^{-1/2}.
\label{larger}
\end{equation}
The detection threshold for merger searches should be approximately
the same as that for inspiral and merger searches, if the number of
template shapes ${\cal N}_{\rm shapes}$ is not too large (see further
discussion below).  Therefore, the gain in event rate over
inspiral/ringdown searches, due to searching for merger waves,
should vary between $1$ and about $4^3=64$, depending on the mass of
the system, if
our estimates of $\epsilon_m$ and $\epsilon_r$ are approximately
correct.  Clearly, the large possible gain in event rate shows the
importance of searching for merger waves \cite{notedavid}.
Moreover, the ratio of $\epsilon_m / \epsilon_r$ of energies
radiated in the merger and ringdown stages, which we have estimated to
be roughly $3$, is quite uncertain; we think that in some cases it could
be rather larger than this (notwithstanding the fact that there has
been at least one wagered bottle of wine in the community that
$\epsilon_m / \epsilon_r \alt 1$ generically~\cite{kipsbet}).  In such
cases the benefit of searching for the merger waves would be even greater.

It is not clear, however, how feasible it will be to produce a set of
numerically generated templates that is complete enough to be used
to successfully implement an optimal filtering search.  There may be
a very large number of distinct waveform shapes, each of which will
require extensive numerical computations.  If both black holes are
spinning rapidly, the waveforms could depend in significant and
nontrivial ways on 6 distinct angular parameters, suggesting that
the number of distinct shapes could be very large.

Turn next to the situation where templates are unavailable.  Consider
first band-pass filtering searches.  From the estimate $\sqrt{{\cal N}_{\rm
bins}} = 4$ of Sec.~\ref{mergernofilters}, combined with
Eq.~(\ref{simp1}), we obtain the relation
\begin{equation}
\left( {S \over N} \right)_{\rm merger,\ band-pass\ search} \simeq {1 \over 4}
\, \left( {S \over N} \right)_{\rm merger,\ optimal}.
\label{notemplates}
\end{equation}
Therefore, from Eq.~(\ref{larger}) it is apparent that the
maximum achievable band-pass filtering SNR for the merger is likely to
be essentially no larger than the SNRs obtained from the
inspiral/ringdown signals.

However, as explained in Secs.~\ref{assumptionsec} and
\ref{nonlinear}, noise monitoring searches will be more efficient than
band-pass filtering searches, and will be almost
as efficient as matched filtering searches in most cases, when
searching for merger waves.  [For the
inspiral waves, by contrast, noise-monitoring searches perform
considerably worse than matched filtering searches, since the parameter
${\cal N}_{\rm bins}$ is much larger for inspiral waves ($\agt 1000$)
than it is for merger waves ($\alt 60$)].  The event-detection rate
from noise-monitoring is a factor
\begin{equation}
{\cal R} = \left( {\rho_* \over \rho_{\rm threshold}} \right)^3
\label{EventRateLoss}
\end{equation}
lower than the event rate from matched filtering.  Here $\rho_*$ is
the noise-monitoring detection threshold, given by Eqs.~(\ref{e3})
and (\ref{e4}) as a function of the parameters $\epsilon$, ${\cal
N}_{\rm start-times}$ and ${\cal N}_{\rm bins}$, and $\rho_{\rm
threshold}$ is 
the matched filtering threshold, given by Eq.~(\ref{MFthreshold0}) as a
function of the parameters ${\cal N}_{\rm shapes}$ and ${\cal N}_{\rm
start-times}$.

Equation (\ref{EventRateLoss}) is based on the assumption that the
statistical properties of the noise are Gaussian.  In reality the
noise in the individual interferometers will have significant
non-Gaussian components.  Coincidencing between several
interferometers is expected to reduce the effect of these non-Gaussian
components, and the efficiency of this reduction may be higher for
matched filtering than for noise-monitoring for the reason discussed
in Sec.~\ref{nonlinear}.  Hence, if coincidencing
does not completely efface non-Gaussian effects, the loss factor in
event rate ${\cal R}$ may be somewhat higher than the estimate
(\ref{EventRateLoss}).

We now estimate the loss factor in event rate ${\cal R}$.  To obtain
the most pessimistic estimate possible of the loss in event rate, we
use the following assumptions: (i) The number of template shapes in
the matched filtering search is ${\cal N}_{\rm shapes}=1$; a more
realistic larger number would yield a smaller ${\cal R}$.  (ii) The
number of frequency bins is ${\cal N}_{\rm bins} = 60$, twice the
upper limit estimated in Sec.~\ref{mergernofilters}.  (iii) The number
of starting times in the data stream is ${\cal N}_{\rm start-times} =
10^8$, corresponding to a sampling time of $0.1 \, {\rm s}$ in a data
set of one third of a year.  Such a large sampling time would only be
appropriate for the largest mass black holes; more realistic sampling
times will be smaller.  Larger values of ${\cal N}_{\rm
start-times}$ give smaller values of ${\cal R}$.  (iv) The parameter
$\epsilon$ in Eqs.~(\ref{e3}) and (\ref{MFthreshold0}) is
$\epsilon=10^{-3}$.   With these
assumptions we obtain $\rho_{\rm threshold}=6.8$, $\rho_*=10.3$ and the
resulting loss in event rate factor is
$$
{\cal R} = 3.5.
$$
Hence, the availability of templates could increase the event
detection rate by a factor of at most $4$, and more typically only
$2$ over noise-monitoring searches.

\def\DISPFIGSEVEN{
{\vskip 0.5cm}
{\psfig{file=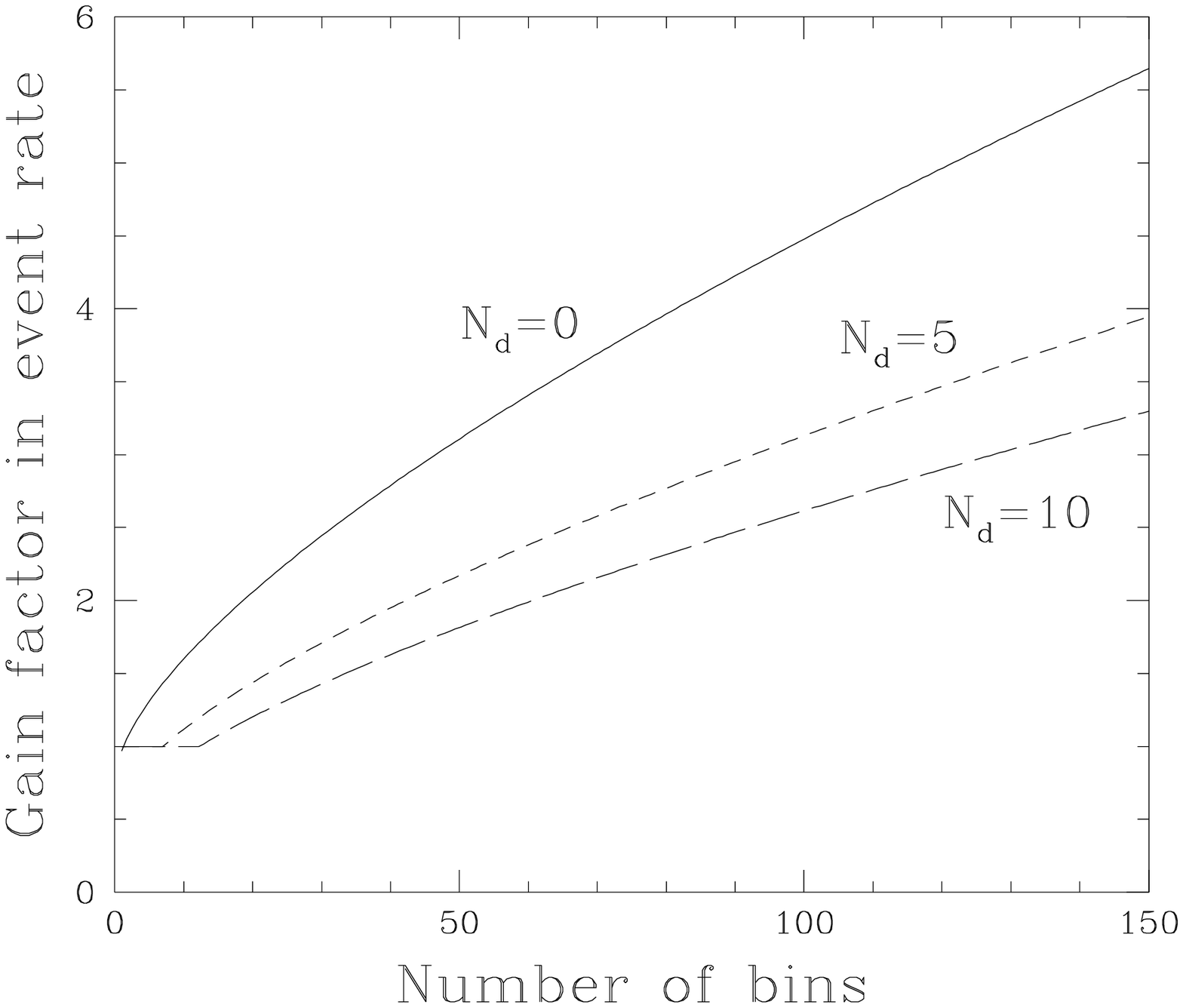,height=9.5cm,width=9cm}}
{\vskip -1.0cm}
\figure{
{This plot shows that factor by which the event detection rate is
increased when a matched filtering search using theoretical template
waveforms is carried out, for gravitational waves from the merger
phase of binary black hole coalescences.  It is assumed that in the
absence of theoretical templates, a search is carried out using the
noise-monitoring, nonlinear filtering method described in the text.
Plotted on the
horizontal axis is the number ${\cal N}_{\rm bins} = 2 T \Delta f$ of
independent frequency bins characterizing the space of signals that
one searches for; $T$ is the maximum expected signal duration and
$\Delta f$ is the frequency bandwidth.  The gain factor ${\cal R}$ in
event rate, which is plotted on the vertical axis, depends on the
number of statistically independent waveform shapes in the set of signals
one is searching for, which is currently unknown.  This number of waveform
shapes can be characterized by the effective dimension ${\cal N}_d$ of
the manifold of signals, as explained in the text.  The upper, solid
line shows the gain factor in the limit when the number of template
shapes is small (${\cal N}_d=0$), and is an upper limit on the gain
factor obtainable from matched filtering.  The lower two dashed lines
show the gain factor when ${\cal N}_d=5$ and ${\cal N}_d=10$.
Our best-guess estimate of the number of frequency bins is ${\cal
N}_{\rm bins} = 30$, corresponding to $T = 50 M$ and $\Delta f = 1 /
(\pi M)$, and ${\cal N}_{\rm bins}$ is unlikely to be much larger than
100 (Sec.~\ref{mergernofilters}).  This plot can be generated
by combining Eqs.~(\ref{e3}), (\ref{e4}), (\ref{MFthreshold0}) and
(\ref{Ndimdef}) of the text, with the assumed parameter values ${\cal
N}_{\rm start-times}=10^8$ and $\epsilon=10^{-3}$.
}
\label{eventgain:fig}}
}

\DISPFIGSEVEN

This discussion assumed that the number of waveform shapes
${\cal N}_{\rm shapes}$ is small.  As this number increases, the
performance of matched filtering searches gradually worsens until
at ${\cal N}_{\rm shapes} \sim {\cal N}_{\rm shapes,max}$, matched
filtering performs about as well as noise monitoring (see
Sec.~\ref{nonlinear}).  From Eqs.~(\ref{numshapes}) and (\ref{e3}), the
critical value of the number of shapes is $\sim 10^{13}$ for
${\cal N}_{\rm bins} = 60$, and $\sim 10^7$ for ${\cal
N}_{\rm bins} = 20$, assuming ${\cal N}_{\rm start-times} = 10^8$.
The number of shapes will vary with the signal-to-noise level
$\rho$.   We can define an effective dimension $N_d$ of the
manifold of signals by the equation
\begin{equation}
\ln\left[{\cal N}_{\rm shapes}(\rho)\right] = {1 \over 2} N_d(\rho)
\ln \left[ 1 + \rho^2 / N_d(\rho) \right].
\label{Ndimdef}
\end{equation}
{}From Eq.~(\ref{numshapes}), we see that $N_d(\rho)$ is the dimension
of the equivalent linear space of signals that has the same number of
distinguishable waveform shapes with SNR $\le \rho$ as the true (nonlinear
and curved) manifold of merger signals.  In Fig.~\ref{eventgain:fig} we
show the gain factor ${\cal R}$ in event rate as a function of ${\cal
N}_{\rm bins}$, for the values $N_d =0$, $5$ and $10$.  The true value
of $N_d$ is quite uncertain and will vary to some extent with $\rho$;
at high signal-to-noise levels it could conceivably be as large as $\sim 10$.

By combining the gain factor of $64$ discussed above with the loss
factor ${\cal R}$, it follows that noise-monitoring searches for
merger waves should increase the number of discovered BBHs by a factor
of up to $\sim 10$ over those found from inspiral and ringdown
searches.

For LISA, the expected SNRs are so high that the availability of
merger templates will likely have no impact on event detection rates.
However, for LISA, as well as for ground-based interferometers,
accurate merger templates will be vitally important for interpreting
the detected waveforms, for extracting all the available information,
and for testing general relativity \cite{paperII}.

\section{CONCLUSIONS}

It seems quite likely that the gravitational waves from merging BBH
systems will be detected by the
ground-based interferometers that are now
under construction.  The initial LIGO interferometers will
be able to detect low mass ($\alt 30 M_\odot$) coalescences of
equal-mass BBHs out to
about $200 \, {\rm Mpc}$ via their inspiral waves, and higher mass
($100 M_\odot \alt M \alt 700 M_\odot$) systems out to about $200 \,
{\rm Mpc}$ via their ringdown waves.  The advanced LIGO
interferometers will be able to detect equal-mass BBH coalescences in the
mass range
$10 M_\odot \alt M \alt 300 M_\odot$ to $z \sim 1/2$ via their
inspiral waves, and higher mass ($200 M_\odot \alt M \alt 3000
M_\odot$ ) systems to $z \sim 1$ via their ringdown waves.
For non-equal mass BBHs, these distances will be reduced by a factor
of $\sim \sqrt{4 \mu/M}$ for inspiral signals and $\sim 4 \mu/M$ for
ringdown signals.

Searches for massive BBHs ($M \agt 50 M_\odot$ for LIGO/VIRGO) based
on merger
waves could increase the range of the interferometers by a additional
factor of $\sim 2$, without requiring detailed knowledge of the
waveform shapes.  It seems likely that BBH coalescences will be
detected early in the gradual process of improvement from the first
interferometers to the advanced interferometers, rather than later,
and there is a strong possibility that they will be the
first sources of gravitational radiation to be detected.

Theoretical template waveforms obtained from numerical relativity
supercomputer simulations will be crucial for analyzing the measured
merger waves.  A match of the detected waveform with a predicted
waveform would be a triumph for the theory of general relativity and
an absolutely unambiguous signature of the existence of black holes.
A complete set of such theoretical templates would also aid the search
for BBHs, but not by a large amount.

The space-based interferometer LISA will be an extremely high
precision instrument for studying the coalescences of supermassive
BBHs.  Coalescences with masses in the range $10^6 M_\odot \alt (1+z)
M \alt 10^9 M_\odot$ should be detectable out to $z \sim 10$ with very
large SNRs ($\agt 10^3$), via their merger and ringdown waves.
Additionally, systems in the mass range $10^4 M_\odot \alt (1+z) M
\alt 3 \times 10^7 M_\odot$ should be detected to similar distances
and with SNRs $\agt 10^2$ via their inspiral waves.

%
%

\acknowledgments

We thank Kip Thorne for suggesting this project to us, for being a
constant source of ideas and encouragement along the way, and for
detailed comments on the paper's content and presentation.  We also thank
Patrick Brady and David Chernoff for some helpful conversations.  This
research was supported by NSF Grants PHY--9220644, PHY--9408378,
PHY--9424337, and PHY--9514726, and by NASA grant NAGW-4268.  S.A.H.\
gratefully acknowledges the support of a National Science Foundation
Graduate Fellowship, and \'E.F.\ likewise acknowledges the support of
an Enrico Fermi fellowship.

\newpage
\widetext
\onecolumn

\appendix{Signal to noise ratio formulae}
\label{appsnr}

In this appendix, we give the details of our calculations of the
signal-to-noise ratios.  The results of this appendix were used to
make Figs.~\ref{initialligosnr}, \ref{advancedligosnr} and
\ref{lisasnr} in the text, and also the approximate SNR formulae
(\ref{lowmassinitial}) and (\ref{lowmassadv}) given in Sec.~\ref{snrsection}.
The following three
subsections give the calculations for the inspiral waves, the merger
waves and the ringdown waves respectively.

Note that throughout this appendix we use ``${\rm M}M_\odot$'' (Mega
solar-mass) as shorthand for $10^6\,M_\odot$.

\subsection{Inspiral}

To calculate the angle-averaged SNR squared for the inspiral, we
insert the inspiral
energy spectrum (\ref{dEdfinspiral}) and our parameterized model
(\ref{noisespec}) of an interferometer's noise spectrum
into Eq.~(\ref{snraveraged}), and integrate from $f=f_s$ to $f={f_{\rm
merge}}$.  The result is
\begin{equation}
\label{inspsnrformula1}
\langle\rho^2\rangle = \left\{ \begin{array}{ll} {\cal F}_i(M,z,D)
	v^3\left[9\alpha^{1/3}-{36\over5}\alpha^{-1/3}-
	{4\over5}\alpha^{-1/3}v^{10/3}-\alpha^3 \left({f_s\over
	f_m}\right)^{8/3}\right] & \alpha f_m \le
	{f_{\rm merge}}/(1+z)\nonumber\\ {\cal F}_i(M,z,D)\left[9\alpha^{1/3}-
	8\left({v\over\alpha}\right)^{1/3}
	-\alpha^3\left({f_s\over f_m}\right)^{8/3}\right] &
	f_m/\alpha \le {f_{\rm merge}}/(1+z) < \alpha f_m \nonumber\\ {\cal
	F}_i(M,z,D) \left[\alpha^{1/3}\left({\alpha^2\over v}\right)
	-\alpha^3\left({f_s\over f_m}\right)^{8/3}\right] &
	f_s \le {f_{\rm merge}}/(1+z) < f_m/\alpha \nonumber\\ 0 &
	{f_{\rm merge}}/(1+z) < f_s,\\ \end{array} \right.
\end{equation}
where
\begin{equation}
v\equiv {(1+z)\alpha f_m \over {f_{\rm merge}}}
\label{vdef}
\end{equation}
and
\begin{equation}
{\cal F}_i(M,z,D) = {{[(1+z)M]^{5/3} [4 \mu/M]}\over 80\pi^{4/3}D(z)^2 h_m^2
f_m^{1/3}}.
\end{equation}
Here $D(z)$ is the luminosity distance to the
source, $f_s$, $\alpha$, $f_m$ and $h_m$ are parameters characterizing
the detector noise spectrum~(\ref{noisespec}), and $f_{\rm merge}$ is
given by Eq.~(\ref{mergefreq}).

Inserting the values of the noise spectrum parameters from
Eq.~(\ref{noisespecLIGOinitial}) for initial LIGO interferometers, we
obtain the following numerical values for the SNR in the equal-mass
case $\mu=M/4$:
\begin{equation}
\label{inspsnrformula2}
\left( {S \over N} \right)_{\rm initial} = \left\{ \begin{array}{ll}
2.8 \left({200\,{\rm Mpc}\over
	D(z)}\right) \left({{(1+z)M}\over18\,M_\odot}\right)^{5/6}
	\left[1 - 0.20\left({{(1+z)M}\over18\,M_\odot}\right)^{10/3}
	\right]^{1/2} & (1+z)M \le 18\,M_\odot \nonumber\\
4.7\left({200\,{\rm Mpc}\over D(z)}\right)
	\left({{(1+z)M}\over18\,M_\odot}\right)^{5/6} \left[1 -
	0.71\left({{(1+z)M}\over18\,M_\odot}\right)^{1/3} \right]^{1/2}
	& 18\,M_\odot < (1+z)M \le 36\,M_\odot \nonumber\\
2.7 \left({200\,{\rm Mpc}\over D(z)}\right)
	\left({{(1+z)M}\over36\,M_\odot}\right)^{-1/2}
	\left[1 - 0.06 \left({{(1+z)M}\over36\,M_\odot}\right)^{8/3}
	\right]^{1/2} & 36\,M_\odot < (1+z)M \le 102\,M_\odot
	\nonumber\\ 0 & 102 \,M_\odot < (1+z)M.  \end{array} \right.
\end{equation}
For the noise curve parameters (\ref{noisespecLIGOadvanced})
appropriate for advanced LIGO interferometers we obtain
\begin{equation}
\label{inspsnrformula3}
\left( {S \over N} \right)_{\rm advanced} = \left\{ \begin{array}{ll}
27 \left({1\,{\rm Gpc}\over D(z)}\right) \left({{(1+z)M}\over
	37\,M_\odot}\right)^{5/6} \left[1 -
	0.16\left({{(1+z)M}\over37\,M_\odot}\right)^{10/3}
	\right]^{1/2} & (1+z)M \le 37\,M_\odot \nonumber\\
43\left({1\,{\rm Gpc}\over D(z)}\right)
	\left({{(1+z)M}\over37\,M_\odot}\right)^{5/6} \left[1 -
	0.65\left({{(1+z)M}\over37\,M_\odot}\right)^{1/3} \right]^{1/2}
	& 37\,M_\odot < (1+z)M \le 95\,M_\odot \nonumber\\
31\left({1\,{\rm Gpc}\over D(z)}\right)
	\left({{(1+z)M}\over95\,M_\odot}\right)^{-1/2}
	\left[1 - .021 \left({{(1+z)M}\over95\,M_\odot}\right)^{8/3}
	\right]^{1/2} & 95\,M_\odot < (1+z)M \le 410\,M_\odot
	\nonumber\\ 0 & 410 \,M_\odot < (1+z)M.  \end{array} \right.
\end{equation}

As explained in Sec.~\ref{egs}, the calculation of
the inspiral SNR for LISA differs from the other SNR calculations in
the following way. If one were to integrate over the whole
frequency domain in the interferometer waveband up to $f= f_{\rm
merge}$ (as is done for the initial and advanced interferometers in
LIGO), in some cases one would obtain the SNR for a measurement of
several hundred years duration, which is obviously irrelevant.
Thus, it is necessary to restrict the integral over frequency in
Eq.~(\ref{snraveraged}) to the domain that corresponds to, say, one
year of observation when calculating inspiral LISA SNRs.  Using the
Newtonian relationship for the rate of frequency sweep, we obtain for
the frequency at time $T$ before merger in the equal-mass case
\begin{equation}
f_{\rm insp}(T) = \left[ f_{\rm merge}^{-8/3} + {64 \over 5}
\pi^{8/3} M^{5/3} (1+z)^{5/3} T \right]^{-3/8}.
\end{equation}
Binaries of redshifted total mass $(1+z) M$ larger than about $5 \times
10^5 M_\odot$ enter the LISA waveband at $f=f_s = 10^{-4} \, {\rm Hz}$
less than one year before merger, while binaries of smaller redshifted
mass spend more than one year in the LISA waveband.
To calculate the SNR, we insert Eq.~(\ref{dEdfinspiral}) into
Eq.~(\ref{snraveraged}) and integrate numerically
from the larger of $f_s$ and $f_{\rm insp}(1 \, {\rm yr})$ to $f_{\rm
merge}$.  The resulting SNR values are shown in Fig.~\ref{lisasnr}.
We also show in Fig.~\ref{lisasnr} the SNR obtained from one year of
observation one hundred years before the final merger, obtained by
integrating from $f_{\rm insp}(100 \, {\rm yr})$ to $f_{\rm insp}(99 \, {\rm
yr})$, as well as a similar curve for one thousand years prior to merger.

Equation (\ref{inspsnrformula1}) applies to LISA only for $(1+z) M
\agt 5 \times 10^5 M_\odot$.  By combining
Eqs.~(\ref{inspsnrformula1}) and (\ref{noisespecLISA}) for $(1+z) M \agt
5 \times 10^5 M_\odot$ together with an approximate fit to
Fig.~\ref{lisasnr} for $(1+z) M \alt 10^5 M_\odot$ we obtain for the
SNR from the last year of inspiral in the equal-mass case
\begin{equation}
\label{inspsnrformula4}
\left( {S \over N} \right)_{\rm LISA}
\approx \left\{ \begin{array}{ll}
	1.5 \times 10^4 \left({1\,{\rm Gpc}\over D(z)}\right)
	\left({{(1+z)M}\over
	0.5\,{\rm M}M_\odot}\right)
	& 100 M_\odot \alt (1+z) M \alt 0.5\,{\rm M}M_\odot \nonumber\\
	1.9\times10^4\left({1\,{\rm Gpc}\over D(z)}\right)
	\left({{(1+z)M}\over0.5\,{\rm M}M_\odot}\right)^{5/6}
	\left[1-0.38\left({{(1+z)M}\over0.5\,{\rm M}M_\odot}
	\right)^{1/3}\right]^{1/2} & 0.5\,{\rm M}M_\odot < (1+z)M
	\le 6.0\,{\rm M}M_\odot\nonumber\\
	5.0\times10^4\left({1\,{\rm Gpc}\over D(z)}\right)
	\left({{(1+z)M}\over6\,{\rm M}M_\odot}\right)^{-1/2}
	\left[1-0.006\left({{(1+z)M}\over6\,{\rm M}M_\odot}
	\right)^{8/3}\right]^{1/2}
	& 6.0\,{\rm M}M_\odot < (1+z) M \le 41 {\rm M}M_\odot \nonumber
	\\
	0 & 41 {\rm M}M_\odot < (1+z) M.\end{array} \right.
\end{equation}

\subsection{Merger}

To calculate the merger SNR we use the energy spectrum
(\ref{dEdfmerger}) and follow the same procedure as above.  The
result is
\begin{equation}
\label{mergesnrformula1}
\langle\rho^2\rangle = \left\{ \begin{array}{ll}
{\cal F}_m(\epsilon_m,M,z,D) v^3
	\left[{{\kappa^3-1}\over\kappa^3}\right] & {f_{\rm merge}}/(1+z) \ge
	\alpha f_m \nonumber\\
{\cal F}_m(\epsilon_m,M,z,D) \left[3\ln
	v-{{v^3-\kappa^3}\over\kappa^3}\right]
	& f_m/\alpha\le{f_{\rm merge}}/(1+z)<\alpha f_m\le{f_{\rm qnr}}/(1+z)
	\nonumber\\
^{\rm I,A}{\cal F}_m(\epsilon_m,M,z,D) \left[2 - {\alpha^6\over v^3}
	- {v^3\over\kappa^3} + 6 \ln \alpha\right] &
	{f_{\rm merge}}/(1+z)\le f_m/\alpha<\alpha f_m \le {f_{\rm qnr}}/(1+z)
	\nonumber\\
^{\rm L}{\cal F}_m(\epsilon_m,M,z,D)\left[3\, \ln \kappa \right] &
	f_m/\alpha \le f_{\rm merge}/(1+z) < f_{\rm qnr}/(1+z) \le \alpha f_m
	\nonumber\\
^{\rm I}{\cal F}_m(\epsilon_m,M,z,D)
	\left[2 - \left({\alpha f_s\over f_m}\right)^3 -
	{v^3\over \kappa^3} + 6 \ln \alpha\right]
	& {f_{\rm merge}}/(1+z) \le f_s < f_m/\alpha < \alpha f_m \le
	{f_{\rm qnr}}/(1+z) \nonumber\\
^{\rm A,L}{\cal F}_m(\epsilon_m,M,z,D) \left[1 +
	3\ln\left({\kappa\alpha^2\over v}\right) -{\alpha^6\over
	v^3}\right] & f_s \le {f_{\rm merge}}/(1+z) < f_m/\alpha \le
	{f_{\rm qnr}}/(1+z) \nonumber\\
^{\rm I,A}{\cal F}_m(\epsilon_m,M,z,D) \left[1 -
	\left({\alpha f_s\over f_m}\right)^3 +
	3\ln\left({\kappa\alpha^2\over v}\right)\right] &
	{f_{\rm merge}}/(1+z)\le f_s<f_m/\alpha\le{f_{\rm qnr}}/(1+z)
	\nonumber\\
^{\rm L}{\cal F}_m(\epsilon_m,M,z,D) \left[\alpha^6 v^{-3}
	(\kappa^3 - 1)\right] &	f_s \le f_{\rm merge}/(1+z) < f_{\rm qnr}/(1+z)
	\nonumber\\
{\cal F}_m(\epsilon_m,M,z,D) \left[\left({\kappa\alpha^2\over
	v}\right)^3 -\left({\alpha f_s\over f_m}\right)^3 \right]
	& {f_{\rm merge}}/(1+z)\le f_s<{f_{\rm qnr}}/(1+z) \le f_m/\alpha
	\nonumber\\
	0 & {f_{\rm qnr}}/(1+z) < f_s. \end{array} \right.
\end{equation}
Here $v$ is given by Eq.~(\ref{vdef}), $\epsilon_m$ is the fraction of
total mass energy radiated during the merger (which we have also denoted by
$\epsilon_{\rm merger}$ in the body of the paper), $\kappa\equiv{f_{\rm qnr}}/
{f_{\rm merge}}$, and
\begin{equation}
{\cal F}_m(\epsilon_m,M,z,D) = {{2\epsilon_m M (1+z)^2 [4 \mu/M]^2}\over
15\pi^2D(z)^2 h_m^2 {f_{\rm merge}}(\kappa-1)}.
\end{equation}
Lines marked with the superscript ``I'' turn out to hold for the initial
LIGO interferometer parameters; those with ``A'' hold for advanced LIGO
interferometer parameters; and those with ``L'' hold for LISA.

Using the numerical values of the noise curve parameters
(\ref{noisespecLIGOinitial}) for initial LIGO interferometers, and
Eqs.~(\ref{mergefreq}), (\ref{fqnrdef}), (\ref{vdef}), and
(\ref{mergesnrformula1}) we find for the initial LIGO interferometers
in the equal-mass case
\begin{equation}
\label{mergesnrformula2}
\left( {S \over N} \right)_{\rm initial} = \left\{ \begin{array}{ll}
1.5 \left({\epsilon_m\over 0.1}\right)^{1/2} \left({200\,{\rm Mpc}\over
	D(z)}\right) \left({{(1+z)M}\over18\,M_\odot}\right)^{5/2}&
	\!\!(1+z)M \le 18\,M_\odot \nonumber\\
1.5 \left({\epsilon_m\over 0.1}\right)^{1/2}
	\left({200\,{\rm Mpc}\over D(z)}\right)
	\left({{(1+z)M}\over18\,M_\odot}\right)
	\nonumber \\	\mbox{~~~~} \times
	\left[1 +
	3\ln\left({{(1+z)M}\over18\,M_\odot}\right)
	-{3.6\times10^{-3}}
	\left({{(1+z)M}\over18\,M_\odot}\right)^3\right]^{1/2}&
	\!\!18\,M_\odot < (1+z)M \le 36\,M_\odot\nonumber\\
6.1\left({\epsilon_m\over 0.1}\right)^{1/2} \left({200\,{\rm Mpc}\over
	D(z)}\right) \left({{(1+z)M}\over36\,M_\odot}\right)
	\nonumber \\	\mbox{~~~~} \times
	\left[1
	+ .23 \left({{(1+z)M}\over36\,M_\odot}\right)^{-3}
	- 0.007\left({{(1+z)M}\over36\,M_\odot}\right)^3\right]^{1/2}&
	\!\!36\,M_\odot < (1+z)M \le 102\,M_\odot\nonumber\\
17.3\left({\epsilon_m\over 0.1}\right)^{1/2} \left({200\,{\rm Mpc}\over
	D(z)}\right) \left({{(1+z)M}\over102\,M_\odot}\right)
	\nonumber \\	\mbox{~~~~} \times
	\left[1 -
	.17 \left({{(1+z)M}\over102\,M_\odot}\right)^3\right]^{1/2}&
	\!\!102\,M_\odot < (1+z)M \le 118\,M_\odot\nonumber\\
9.9 \left({\epsilon_m\over 0.1}\right)^{1/2} \left({200\,{\rm Mpc}\over
	D(z)}\right) \left({{(1+z)M}\over118\,M_\odot}\right) \left[1 -
	3.1\ln\left({{(1+z)M}\over230\,M_\odot}\right)\right]^{1/2}&
	\!\!118\,M_\odot < (1+z)M \le 230\,M_\odot\nonumber\\
20 \left({\epsilon_m\over 0.1}\right)^{1/2} \left({200\,{\rm Mpc}\over
	D(z)}\right) \left({{(1+z)M}\over230\,M_\odot}\right)^{-1/2}
	\left[1 - 0.04\left({{(1+z)M}\over230\,M_\odot}\right)^3
	\right]^{1/2}&
	\!\!230\,M_\odot < (1+z)M \le 660\,M_\odot\nonumber\\
	0 & 660\,M_\odot < (1+z) M. \end{array}
	\right.
\end{equation}

Similarly using Eq.~(\ref{noisespecLIGOadvanced}) we find for
advanced LIGO interferometers
\begin{equation}
\label{mergesnrformula3}
\left( {S \over N} \right)_{\rm advanced} = \left\{ \begin{array}{ll}
13 \left({\epsilon_m\over 0.1}\right)^{1/2}
	\left({1\,{\rm Gpc}\over D(z)}\right)
	\left({{(1+z)M}\over37\,M_\odot}\right)^{5/2} & \!\!(1+z)M \le
	37\,M_\odot \nonumber\\
13\left({\epsilon_m\over 0.1}\right)^{1/2}
	\left({1\,{\rm Gpc}\over D(z)}\right)
	\left({{(1+z)M}\over37\,M_\odot}\right)
	\nonumber \\	\mbox{~~~~} \times
	\left[1 +
	3\ln\left({{(1+z)M}\over37\,M_\odot}\right)
	-3.6\times10^{-3}\left({{(1+z)M}\over37\,M_\odot}\right)^3\right]^{1/2}
	& \!\!37\,M_\odot < (1+z)M \le 95\,M_\odot\nonumber\\
76 \left({\epsilon_m\over 0.1}\right)^{1/2}
	\left({1\,{\rm Gpc}\over
	D(z)}\right) \left({{(1+z)M}\over95\,M_\odot}\right)
	\nonumber \\	\mbox{~~~~} \times
	\left[1
	- {0.21}\left({{(1+z)M}\over95\,M_\odot}\right)^{-3}
	-{0.013}\left({{(1+z)M}\over95\,M_\odot}\right)^3\right]^{1/2}
	& \!\!95\,M_\odot < (1+z)M \le 240\,M_\odot\nonumber\\
88 \left({\epsilon_m\over 0.1}\right)^{1/2} \left({1\,{\rm Gpc}\over
	D(z)}\right) \left({{(1+z)M}\over240\,M_\odot}\right)
	\nonumber \\	\mbox{~~~~} \times
	\left[1 -
	3\ln\left({{(1+z)M}\over620\,M_\odot}\right)
	-0.061\left({{(1+z)M}\over240\,M_\odot}\right)^{-3}\right]^{1/2}
	& \!\!240\,M_\odot < (1+z)M \le 410\,M_\odot\nonumber\\
150\left({\epsilon_m\over 0.1}\right)^{1/2} \left({1\,{\rm Gpc}\over
	D(z)}\right) \left({{(1+z)M}\over410\,M_\odot}\right) \left[1 -
	3.0\ln\left({{(1+z)M}\over620\,M_\odot}\right)\right]^{1/2} &
	\!\!410\,M_\odot < (1+z)M \le 620\,M_\odot\nonumber\\
220 \left({\epsilon_m\over 0.1}\right)^{1/2} \left({1\,{\rm Gpc}\over
	D(z)}\right) \left({{(1+z)M}\over620\,M_\odot}\right)^{-1/2}
	\left[1 - 0.013 \left({{(1+z)M}\over620\,M_\odot}\right)^3
	\right]^{1/2} & \!\!620\,M_\odot < (1+z)M \le
	2600\,M_\odot\nonumber\\
	0 & \!\!2600\,M_\odot < (1+z) M. \end{array}
	\right.
\end{equation}
Finally, using the parameters (\ref{noisespecLISA}) appropriate for
LISA, we obtain
\begin{equation}
\label{mergesnrformula4}
\left( {S \over N} \right)_{\rm LISA}
= \left\{ \begin{array}{ll}
1.9\times10^3 \left({\epsilon_m\over 0.1}\right)^{1/2} \left({1\,{\rm Gpc}\over
	D(z)}\right)\left({{(1+z)M}\over2.0\times 10^5 M_\odot}
	\right)^{5/2} & \!\!(1+z)M \le 2.0\times 10^5\,M_\odot \nonumber\\
1.9\times10^3 \left({\epsilon_m\over 0.1}\right)^{1/2}
	\left({1\,{\rm Gpc}\over D(z)}\right)
	\left({{(1+z)M}\over2.0\times10^5M_\odot}\right)
	\nonumber \\	\mbox{~~~~} \times
	\left[1 +
	3\ln\left({{(1+z)M}\over2.0\times10^5M_\odot}\right)
	-3.6\times10^{-3}\left({{(1+z)M}\over2.0\times10^5M_\odot}
	\right)^3\right]^{1/2} & \!\!2.0\times10^5M_\odot < (1+z)M \le
	1.3\,{\rm M}M_\odot\nonumber\\
2.8\times10^4 \left({\epsilon_m\over 0.1}\right)^{1/2}
	\left({1\,{\rm Gpc}\over D(z)}\right)
	\left({{(1+z)M}\over1.3\,{\rm M}M_\odot}\right) &
	\!\!1.3\,{\rm M}M_\odot < (1+z)M \le 6.0\,{\rm M}M_\odot
	\nonumber\\
3.4\times10^4 \left({\epsilon_m\over 0.1}\right)^{1/2} \left({1\,{\rm
	Gpc}\over D(z)}\right) \left({{(1+z)M}\over6.0\,{\rm
	M}M_\odot}\right) \nonumber \\
	\mbox{~~~~} \times \left[1 - 3\ln\left({{(1+z)M}\over39\,{\rm
	M}M_\odot}\right) - \left({{(1+z)M}\over6.0\,{\rm M}M_\odot}
	\right)^{-3}\right]^{1/2} & \!\!6.0\,{\rm M}M_\odot < (1+z)M \le
	39\,{\rm M}M_\odot\nonumber\\
3.6\times10^5 \left({\epsilon_m\over 0.1}\right)^{1/2} \left({1\,{\rm
	Gpc}\over D(z)}\right) \left({{(1+z)M}\over39\,{\rm
	M}M_\odot}\right)^{-1/2} &
	\!\!39\,{\rm M}M_\odot < (1+z)M \le 41\,{\rm M}M_\odot\nonumber\\
3.4\times10^5 \left({\epsilon_m\over 0.1}\right)^{1/2} \left({1\,{\rm
	Gpc}\over D(z)}\right) \nonumber \\
	\mbox{~~~~} \times \left({{(1+z)M}\over41\,{\rm M}M_\odot}
	\right)^{-1/2} \left[1 - 3.8 \times 10^{-3}
	\left({(1+z)M\over 41\,{\rm M}M_\odot}\right)^{3} \right]^{1/2} &
	\!\!41\,{\rm M}M_\odot < (1+z)M \le 260\,{\rm M}M_\odot\nonumber\\
	0 & \!\!260 \,{\rm M} M_\odot < (1+z)M.
	\end{array}
	\right.
\end{equation}

\subsection{Ringdown}

The ringdown SNRs are calculated a little differently from the inspiral
and merger SNRs.  First, we use the effective energy spectrum
(\ref{dEdfqnr}) which yields an estimate of the true SNR obtainable
from the model waveform (\ref{qnrwaveform}) that is accurate to within
a few tens of percent (see Appendix \ref{ringdown3}).  Second, the
integral over frequency in the SNR formula (\ref{snraveraged}) with the
noise spectrum (\ref{noisespec}) and the energy spectrum (\ref{dEdfqnr})
cannot easily be evaluated analytically.  Hence, we calculated this integral
numerically to produce the plots of ringdown SNR versus BBH mass shown in
Figs.~\ref{initialligosnr}, \ref{advancedligosnr} and \ref{lisasnr}.

In the remainder of this appendix we derive approximate formulae for
the ringdown SNR as a function of mass, by approximating the ringdown
energy spectrum as a delta function at the ringdown frequency
[{\it cf}.~Eq.~(\ref{dEdfqnrapprox})].  This approximation yields (see
Appendix \ref{ringdown3} and Ref.~\cite{noteA})
\begin{equation}
\langle\rho^2\rangle =
{(1+z)^3 M^2\,{\cal A}^2 Q [4 \mu/M]^2 \over 20 \pi^2 D(z)^2 {f_{\rm qnr}}
S_h[{f_{\rm qnr}}/(1+z)]}.
\label{qnrsnr}
\end{equation}
Using Eq.~(\ref{fqnrandQ}) and the relation (\ref{qnrenergy}) between
the dimensionless coefficient ${\cal A}$ and the radiated energy we can
rewrite formula (\ref{qnrsnr}) as
\begin{equation}
\langle\rho^2\rangle =
{8\over5} \, {1 \over F(a)^2} \,
\epsilon_r \,
{ (1+z) M \over S_h[f_{\rm qnr} / (1+z)]} \,
\left[ {(1+z) M \over D(z) }\right]^2 \,
\left[ {4 \mu \over M }\right]^2,
\label{qnrsnrII}
\end{equation}
where $\epsilon_r$ is the fraction of the total mass
energy radiated in the ringdown, and
\begin{equation}
F(a) = 1 - {63\over100} (1 - a)^{3/10}.
\end{equation}
An equivalent formula was previously obtained by Finn {\cite{compare}}.

We find the following numerical result when we insert our assumed values
$\epsilon_r = 0.03$ and $a=0.98$ for the ringdown signal together with
the parameters for the initial LIGO interferometer noise curve in the
equal-mass case:
\begin{equation}
\label{qnrsnr2}
\left( {S \over N} \right)_{\rm initial} = \left\{ \begin{array}{ll} 0.08
	\,\left( {\epsilon_r \over 0.03} \right)^{1/2} \,
	\left({200\,{\rm Mpc}\over
	D(z)}\right) \left({{(1+z)M}\over18\,M_\odot}\right)^{5/2} &
	\!\!(1+z)M \le 118\,M_\odot \nonumber\\
	8.8
	\,\left( {\epsilon_r \over 0.03} \right)^{1/2} \,
	\left({200\,{\rm Mpc}\over
	D(z)}\right) \left({{(1+z)M}\over118\,M_\odot}\right) &
	\!\!118\,M_\odot < (1+z)M \le 230\,M_\odot \nonumber\\
	17
	\,\left( {\epsilon_r \over 0.03} \right)^{1/2} \,
	\left({200\,{\rm Mpc}\over D(z)}\right)
	\left({{(1+z)M}\over230\,M_\odot}\right)^{-1/2} & \!\!230\,M_\odot <
	(1+z)M \le 660\,M_\odot \nonumber\\ 0 & \!\!660\,M_\odot < (1+z) M.
	\end{array} \right.
\end{equation}
The corresponding formulae for advanced LIGO interferometers are
\begin{equation}
\label{qnrsnr3}
\left( {S \over N} \right)_{\rm advanced}
= \left\{ \begin{array}{ll} 0.71
	\,\left( {\epsilon_r \over 0.03} \right)^{1/2} \,
	\left({1\,{\rm Gpc}\over
	D(z)}\right) \left({{(1+z)M}\over37\,M_\odot}\right)^{5/2} &
	\!\!(1+z)M \le 240\,M_\odot \nonumber\\
	77
	\,\left( {\epsilon_r \over 0.03} \right)^{1/2} \,
	\left({1\,{\rm Gpc}\over
	D(z)}\right) \left({{(1+z)M}\over240\,M_\odot}\right) &
	\!\!240\,M_\odot < (1+z)M \le 620\,M_\odot \nonumber\\
	200
	\,\left( {\epsilon_r \over 0.03} \right)^{1/2} \,
	\left({1\,{\rm Gpc}\over D(z)}\right)
	\left({{(1+z)M}\over620\,M_\odot}\right)^{-1/2} & \!\!620\,M_\odot <
	(1+z)M \le 2600\,M_\odot \nonumber\\ 0 & \!\!2600\,M_\odot < (1+z) M.
	\end{array} \right.
\end{equation}
Finally, the corresponding formulae for LISA are
\begin{equation}
\label{qnrsnr4}
\left( {S \over N}\right)_{\rm LISA} = \left\{ \begin{array}{ll} 96
	\,\left( {\epsilon_r \over 0.03} \right)^{1/2} \,
	\left({1\,{\rm Gpc}\over
	D(z)}\right) \left({{(1+z)M}\over0.2\,{\rm M}M_\odot}
	\right)^{5/2} &	\!\!(1+z)M \le 1.3\,{\rm M}M_\odot \nonumber\\
	1.0\times10^4
	\,\left( {\epsilon_r \over 0.03} \right)^{1/2} \,
	\left({1\,{\rm Gpc}\over
	D(z)}\right) \left({{(1+z)M}\over1.3\,{\rm M}M_\odot}\right) &
	\!\!1.3\,{\rm M}M_\odot < (1+z)M \le 39\,{\rm M}M_\odot
	\nonumber\\
	3.1\times10^5
	\,\left( {\epsilon_r \over 0.03} \right)^{1/2} \,
	\left({1\,{\rm Gpc}\over D(z)}\right)
	\left({{(1+z)M}\over39\,{\rm M}M_\odot}\right)^{-1/2} &
	\!\!39 \, {\rm M} M_\odot < (1+z)M \le 260\,{\rm M}M_\odot
	\nonumber \\
	0 & \!\!260 \, {\rm M} M_\odot < (1+z) M.
	\end{array} \right.
\end{equation}
By comparing Eqs.~(\ref{qnrsnr2}) -- (\ref{qnrsnr4}) with
Figs.~\ref{initialligosnr} -- \ref{lisasnr} it can be seen that the
delta-function energy spectrum approximation is fairly good except for
$M \agt 3000 M_\odot$ for advanced LIGO interferometers and $M \agt
3\times 10^8 M_\odot$ for LISA.  The approximation fails to capture
the high mass tails of the SNR curves.

\newpage
\narrowtext
\twocolumn

\appendix{Energy spectrum for ringdown waves}
\label{ringdown3}

There is a subtlety in calculating the SNR for the
ringdown waves, related to the fact that the SNR squared does not
accumulate locally in the time domain.  In order to explain this
subtlety, let us focus not on the angle-averaged SNR squared which was
our main concern in the body of the paper, but rather on the SNR
squared obtained in one interferometer from a specific source with  specific
relative angular orientations.  In this case
the waveform $h(t)$ seen in the interferometer, for $t>0$,  is of the form
\begin{equation}
h(t) = h_0 \cos(2 \pi f_{\rm qnr} t + \psi_0) e^{-t/\tau}
\label{ringdownbasic}
\end{equation}
for some
constants $h_0$ and $\psi_0$, while $h(t)$ is the (unknown)
merger waveform for $t<0$.

Let us also focus first on the simple, idealized case of white noise,
$S_h(f) = S_h = {\rm const}$.  Then, the SNR squared (\ref{snr})
accumulates locally in time:
\begin{equation}
\rho^2 = {2 \over S_h} \int_{-\infty}^\infty dt \,  h(t)^2.
\end{equation}
Hence, for white noise, the SNR squared from the ringdown is clearly
unambiguously given by
\begin{eqnarray}
\rho^2_{\rm ringdown} &=& {2 \over S_n} \int_0^\infty h_0^2 \cos^2(2
\pi f_{\rm qnr} t + \psi_0)^2 e^{-2 t / \tau} \nonumber \\
\mbox{} &=& {h_0^2 \tau \over 2 S_h} \left[ 1 + { \cos(2  \psi_0) - Q \sin(2
\psi_0) \over 1 + Q^2} \right] \nonumber \\
\mbox{} &\approx & {h_0^2 \tau \over 2 S_h} \left[ 1 + O(1/Q)\right],
\label{exactwhite}
\end{eqnarray}
where $Q =  \pi f_{\rm qnr} \tau$.  Now consider the case when the
noise is not exactly white.  Naively, we expect that in the Fourier
domain the energy spectrum of the ringdown signal will be a resonance
curve that peaks at $f = f_{\rm qnr}$ with width $\sim f_{\rm qnr}/Q$.
Thus, for large $Q$ we would expect that most of the SNR squared will be
accumulated near $f = f_{\rm qnr}$, unless
the noise spectrum varies very
strongly with frequency.  Moreover, if the noise spectrum
$S_h(f)$ does not vary much over the bandwidth $\sim f_{\rm qnr}/Q$ of
the resonance peak, then we would expect the formula
(\ref{exactwhite}) to be valid to a good approximation, with $S_h$
replaced by $S_h(f_{\rm qnr})$.  We show below that this is indeed the
case: under such circumstances, Eq.~(\ref{exactwhite}) is fairly
accurate, and the resulting approximate ringdown SNR is
embodied in our approximate delta-function energy spectrum
(\ref{dEdfqnrapprox}) and
in Eqs.~(\ref{qnrsnr}) - (\ref{qnrsnr3}) of Appendix \ref{appsnr}
\cite{noteA}.

In many cases of interest, it will indeed be true that most of the SNR
squared for ringdown waves will be accumulated in the vicinity of the
resonance peak, so that the SNR will approximately be given by
Eq.~(\ref{exactwhite}).  However, this will not always be the case.
For instance, suppose that we were lucky enough that two $10^5
M_\odot$ black holes were to merge at the center of our own galaxy.
Would such an event be detectable by advanced LIGO interferometers?
Clearly, most of the power in the ringdown waves in this case would be
far below the LIGO/VIRGO waveband.  However, given that the merger is only
at $\sim 10 \, {\rm kpc}$, one might hope to be able to detect the tail of
the ringdown waves that extends upwards in frequency into the LIGO/VIRGO
waveband.  Or, consider the detectability of a ringdown of a nearby
$10^3 M_\odot$ black hole by LISA.  In this case most of the ringdown
power is concentrated at frequencies above LISA's optimum waveband, and
the detectability of the signal is determined by the amount of power
in the low frequency tail of the ringdown.  In such cases, it is clearly
necessary to understand the power carried in the ringdown waves at
frequencies far from the resonant frequency.

Normally, such an understanding is obtained simply by taking a Fourier
transform of the waveform $h(t)$.  In the case of ringdown waves from
BBH mergers, however, the waveform for $t<0$ is the unknown, merger
waveform.  In order to obtain the SNR squared from the ringdown signal
alone, one might guess that the appropriate thing to do is to
take $h(t)=0$ for $t<0$, and insert this together
with Eq.~(\ref{ringdownbasic}) into the standard formula (\ref{snr})
for the signal to noise squared.  However, the resulting energy
spectrum has unrealistic high frequency behavior due to a
discontinuity in $h(t)$ at $t=0$ (or a discontinuity in $h^\prime(t)$
at $t=0$ in the case $\psi_0=\pi/2$), and the resulting SNRs can
in some cases differ from the correct values (see below) by
factors $\agt 10$.  Other choices for $h(t)$ for $t<0$ [for instance
$h(t)=h(0)$] get around this problem but instead have unrealistic low
frequency behavior.  In any case, it is clear that these choices are
somewhat {\it ad hoc} and that there should be some more fundamental,
unique way to calculate the SNR.

We now explain the correct method to calculate the SNR.  The question
that effectively is being asked is: What is the probability that there is a
ringdown waveform present in the data stream, starting at (say) $t=0$?
This probability is to be calculated given only the data from $t>0$,
without using the measured data from $t<0$ which is contaminated by
the unknown merger waveform.  To do this one must effectively
integrate over all possible realizations of the noise for $t<0$.
The necessity for such an integration is illustrated by the following
simple analogy.  Suppose that one
is measuring two real variables, $h_+$ (``waveform for positive $t$'') and
$h_-$ (``waveform for negative $t$''), and that the measured values of
these variables are ${\bar h}_+$ and ${\bar h}_-$.  Suppose that because of the
noise in the measurement process, the probability distribution for the
true values of these parameters given their measured values is
\begin{eqnarray}
p(h_+,h_-) &=& {1 \over 2 \pi \sigma^2} \exp\bigg\{ -{1 \over 2 \sigma^2}
\big[ (h_+ - {\bar h}_+)^2
\nonumber \\
\mbox{} && + (h_- - {\bar h}_-)^2 +
2 \varepsilon (h_+
- {\bar h}_+) (h_- - {\bar h}_-) \big] \bigg\}.
\label{pbasic}
\end{eqnarray}
Thus, $h_+$ and $h_-$ are Gaussian distributed about their means
${\bar h}_+$ and ${\bar h}_-$, and they are correlated.
If we assume that $h_-=0$ [analogous to assuming $h(t) =0$ for $t<0$],
we obtain for the probability distribution for $h_+$
\begin{equation}
p(h_+ | h_-=0) = {1 \over \sqrt{2 \pi} \sigma} e^{-(h_+ - {\bar
h}_+^\prime)^2 / (2 \sigma)^2},
\label{pbad}
\end{equation}
where ${\bar h}_+^\prime = {\bar h}_+ - \varepsilon {\bar h}_-$.
By contrast, if we instead calculate the probability distribution for
$h_+$ alone by integrating over $h_-$ we find
\begin{equation}
p(h_+) = {1 \over \sqrt{2 \pi} \sigma^*} e^{-(h_+ - {\bar
h}_+)^2 / (2 \sigma^*)^2},
\label{pgood}
\end{equation}
where $\sigma^* = \sigma / \sqrt{1-\varepsilon^2}$.  It is clear in
this simple example that one should use the reduced distribution
(\ref{pgood}) rather than the distribution (\ref{pbad}).
Note also that the widths of the probability distributions (\ref{pbad}) and
(\ref{pgood}) are different, and that the correct distribution
(\ref{pgood}) could not have been obtained from the joint distribution
(\ref{pbasic}) for any assumed choice of $h_-$.

Turn now to the analogous situation for random processes.  If $n(t)$
is the interferometer noise, let $C_n(\Delta t) \equiv \langle n(t)
n(t +\Delta t) \rangle$ denote the autocorrelation function.
Define the inner product
\begin{equation}
\langle h_1 | \, h_2 \rangle \equiv
\int_0^\infty dt \int_0^\infty dt^\prime \,K(t,t^\prime) \,h_1(t)
h_2(t^\prime)
\label{innerhalfline}
\end{equation}
on the space of functions $h(t)$ for $t>0$,
where the kernel $K(t,t^\prime)$ is determined from
\begin{equation}
\int_0^\infty dt^{\prime\prime} K(t,t^{\prime\prime})
C_n(t^{\prime\prime} - t^\prime) = \delta(t - t^\prime)
\end{equation}
for $t,t^\prime \ge 0$.  The quantity
$K(t,t^\prime)$ is analogous to the modified width $\sigma^*$ in
Eq.~(\ref{pgood}) above.  Using the inner product
(\ref{innerhalfline}), the usual theory of optimal filtering
\cite{wainandzub,cutlerflan1} can be applied to random processes on
the half line $t>0$.
Thus, if $s(t)$ is the interferometer output and $h(t)$ is the
waveform (\ref{ringdownbasic}), the detection statistic is $Y = \langle s|
h \rangle$, and the SNR squared for the measurement is
\begin{eqnarray}
\rho^2 & \equiv&  {E[Y]^2 \over E[Y^2] - E[Y]^2} \nonumber \\
\mbox{} &=& \langle h | h \rangle \nonumber \\
\mbox{} &=& \int_0^\infty dt \int_0^\infty dt^\prime K(t,t^\prime) h(t)
h(t^\prime),
\label{fullans}
\end{eqnarray}
where $E[...]$ means expectation value.  If we define
\begin{equation}
G(f,f^\prime) = \int_0^\infty dt \, \int_0^\infty dt^\prime \, e^{2
\pi i f t} \, e^{-2 \pi i f^\prime t^\prime}  K(t,t^\prime)
\end{equation}
and
\begin{equation}
{\tilde h}(f) = \int_0^{\infty} e^{2 \pi i f t}\, h(t) \, dt,
\end{equation}
the SNR squared can be rewritten as
\begin{equation}
\rho^2 = \int_{-\infty}^\infty df \, \int_{-\infty}^\infty df^\prime \,
{\tilde h}(f)^* \, G(f,f^\prime) {\tilde h}(f^\prime).
\label{finalans}
\end{equation}
Note that the Fourier transform $G(f,f^\prime)$ of $K(t,t^\prime)$ is
not proportional to $\delta(f - f^\prime) / S_h(f)$ but instead is
in general {\it non-diagonal} in frequency.

The formula (\ref{finalans}) resolves the ambiguities discussed above
in the method of calculating the ringdown SNR; the result does not
require a choice of the waveform $h(t)$ for $t<0$.
Unfortunately, the final answer (\ref{finalans}) is complicated in the sense
that it cannot be expressed in the form (\ref{snraveraged}) for any effective
energy spectrum $dE/df$.  This is somewhat inconvenient for the
purposes of this paper: the wave's energy spectrum is a useful and key
tool for visualizing and understanding the SNRs.  Clearly, an
approximate, effective energy spectrum (to the extent that one exists)
would be very useful.  We now turn to a derivation of
such an approximate, effective energy spectrum, namely the spectrum
(\ref{dEdfqnr}) which is used throughout the body of this paper.

We start our derivation by describing
an alternative method of calculating the exact ringdown SNR given by
Eqs.~(\ref{ringdownbasic}) and (\ref{fullans}).  It is straightforward
to show that the quantity
(\ref{fullans}) can be obtained by (i)  choosing {\it
any} waveform $h(t)$ for $t<0$,  (ii) calculating the
SNR from the usual formula (\ref{snr}), and (iii) minimizing over all
choices of the function $h(t)$ on the negative real axis.
We have experimented with several choices of $h(-t)$ for $t>0$, namely
$h(-t) = 0$, $h(-t) = h(0)$, $h(-t) = h(t)$.  We found that the SNR
obtained by minimizing over these choices is always (for the entire
black hole mass ranges discussed in Sec.~\ref{snrsection})
within a few tens of percent of the SNR obtained from the following
prescription: (i) Assume that $h(t)$ for negative $t$ is identical to
the waveform for positive $t$ except for the sign of $t/\tau$; {\it i.e.},
that
\begin{equation}
h(t) = h_0 \cos(2 \pi f_{\rm qnr} t + \psi_0) e^{-|t|/\tau}
\label{ringdownansatz}
\end{equation}
for positive and negative $t$.  (ii) Calculate the
total SNR using the standard formula (\ref{snr}). (iii) Divide by
a correction factor of $\sqrt{2}$ in amplitude to compensate for the
doubling up.  This prescription gives the correct, exact result
(\ref{fullans}) for white noise.  For more realistic noise curves,
the errors of a few tens of percent resulting from this prescription are
unimportant compared to the uncertainty in the overall amplitude ${\cal A}$
of the ringdown signal.  Moreover, the resulting SNR values multiplied by
$\sqrt{2}$ are an upper bound for the true SNR (since if our {\it ad hoc}
choice of $h(t)$ for $t < 0$ happened to be exactly right, then the
prescription would underestimate the SNR by $\sqrt{2}$).

We now explain how to obtain the energy spectrum (\ref{dEdfqnr}) from the
above approximate prescription.  From Eqs.~(\ref{decompos}) and
(\ref{qnrwaveform}) it can be seen that
the waveform as seen in one interferometer, before angle averaging, is
given by Eq.~(\ref{ringdownbasic}) with
\begin{equation}
h_0 e^{i \psi_0} = {{\cal A} M \over r} \left[F_+(\theta,\varphi,\psi) + i
F_\times(\theta,\varphi,\psi)\right] \, {_2S^2_2(\iota,\beta,a)} e^{i
\varphi_0}.
\label{ringdownconsts}
\end{equation}
Here the angles $\theta$, $\varphi$, $\psi$, $\iota$ and $\beta$ have the
meanings explained in Sec.~\ref{derivesnrformula}.  Let us now insert
the waveform (\ref{ringdownbasic}) into the formula (\ref{fullans}) for
the exact SNR, and then average over the angles $\theta$, $\varphi$,
$\psi$, $\iota$ and $\beta$ using Eqs.~(\ref{angleaveragedefine}) and
(\ref{spheroidal}). This yields for the angle-averaged, exact SNR squared
\begin{equation}
\langle \rho_{\rm exact}[h(t)]^2 \rangle = {1 \over 20 \pi} \left[
\rho_{\rm exact}[h_{+,0}(t)]^2 + \rho_{\rm
exact}[h_{\times,0}(t)]^2 \right],
\label{angleaveragedexact}
\end{equation}
where $\rho_{\rm exact}[h(t)]$ denotes the exact SNR functional
(\ref{fullans}) and
\begin{eqnarray}
h_{+,0}(t) &=& {{\cal A} M \over r} \, \cos(2 \pi f_{\rm qnr} t) \,
e^{-t/\tau} \nonumber \\
\mbox{}
h_{\times,0}(t) &=& {{\cal A} M \over r} \, \sin(2 \pi f_{\rm qnr} t) \,
e^{-t/\tau},
\end{eqnarray}
for $t>0$.  Now, for each of the two terms on the right hand side of
Eq.~(\ref{angleaveragedexact}), we make the approximation discussed
above consisting of using Eqs.~(\ref{snr}) and (\ref{ringdownansatz})
and dividing by 2.  This yields
\begin{equation}
\langle \rho_{\rm exact}[h(t)]^2 \rangle \approx {1 \over 10 \pi}
\int_0^\infty df \, { \left[ | {\tilde
h}_{+,0}(f)|^2 + | {\tilde h}_{+,0}(f)|^2 \right]
\over S_h(f)},
\end{equation}
where it is understood that $h_{+,0}$ and $h_{\times,0}$ have been
extended to negative $t$ in the manner of Eq.~(\ref{ringdownansatz}).
Finally, evaluating the Fourier transforms yields an angle averaged
SNR squared of the form (\ref{snraveraged}), with energy spectrum
given by Eq.~(\ref{dEdfqnr}).



\begin{references}
\bibitem{ordercount} By  post-$n$-Newtonian order, we mean
	$O[(v/c)^{2 n}]$ in secular terms and $O[(v/c)^{5+2 n}]$ in
	dissipative, radiation reaction terms ($2 n$ orders beyond the
	leading quadrupole radiation reaction term).  This is the
	standard counting method for post-Newtonian waveforms.
\bibitem{IBBH1} Even if such methods are not found, post-Newtonian templates
	should be adequate for searching for inspiral waves even in the IBBH
	regime; IBBH templates will likely be needed only as a
	foundation for extracting the waves' information.  We also
	note that the method of Pad\'e
	approximants has recently been found to improve significantly the
	convergence of the post-Newtonian series \cite{Sathya}; the
	implications of this for the IBBH regime are not yet clear.
\bibitem{notsoclean} One can convince oneself of the likelihood of
	this cleanliness by considering various possible sources of
	perturbations to the gravitational-wave signal.  Accretion disks will
	have masses much smaller than the black hole masses (for the equal
	mass coalescences that we are considering); and moreover for solar
	mass BBHs any accretion disks from previous phases in the
	binary's evolution will be long gone before the black holes coalesce.
	However, one case where a
	possible perturbing effect on the gravitational-wave signal may be
	observable is the stochastic influence of a surrounding stellar
	cluster on the orbits of members of a SMBH binary during the early
	stages of the inspiral.  While strictly speaking this could make our
	assumption of the applicability of Wiener optimal filtering invalid
	for some SMBH inspirals, such inapplicability is not a cause
	for concern: the SNRs are typically so high anyway that it
	should be possible to easily detect both the coherent portions of the
	inspiral radiation and any isolated changes in phase due to
	close-encounter perturbations.
\bibitem{twosided} Our convention for the normalization of the
	spectral noise density $S_h(f)$ is such that
	$$
	\langle n(t)^2 \rangle = \int_0^\infty \,S_h(f) \, df,
	$$
	where $n(t)$ is the detector noise (``One-sided spectral noise
	density'').  This determines the constant
	$4$ appearing in Eq.~(\ref{snr}).  Note that the alternative
	convention used in Refs.~\cite{helstrom,wainandzub} results in
	a factor $2$ instead of $4$ in Eq.~(\ref{snr}).  In Eq.~(29)
	of Ref.~\cite{300yrs}, the spectral noise density that is used
	is the one-sided spectral noise density (as in this paper),
	and thus the overall factor in front of the integral should be $4$,
	not $2$.
\bibitem{importantnote} Note that we distinguish between the number of
	statistically independent waveform shapes ${\cal N}_{\rm shapes}$
	in the manifold of waveforms that one is trying to detect
	using the matched filtering process, and the actual number of
	templates ${\cal N}_{\rm templates}$ used in the computations.
	The number ${\cal N}_{\rm shapes}$ is the volume of the subset
	of the manifold of signals with SNR $\le \rho$, in the natural
	metric induced by the statistical properties of the noise; thus ${\cal
	N}_{\rm shapes}$ depends on $\rho$.  The number ${\cal N}_{\rm
	templates}$ is defined in Sec.~\ref{qnrsearches}.  In general
	${\cal N}_{\rm templates} < {\cal N}_{\rm shapes}$ since
	the optimum computational method will have less than one template per
	statistically independent waveform shape.  For the purposes
	of calculating SNR thresholds, $\ln({\cal N}_{\rm templates})
	\approx \ln({\cal N}_{\rm shapes})$ to a good approximation.
\bibitem{explainorb} This is the value predicted by the numerical
	initial data sets of Ref.~\cite{cook} at $f = f_{\rm merge} = 0.02/M$;
	see also the arguments concerning angular momentum in
	Sec.~\ref{mergerphase}.
\bibitem{rangenote} When we say that a source is visible out to a
	distance $D$, we mean that the rate of detection of events is
	the roughly the same
	as the rate of occurrence of events within a sphere of radius $D$.
	Note that some events within this sphere will be missed, and some
	outside this sphere will be detected, due to the beaming and
	orientation effects discussed in Sec.~\ref{derivesnrformula}.
	Rare, optimally oriented sources will be visible out to
	several times $D$.
\bibitem{freqtime} During the inspiral phase, the timescale over which
	the frequency of the waves changes is much longer than the
	instantaneous period, so it makes sense to talk about the frequency of
	the waves at a certain time.  In particular, we can (roughly) talk
	either about the time or the frequency at which the inspiral ends ---
	the notion of instantaneous frequency is just beginning to break down
	then.
\bibitem{noteguage} Actually the QNR metric perturbation will vanish
	at the event horizon; one could instead use the boundary of the
	ergo-region.   Also the ratio ${\cal R}$ of the circumferences
	will actually be gauge dependent.  However, if we expand ${\cal R}$ as
	${\cal R} = 1 + \alpha {\cal A} + O({\cal A}^2)$, then the coefficient
	$\alpha$ will be gauge invariant and we could take ${\cal A}_2 =
	1/\alpha$.
\bibitem{Aestimate} More specifically, an ellipsoidal solid body of
	mass $M$ of uniform density whose axes are given by $l^2, l^2 (1
	\pm (3/5) \sin(\omega t))$ gives rise to a luminosity $L=9 M^2
	l^4 \omega^6 / 5^5$; comparing this to the luminosity ${\cal
	A}^2 M^2 \omega_{\rm qnr}^2 / (16 \pi)$ for the waveform
	(\ref{qnrwaveform}) at 	$t=0$, and identifying 	$\omega =
	\omega_{\rm qnr} = 2 \pi\,0.13 / M$ and taking $l =
	r_+ = M + \sqrt{M^2 - a^2}$ for $a = 0.97 M$ say yields
	${\cal A} = 0.4$.  This crude estimation method clearly does
	not take into account the potential barrier through which the
	radiation must propagate (among other things).  We imagine that
	a more precise calculation will find a somewhat smaller value
	of ${\cal A}$.
\bibitem{hnexplain} See Eq.~(32) of Ref.~\cite{300yrs}.  Note also that our
	$h_n(f)$ is similar to the
	quantity $h_{\rm 3/yr}$ defined in Ref.~\cite{300yrs}, which
	is the same as the
	quantity denoted $h_{\rm SB}(f)$ in Refs.~\cite{ligoscience,Kipreview}.
	These quantities differ from $h_n(f)$ by the
	factor $\sqrt{\ln {\cal N}} \sim 5$, where ${\cal N}$ is the
	number of independent starting times per third of a year that
	must be tried
	in the optimal filtering search process; see Eq. (34) of
	Ref.~\cite{300yrs}.  We omit this factor in our definition of
	$h_n$ since in paper II \cite{paperII} we will be discussing
	SNRs in situations when the waves have already been detected
	and where the factor $\sqrt{\ln {\cal N}}$ is not relevant.
	The factor instead appears in our SNR thresholds
	discussed in Sec.~\ref{detection}.
	Note also that our signal-to-noise ratios differ from those in
	Refs.~\cite{ligoscience,Kipreview} by an overall factor of
	$\sqrt{3/2}$, as discussed after Eq.~(\ref{snraveraged}) above.
\bibitem{details} We obtained the metric (\ref{metric}) in the
	following way:  (i) We
	took the large $Q$ limit of the matrix ${\cal C}^{\prime \, -1}_{ij}$
	of Ref.~\cite{finnmeasure}, which was calculated assuming white noise,
	(ii) We inverted this matrix, took a $2
	\times 2$ sub-block and re-inverted, thus eliminating (effectively
	integrating out) the variables $V$ and $T$.  This procedure for
	eliminating the time-of-arrival variable $T$ is equivalent to
	the procedure embodied in Eq.~(2.28) of Ref.~\cite{owen}.  (iii) We
	divided by 2 in order to accord with the definition (2.12) of
	Ref.~\cite{owen}.
\bibitem{notedavid} In the situation where merger templates are
	available, the optimum search method would in fact be a combination
	of inspiral, merger and ringdown searches, in which the entire
	inspiral--merger--ringdown signal as a whole is searched for using
	matched filtering.
\bibitem{compare} The small $z$, equal-mass limit of
	Eq.~(\ref{qnrsnrII}) agrees
	with the large $Q$ limit of Eq.~(4.7) of
	Ref.~\cite{finnmeasure} when that equation is multiplied by a
	correction factor of 2.
\bibitem{noteA} To see that angle-averaging the approximate formula
	(\ref{exactwhite})
	yields the formula (\ref{qnrsnr}) for the ringdown SNR, corresponding
	to the approximate delta function energy spectrum
	(\ref{dEdfqnrapprox}), proceed as follows.  Starting from
	Eq.~(\ref{ringdownconsts}), average over the angles
	$\theta$,$\varphi$,$\psi$,$\iota$ and $\beta$ using
	Eqs.~(\ref{angleaveragedefine}) and (\ref{spheroidal}).  This
	yields
$$
\langle h_0(\theta,\varphi,\psi,\iota,\beta)^2 \rangle = {1 \over 10
\pi} {{\cal A}^2 M^2 \over r^2}.
$$
	Substitute this into Eq.~(\ref{exactwhite}), make the replacement
	$S_h \to S_h(f_{\rm qnr})$, and make the
	appropriate substitutions to account for cosmological effects as
	explained in Sec.~\ref{derivesnrformula} to obtain Eq.~(\ref{qnrsnr}).
\end{references}
\end{document}